%% file: main.tex
\documentclass[lettersize, onecolumn, draftclsnofoot]{IEEEtran}

\usepackage{amsfonts}
\usepackage{algorithmic}
\usepackage{algorithm}
\usepackage{array}
\usepackage{textcomp}
\usepackage{stfloats}
\usepackage{verbatim}
\usepackage{graphicx}
\usepackage{jdhmacros}
\usepackage{comment}

\usepackage{enumitem}
\usepackage{varwidth}
\usepackage{graphbox}
\usepackage{wrapfig}

\newtheorem{theorem}{Theorem}[section]
\newtheorem{lemma}{Lemma}
\newtheorem{corollary}{Corollary}[lemma]
\newtheorem{claim}{Claim}
\newtheorem{remark}{Remark}

\renewcommand\qedsymbol{$\blacksquare$}

\newcommand{\RNum}[1]{\uppercase\expandafter{\romannumeral #1\relax}}

\newcommand{\MSE}{\textup{MSE}}
\newcommand{\HCR}{\textup{HCR}}
\newcommand{\ML}{\textup{ML}}
\newcommand{\CR}{\textup{CR}}

\usepackage[noabbrev]{cleveref}
\def\BibTeX{{\rm B\kern-.05em{\sc i\kern-.025em b}\kern-.08em
    T\kern-.1667em\lower.7ex\hbox{E}\kern-.125emX}}

\begin{document}

\title{
	\huge A Renderer-Enabled Framework for Computing Parameter Estimation Lower Bounds in Plenoptic Imaging Systems
}

\author{Abhinav V. Sambasivan, Liam J. Coulter \emph{Student Member, IEEE}, Richard G. Paxman, and Jarvis D. Haupt, \emph{Senior Member, IEEE}
\thanks{This paper was presented in part at the 2019 53rd Asilomar Conference on Signals, Systems, and Computers. Abhinav Sambasivan's contributions to this work took place while he was a student at the University of Minnesota.  Liam Coulter and Jarvis Haupt are with the Department of Electrical and Computer Engineering at the University of Minnesota -- Twin Cities; Liam Coulter is also with RTX BBN Technologies; Abhinav Sambasivan is with Shopify; Richard Paxman is with ARKA Group, LP - Ypsilanti, MI. Emails: {abhinav.vs@gmail.com; \{coult099, jdhaupt\}@umn.edu; RPaxman@arka.org}. The authors graciously acknowledge support from the DARPA REVEAL program, Contract No. HR0011-16-C-0024, and the Raytheon Employee Scholar Program. The authors are also grateful to the Minnesota Supercomputing Institute (MSI) at the University of Minnesota for providing computational resources for rendering some of the scenes used in Section \ref{subsec:error_analysis_results}. This work has been submitted to the IEEE for possible publication. Copyright may be transferred without notice, after which this version may no longer be accessible.
}}
\markboth{}
{}

\maketitle

\begin{abstract}
\input{abstract.tex}

\end{abstract}


\input{intro.tex}

\input{problem_statement.tex}

\input{rec_lb.tex}

\input{inexact_renderings.tex}

\input{conclusion.tex}

\input{acknowledgment.tex}

\input{appendix.tex}



\bibliographystyle{IEEEbib}
\bibliography{NLOS_lb_ref}

\end{document}

%% file: abstract.tex
This work focuses on assessing the information-theoretic limits of scene parameter estimation in plenoptic imaging systems.
A general framework to compute lower bounds on the parameter estimation error from noisy plenoptic observations is presented, with a particular focus on passive indirect imaging problems, where the observations do not contain line-of-sight information about the parameter(s) of interest. Using computer graphics rendering software to synthesize the often-complicated dependence among parameter(s) of interest and observations, i.e. the forward model, the proposed framework evaluates the Hammersley-Chapman-Robbins bound to establish lower bounds on the variance of any unbiased estimator of the unknown parameters. The effects of inexact rendering of the true forward model on the computed lower bounds are also analyzed, both theoretically and via simulations. Experimental evaluations compare the computed lower bounds with the performance of the Maximum Likelihood Estimator on a canonical object localization problem, showing that the lower bounds computed via the framework proposed here are indicative of the true underlying fundamental limits in several nominally representative scenarios. 

%% file: intro.tex
\section{Introduction}\label{sec:intro:HCR}
Conventional imaging systems are modeled after human vision and provide information about a scene via a two-dimensional image, where each pixel is described by a tristimulus value (e.g., RGB). The \emph{plenoptic function}, in contrast, captures the intensity of light at every location in space as a function of a number of latent parameters including, for example, all possible incidence angles, wavelengths, and time instances, providing much more information about a given scene than a conventional camera \cite{adelson_plenoptic}. Formally, a plenoptic function in this case could take the form
\begin{equation*} 
	\Lb = \Lb(\rb,\bvarphi,\nu,t),
\end{equation*}
where $\rb\in\RR^3$ refers to the location at which the plenoptic observation is made, $\bvarphi\in[0,2\pi)\times [0,\pi)$ denotes the angle pair (direction of arrival of light) in polar coordinates,  $\nu$ is the wavelength, and $t$ is the time index. 
Plenoptic imaging has been used for a wide range of applications including stereoscopy \cite{adelson_stereo, levoy1996light}, microscopy \cite{microscopy_prevedel,levoy_microscopy}, and non-line-of-sight (NLOS) imaging \cite{sasaki2018light,corner_cam,light_field_from_shadows,yedidia2019using}.

While plenoptic imaging systems have a variety of applications in computer vision and image processing, we focus here on NLOS or indirect imaging. NLOS imaging corresponds to the scenario where the scene of interest is hidden from the observer (or imaging system); our overarching aim here is to understand how well one can recover parameters of the hidden scene from indirect reflections off various surfaces in the direct LOS of the observer (or imaging system). The parameters of interest in these problems could be as simple and high-level as the location of hidden objects (NLOS object tracking) or as complex as the entire hidden scene (NLOS image reconstruction). NLOS imaging, or colloquially ``imaging around corners," has numerous potential applications in defense, autonomous navigation, and imaging inaccessible tissues in endoscopy for better diagnostics, to name a few. 

More formally, we consider the problem of determining the fundamental estimation performance limits of parameter estimation from noisy plenoptic observations in NLOS imaging tasks from an estimation-theoretic perspective. The primary difficulty in establishing lower bounds for plenoptic imaging problems lies in the complexity of the forward imaging model, which codifies the functional dependence of the nominal (noiseless) observations on the scene parameters. In this paper, we build on our previous work \cite{avs2019asilomar}, where we first proposed combining analytical tools from classical estimation theory \cite{cramer_original, rao_original, hammersley, chapman_robbins} with computer graphics rendering engines \cite{Mitsuba,redner} that help us simulate the often complicated forward model, in order to compute lower bounds on the estimability of scene parameters from noisy plenoptic observations.
In doing so, we can provide a benchmark of optimality against which various estimation strategies could be compared, and also obtain useful insights on where information about the NLOS scene parameters is localized in a given set of observations. 

\subsection{Prior Art}\label{subsec:prior_art}
NLOS imaging methods can be broadly classified into two categories: (1) active imaging methods, and (2) passive imaging methods. Active imaging typically involves illuminating the hidden scene using external stimuli, e.g., pulsed lasers, LIDARs, etc., and using the information from the returning photons such as time-of-flight information and other transients, (sometimes measured by specialized detectors, e.g., single-photon avalanche diodes or SPADs), to reconstruct the shape and albedo of hidden objects. These methods were initially demonstrated in \cite{velten2012ToF_nlos,velten2013femto} and have recently gained attention \cite{arellano2017fast,thrampoulidis2018active,heide2019non,chen-steady-19,klein_tracking_2016,tsai_cvpr_2019,Pérez_Redo-Sanchez_2024,Yu-24,ye_pnp_2024}. However, active imaging can require the use of expensive and substantial hardware, such as femtosecond lasers and specialized detectors, which make it less amenable to field deployment in real-life applications; there are also cases for which the observer is conducting a stealth operation and active illumination can aid the target in detecting that they are being observed. 

In this work, we are particularly interested in passive imaging methods that rely on the intensity or radiance measurements obtained at the imaging device (by sources that are often not under the observer's control) to estimate the hidden scene parameters. The main challenge in passive NLOS imaging is extremely low signal-to-noise ratio (SNR): the indirect signal photons reflecting off diffuse (non-mirror-like) surfaces have very low signal strength compared to ambient lighting and direct illumination, often making the passive imaging problem ill-conditioned. In a pioneering work \cite{torralba2012accidental} demonstrated that the presence of occluders like sharp edges and corners facilitates the NLOS recovery problem. This led to many follow-up works that exploit occluders (and motion) in the hidden scene to perform NLOS imaging \cite{corner_cam,light_field_from_shadows,sasaki2018light,yedidia2019using,saunders2019computational,seidel2019corner, fadlullah_2024_icassp}. Other passive imaging methods use \emph{coherence-based} techniques for NLOS reconstruction \cite{katz2014coherence,smith2018coherence,batarseh2018coherence,viswanath2018coherence}. A comprehensive survey of various NLOS imaging methods can be found in the survey  \cite{nlos_survey} or the chapter  \cite{Ji2024}. 

More recently, \cite{murray2019occlusion,seidel2020two} proposed using the Cramer-Rao bound and Fisher information as a proxy to study the feasibility and conditioning of their NLOS problem, and \cite{czajkowski_two-edge-resolved_2024} used Fisher information to equalize a hidden scene discretization and enhance their reconstruction . Additionally, several authors have investigated the analysis-by-synthesis problem \cite{klein_tracking_2016,tsai_cvpr_2019,zhang_differential_2019}, making use of differentiable rendering engines to estimate scene parameters, while \cite{iseringhausen_non-line--sight_2020} examined errors introduced by approximations in the renderer. The aforementioned works have shown significant promise in ``imaging around corners," but a thorough information-theoretic treatment of the NLOS parameter estimation problem for more general and realistic scenes is still lacking.

\subsection{Our Contributions}\label{subsec:contribution}
We present a general-purpose framework for establishing fundamental limits for NLOS parameter estimation problems using noisy \emph{passive} plenoptic observations. 
In contrast to the aforementioned efforts, which develop simplified/approximate (linear) forward models for controlled experimental environments to study NLOS imaging, our framework can handle complicated (non-linear) forward models that describe realistic scenes. In \cite{avs2019asilomar}, we proposed using rendering engines to simulate the forward model, enabling  numerical evaluation of the Hammersley-Chapman-Robbins (HCR) bound to identify a lower bound on the variance of any unbiased estimator of the parameter(s) of interest. 
Our framework also enables us to localize the (Fisher) information content in the observations, which can be used to further validate some of the conclusions about the benefits of occluding objects for NLOS imaging \cite{corner_cam,yedidia2019using,saunders2019computational,seidel2019corner,seidel2020two,murray2019occlusion}. 

One important assumption made in our previous work \cite{avs2019asilomar} is that the rendering engine simulates the forward model \emph{exactly}, i.e., provides an \emph{error-free} version of the true plenoptic observations for a given set of scene parameter values. However, in practice this assumption might not hold as the rendering engine might yield a close though inexact estimate of the true plenoptic observations. In this work, we consider \emph{unbiased and progressive} rendering techniques, which produce unbiased estimates of the true plenoptic observations with continually decreasing error as we let the renderer run indefinitely (see documentation of {\tt Mitsuba} \cite{Mitsuba}).
We extend the efforts in our previous work \cite{avs2019asilomar} by analyzing effects of using such \emph{inexact renderings} in our lower bounding framework and provide a simple method to estimate intervals for the true HCR lower bounds. 

We instantiate the HCR lower bounds for Poisson noise and additive white Gaussian noise (AWGN) models, and demonstrate the utility of our framework using a few canonical estimation problems. Finally, we compare our lower bounds with the performance of Maximum Likelihood Estimators for the problem of object localization. 

This paper is organized as follows. In Section \ref{sec:problem_setup} we provide an overview of the \emph{rendering equation} \cite{kajiya_rendering_eq}, which explains the forward model for our settings, along with the formal statement of the problem we aim to solve; we conclude this section with a discussion of the image gradients used in our analyses.  In Section \ref{sec:exact_lower_bounds} we present our renderer-enabled lower bound computation framework (assuming the rendering engine is noise-free), and discuss the information-theoretic tools used. The second half of this section is dedicated to numerical experiments with a realistic scene that show the practical utility of these methods. In Section \ref{sec:rendering_errors} we extend the previous analysis to take into account the effect of rendering errors on the lower bound computation; we derive analytical results and show numerical experiments conducted with an idealized scene, to highlight the effect of rendering error on the bounds. We also compare our bounds for localizing an object with a maximum likelihood estimator (MLE).  We finish with some concluding remarks about future work in Section \ref{sec:conclusion}.

%% file: problem_statement.tex
\section{Preliminaries and Problem Statement}\label{sec:problem_setup}
Throughout this paper, $\Theta \subseteq \RR^d$ denotes the parameter class containing all possible values of the parameter of our interest. For a given scene with some unknown (deterministic) parameter ${\btheta^*} \in \Theta$, we denote the set of plenoptic samples (obtained by sampling the full plenoptic field) associated with the scene by, $\Lb_{\btheta^*} \triangleq \cbr{\Lb_{\btheta^*}(\bomega)}_{\bomega\in\Omega},$ 
where we introduce the shorthand $\bomega:= [\rb,\bvarphi,\nu,t],$ to denote the arguments of the plenoptic function, and $\Omega$ is the observation space, which is a subset of the domain over which the plenoptic function is defined.

\subsection{The Rendering Equation}\label{subsec:rendering}
Information-theoretic treatment of this estimation problem requires understanding the \emph{forward model} $\btheta^*\mapsto\Lb_{\btheta^*}$ that captures the functional dependence of the plenoptic observations on the scene parameters. This can be mathematically explained by the rendering equation \cite{kajiya_rendering_eq}, which models the behavior of light rays as they originate from a light source, bounce off the objects in the scene and ultimately reach the detector (e.g., an imaging device). This forward mapping is commonly referred to as ``light-transport" in the computer-graphics literature.

For fixed $(\nu,t)$, light \emph{incident} on a surface point $\rb$ of a scene, along direction $-\bvarphi_i$, denoted by $\Lb^{\rm in}_{\btheta^*}(\rb,-\bvarphi_i)$, is typically scattered along all directions, as shown in Figure~\subref*{fig:scattering}. The proportion of this light reflected along $\bvarphi_o$ is determined by a surface-dependent quantity known as the bi-directional reflectance distribution function (BRDF) $f(\rb,\cdot,\cdot)$. 

\begin{figure*}[t]
	\centering
	\subfloat[][]
	{
		\includegraphics[width=0.415\columnwidth]{figures/NLOS_lb/scattering.pdf}
		\label{fig:scattering} 	
	}
\hspace{0.01\columnwidth}
	\subfloat[][]
	{
		\includegraphics[width=0.415\columnwidth]{figures/NLOS_lb/rendering_eqn.pdf} 
		\label{fig:rendering_eqn}	
	}
	\caption{The rendering equation, explained graphically: (a) The proportion of incident light coming in from direction $\bvarphi_i$ that gets reflected along direction $\bvarphi_o$ is determined by the BRDF of the surface; (b) Light \emph{incident} on a surface point $\rb$, can be seen as light \emph{leaving} from another point in the scene $g(\rb,\bvarphi_i),\ \Rightarrow\Lb^{\rm in}_{\btheta^*}(\rb,-\bvarphi_i) = \Lb_{\btheta^*}^{\rm out}(g(\rb,\bvarphi_i),-\bvarphi_i)$.}
\end{figure*}

The overall radiance \emph{leaving} a surface point $\rb$, along direction $\bvarphi_o$, denoted by $\Lb_{\btheta^*}^{\rm out}(\rb,\bvarphi_o)$, can hence be obtained by aggregating the contributions of $\Lb^{\rm in}_{\btheta^*}(\rb,-\bvarphi_i)$ from all possible incident directions $-\bvarphi_i$, and also any light emitted by the surface (if it is an emitter). Mathematically, 
\begin{eqnarray*}
    \Lb_{\btheta^*}^{\rm out}(\rb,\bvarphi_o) & = \ \Lb_{\btheta^*}^{\rm e}(\rb,\bvarphi_o) + \int_{S^2_+(\rb)}\hspace{-0.5em} \Lb_{\btheta^*}^{\rm in}(\rb,-\bvarphi_i) f(\rb,\bvarphi_o,\bvarphi_i) (\bvarphi_i \cdot \nbb) d\bvarphi_i, 
\end{eqnarray*}
where $\Lb_{\btheta^*}^{\rm e}(\rb,\bvarphi_o)$ is the emitted radiance at $\rb$ along the direction $\bvarphi_o$, $\nbb$ is the surface normal at $\rb$, and $S^2_+(\rb)$ is the unit hemisphere at $\rb$ containing all outgoing directions. 

From Figure \ref{fig:rendering_eqn}, we can see that $\Lb^{\rm in}_{\btheta^*}(\rb,-\bvarphi_i) = \Lb_{\btheta^*}^{\rm out}(g(\rb,\bvarphi_i),-\bvarphi_i),$ where $g(\rb,\bvarphi_i)$ is a scene geometry dependent operator that essentially finds the first surface point reached when traveling outward from $\rb$, along direction $\bvarphi_i$. Using this we can obtain the rendering equation as
\begin{eqnarray}\label{eq:rendering_eq}
    \Lb_{\btheta^*}^{\rm out}(\rb,\bvarphi_o) 
    & = \ \Lb_{\btheta^*}^{\rm e}(\rb,\bvarphi_o) +\int_{S^2_+(\rb)}\hspace{-0.8em} \Lb_{\btheta^*}^{\rm out}(g(\rb,\bvarphi_i),-\bvarphi_i) f(\rb,\bvarphi_0,\bvarphi_i) (\bvarphi_i \cdot \nbb) d\bvarphi_i,
\end{eqnarray}
where $\Lb^{\rm out}_{\btheta^*}$ is the plenoptic function of interest in the forward model alluded to above.

Equation (\ref{eq:rendering_eq}) is a Fredholm integral equation of the second kind, and is difficult to solve in closed form for all but simplest imaging settings \cite{polyanin_handbook_1998,arfken_integral_2005}. 
To overcome this hurdle, ray-tracing/rendering engines are typically utilized to approximately solve the integral equation in (\ref{eq:rendering_eq}) using Monte Carlo methods. Ray-tracing engines are widely used in the computer graphics community to generate photo-realistic images by tracing the path of light rays as they interact with objects in a scene. Here, we rely upon this (mature) technology for our purposes; for a given parameter value $\btheta^*$, we use ray-tracing packages to approximately solve (\ref{eq:rendering_eq}) and synthesize the plenoptic observations $\Lb_{\btheta^*}$.

\subsection{Formal Problem Statement}\label{subsec:formal_prob_statement}
Here we formalize the problem statement. We model the noisy observations $\Yb_{\bomega}$ as independent draws from a known class of probability distributions $p(\Yb_{\bomega};\btheta^*)$, whose parameters depend on $\btheta^*$ through the plenoptic function. Letting $\Yb_{\Omega} \triangleq \cbr{\Yb_{\bomega}}_{\bomega\in\Omega}$ be shorthand for the entire collection of noisy plenoptic observations, the likelihood of the observations can be written as
\begin{equation}\label{eq:likelihood}
 p(\Yb_\Omega;\btheta^*) = \prod_{{\bomega} \in\Omega} p(\Yb_{\bomega};\btheta^*) \triangleq \PP_{\btheta^*}.  
\end{equation}

Let $\cY$ denote the set of all possible observations for our problem. We are then interested in determining the fundamental limits of imaging a given scene with unknown parameter ${\btheta}^*\in\Theta$, from noisy plenoptic measurements $\Yb_\Omega \in\cY$.  Specifically, we seek to evaluate the performance of estimators, $\hat{{\btheta}}(\Yb_\Omega): \cY\rightarrow\Theta,$ which have a finite measure with respect to $\PP_{\btheta^*}$, as a function of the true underlying parameter of interest ${\btheta}^*$. The accuracy (or lack thereof) of an estimator $\hat{{\btheta}}$ in estimating the true scene parameter ${\btheta}^*$ can be measured in terms of the Mean Squared Error (MSE), 
\begin{equation}\label{eq:MSE}
	\MSE({\btheta}^*) = \EE \left[ \bignorm{\hat{{\btheta}}(\Yb_\Omega) - {\btheta}^*}^2_2 \right], 
\end{equation}
where the expectation is taken with respect to the randomness in the noisy observations $\Yb_\Omega$.

In order to derive meaningful \emph{local lower bounds} (bounds as a function of ${\btheta}^*$) on the MSE of estimators, we further need to make assumptions about the class of estimators. Here, we restrict our attention to the class of \emph{unbiased estimators}, for which the MSE reduces to variance (or trace of the covariance matrix, to be precise).  Hence we seek lower bounds on the variance $\Var(\btheta^*_j)$ of an individual parameter $\btheta^*_j$, or on $\sum_{j=1}^{J} \Var (\btheta^*_j) \triangleq \Var(\btheta^*)$.

\subsection{Preview: Differentiation Considerations}\label{subsec:fd_vs_ad}
Parts of the analysis in this work rely on accurate estimates of image gradient information with respect to scene parameters (such as an object's shape, size, color, etc.). In general for rendering tasks, the gradient can be estimated using finite differences (FD) or recently developed \textit{differentiable renderers} such as {\tt Redner} \cite{redner} or {\tt Mitsuba 3.0} \cite{jakob2022mitsuba3} that enable direct rendering of gradients (of some function defined on the scene) with respect to the scene parameters of interest. Differentiable rendering makes use of automatic differentiation (AD) computation graphs which break down complex interactions into constituent simple mathematical operations, making gradient estimation relatively straightforward. For this reason we refer to gradients produced using differentiable rendering as AD gradients. 

We chose to use FD gradients in this work primarily because FD gradients gave rise to fewer spurious artifacts in the gradient \textit{for the parameters of interest} we considered here.  In particular, in the evaluations that follow we consider the location and size of an object in the scene as the unknown parameters to be estimated. For these and other location-dependent parameters, the gradients in the image are produced by changing scene geometry, and are induced by visibility discontinuities. To see why this is the case, consider how changing a visible object's size will affect the rendered image: if we slightly increase the size of an object, the scene around the object's boundary will become invisible to the sensor. Likewise, new parts of the scene will become visible if we slightly decrease its size. 

Computing derivatives with respect to such parameters can be a difficult problem for differentiable renderers, which work by running full reverse Monte Carlo simulations to produce gradients. Similar to rendering in the forward direction, producing an AD gradient with a differentiable renderer involves tracing paths from light sources through the scene to the sensor, but these paths contribute to the gradient only if they are affected by the parameter with respect to which we are differentiating. If we wish to estimate a gradient with respect to a visible object's color parameters, for example, the space of paths that produce gradients can be comparatively large, since many rays from light sources in the scene will interact with a visible object's color. However, since we are differentiating here with respect to parameters whose gradients are induced by changing scene visibility, the space of paths that produce gradients in this case is extremely sparse (essentially consisting of rays which are affected by changing visibility along the object's boundary, a small subset of rays). Because of this, the AD gradient images for an object's size and location may contain unexpected noise characteristics and image artifacts. 

\begin{figure}[t]
    \centering
    \includegraphics[width=0.375\columnwidth]{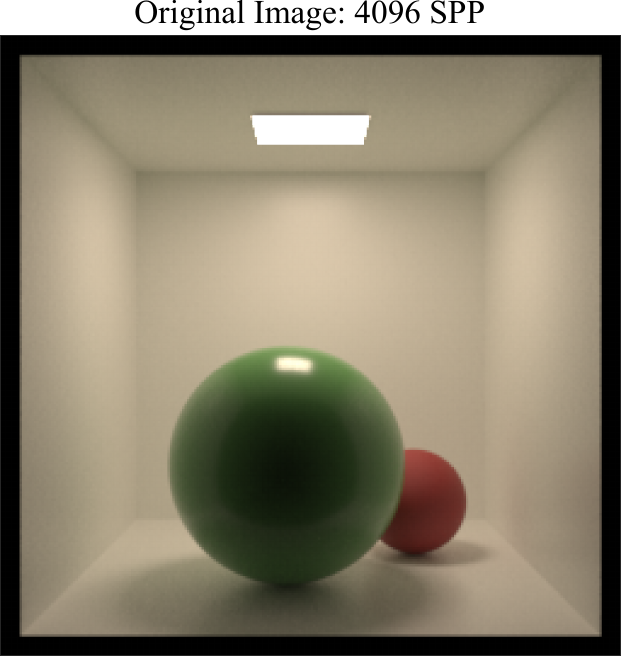}
    \caption{A modified version of the Cornell Box scene with white walls and a single overhead light emitting white light uniformly across all wavelengths, rendered using $2^{12} = 4096$ samples per pixel (SPP). The walls and both spheres use the Mitsuba BSDF plugin \texttt{roughplastic}; the walls have $\alpha=0.1$, the green sphere has $\alpha=0.05$, and the red sphere has $\alpha = 0.4862$.}
    \label{fig:cbox_original}
\end{figure}

To see an example of this, consider the scene in Figure~\ref{fig:cbox_original}, a modified version of the Cornell Box scene. Here, the reflectance properties of the walls and spheres in the scene are modeled using the Mitsuba plugin \texttt{roughplastic} which makes use of a rough Fresnel transmittance function to model a roughly dielectric material with internal scattering \cite{Mitsuba, jakob2022mitsuba3, walter_microfacet_2007}. For this plugin, a parameter $\alpha$ specifies the roughness of the surface micro-geometry along the tangent and bitangent directions \cite{jakob2022mitsuba3}: as $\alpha$ approaches 0, the surface gets smoother and reflections more specular; as $\alpha$ increases, the surface becomes rougher and reflections more diffuse.

\begin{figure}[t]
    \centering
	\subfloat[][]
	{
		\includegraphics[width=0.15\columnwidth]{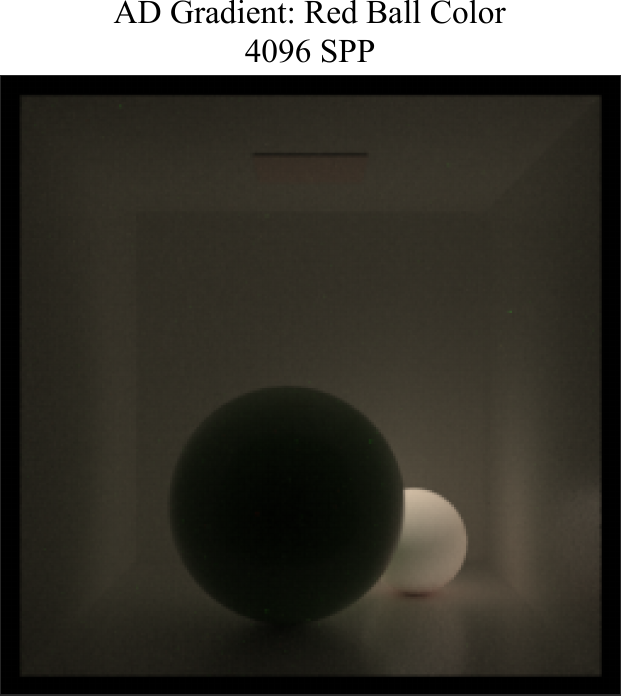}
		\label{fig:ad_color_4096}
	}
	\subfloat[][]
	{
		\includegraphics[width=0.15\columnwidth]{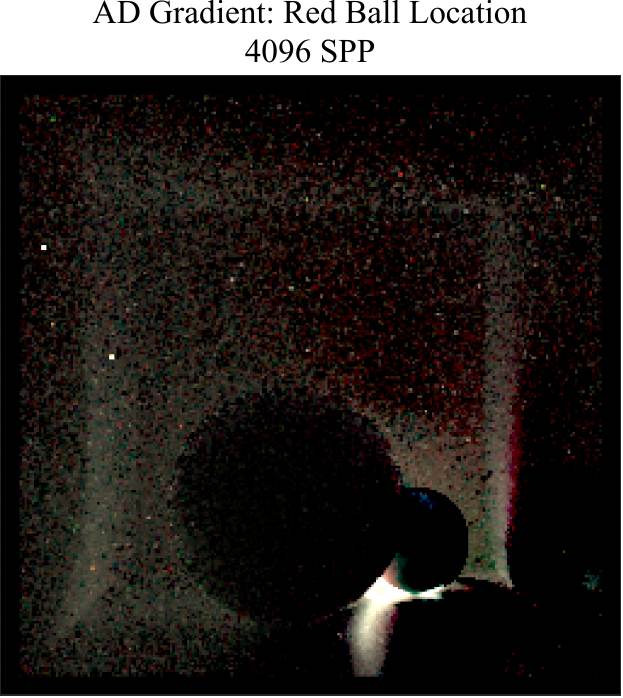}
		\label{fig:ad_location_4096}
	}
	\subfloat[][]
	{
		\includegraphics[width=0.15\columnwidth]{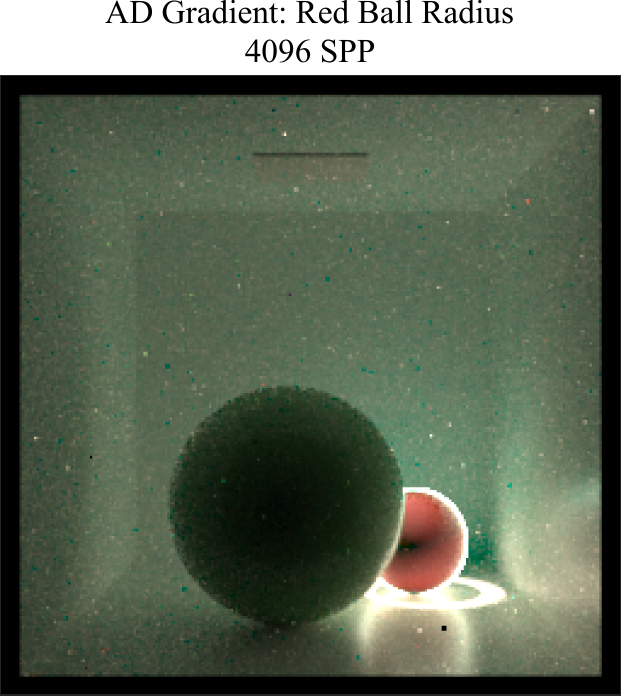}
		\label{fig:ad_size_4096}
	} 
    \subfloat[][]
	{
		\includegraphics[width=0.15\columnwidth]{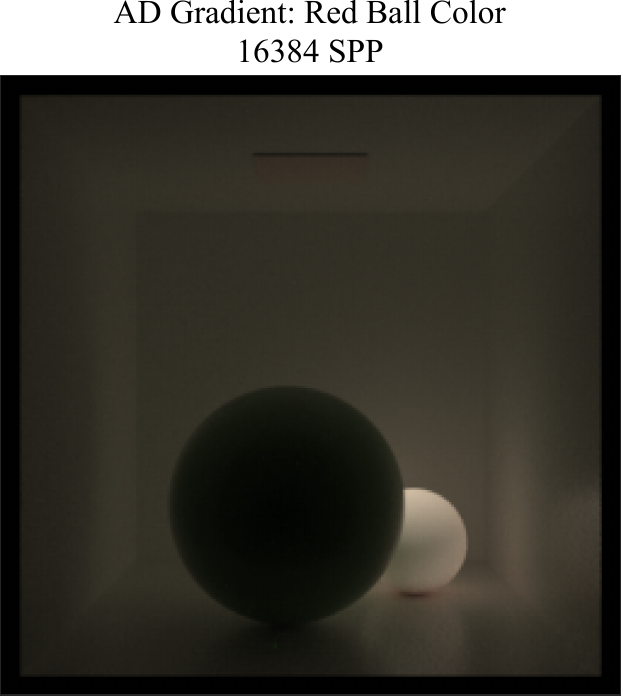}
		\label{fig:ad_color_16384}
	}
	\subfloat[][]
	{
		\includegraphics[width=0.15\columnwidth]{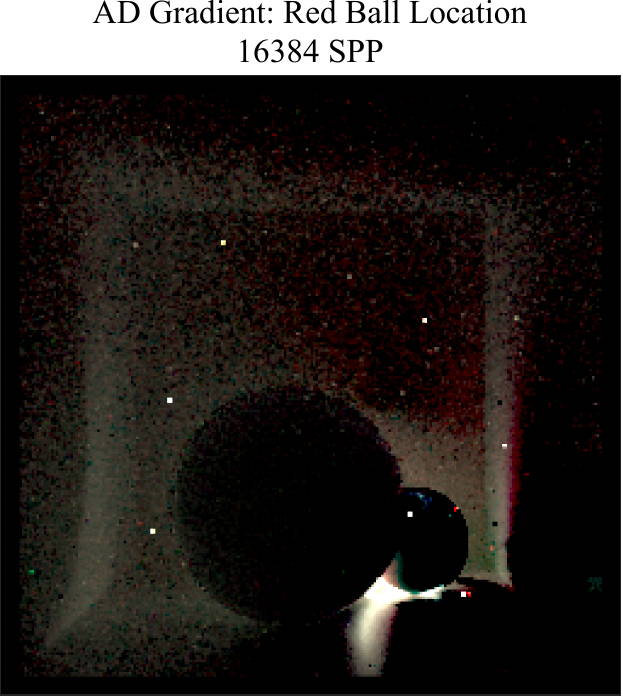}
		\label{fig:ad_location_16384}
	}
	\subfloat[][]
	{
		\includegraphics[width=0.15\columnwidth]{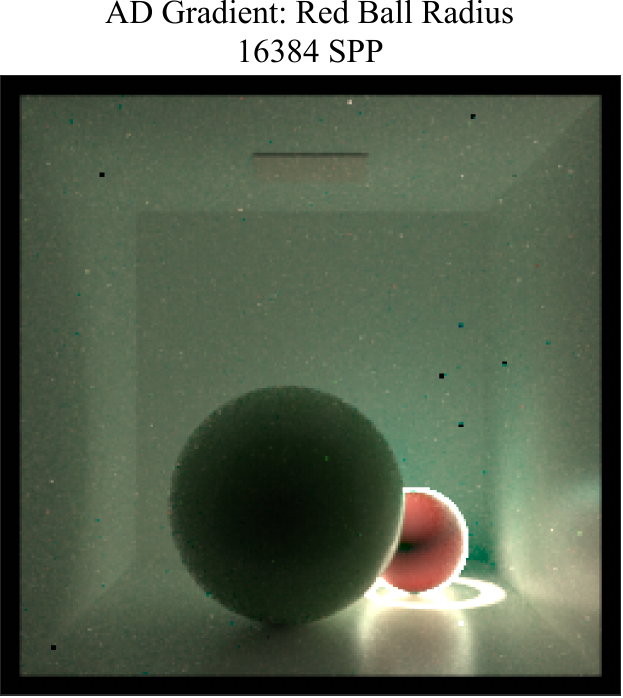}
		\label{fig:ad_size_16384}
	} \\
	\subfloat[][]
	{
		\includegraphics[width=0.15\columnwidth]{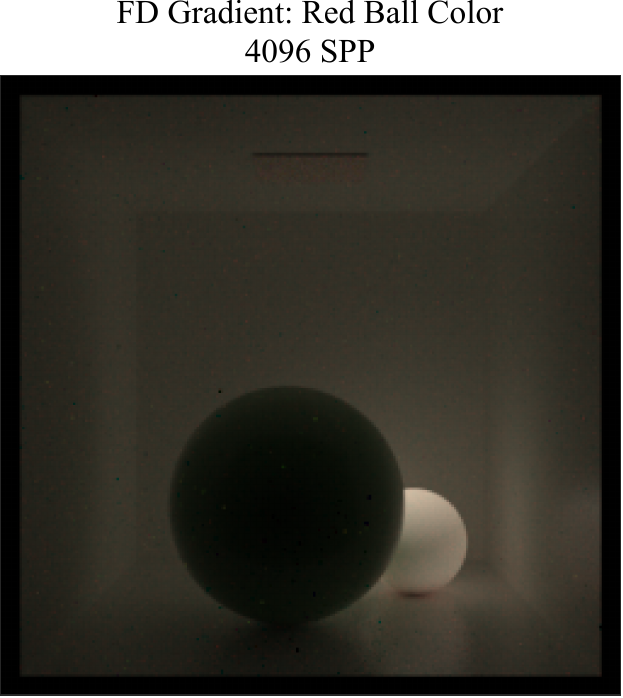}
		\label{fig:fd_color_4096}
	}
	\subfloat[][]
	{
		\includegraphics[width=0.15\columnwidth]{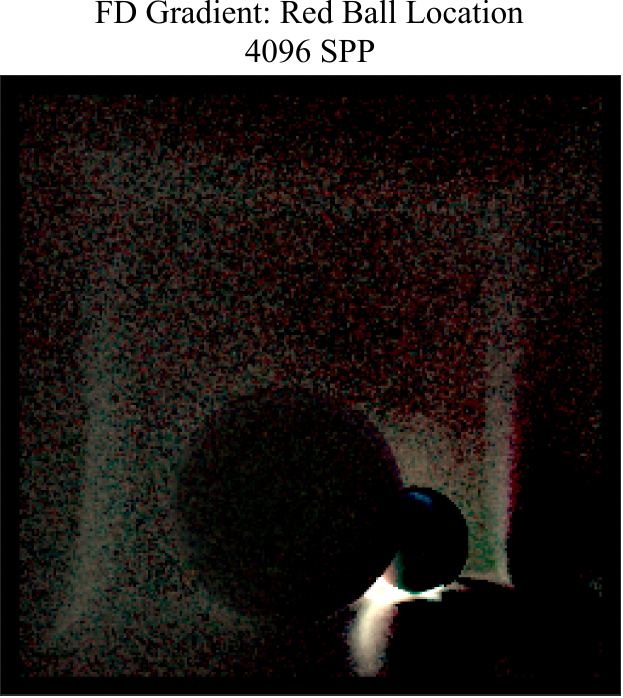}
		\label{fig:fd_location_4096}
	}
	\subfloat[][]
	{
		\includegraphics[width=0.15\columnwidth]{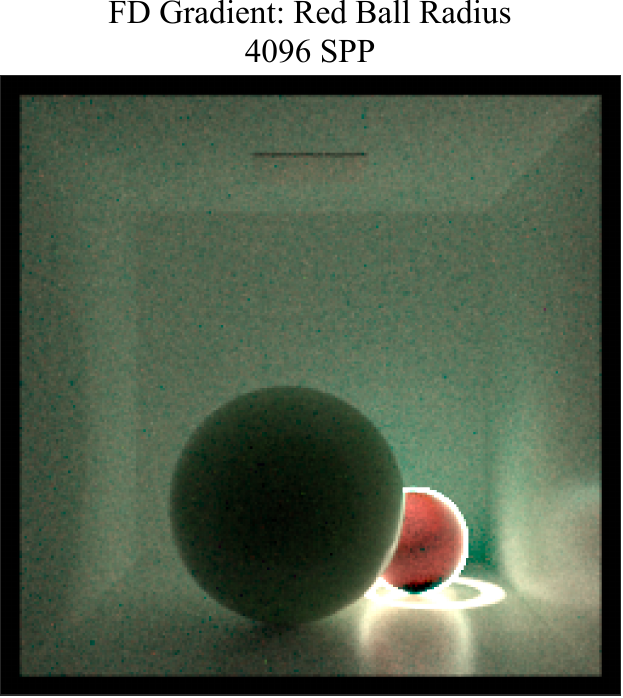}
		\label{fig:fd_size_4096}
	}
    \subfloat[][]
	{
		\includegraphics[width=0.15\columnwidth]{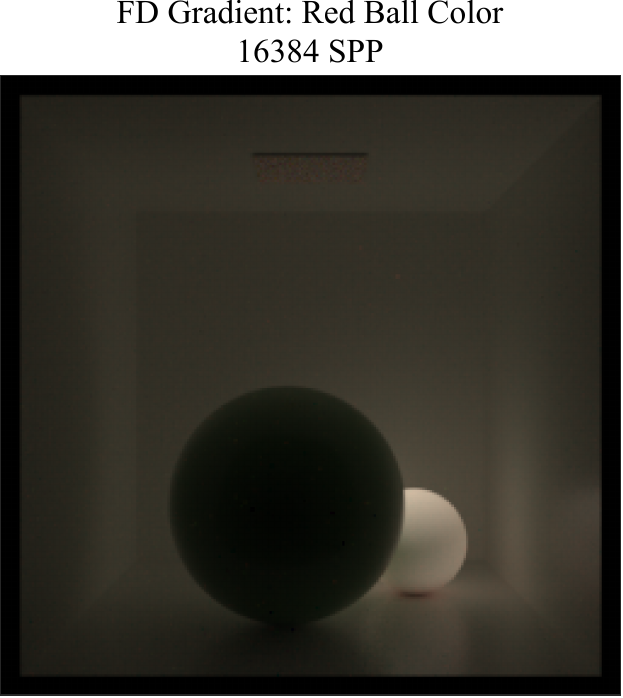}
		\label{fig:fd_color_16384}
	}
	\subfloat[][]
	{
		\includegraphics[width=0.15\columnwidth]{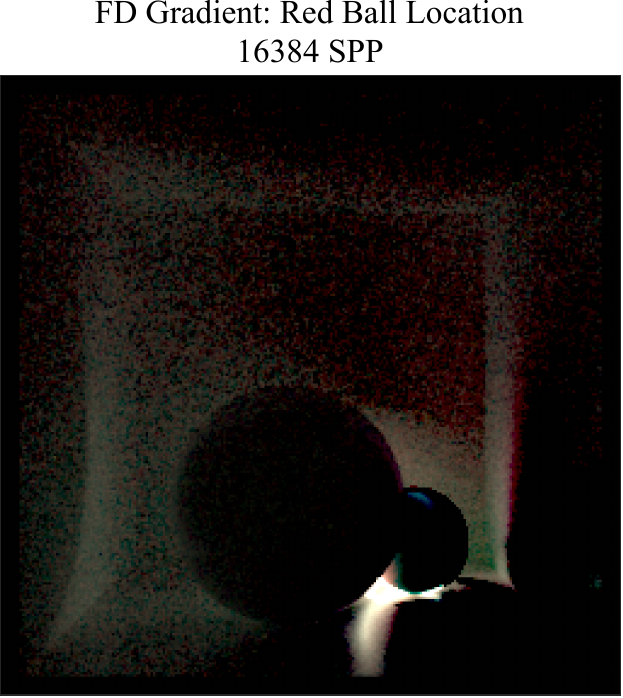}
		\label{fig:fd_location_16384}
	}
	\subfloat[][]
	{
		\includegraphics[width=0.15\columnwidth]{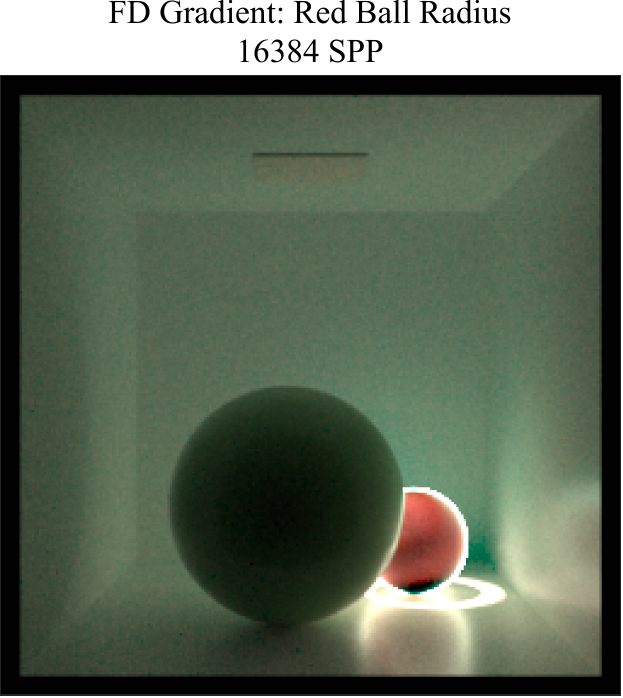}
		\label{fig:fd_size_16384}
	}
    
    \caption{Rendered gradients with $2^{12}=4096$ SPP ((a) - (c) and (g) - (i)) and $2^{14}=16384$ SPP ((d) - (f) and (j) - (l)), respectively. The top row shows gradients from differentiable rendering (AD) with respect to the red ball's: (a), (d) color, (b), (e) location, and (c), (f) radius. The bottom row shows the FD gradients with respect to the red ball's: (g), (j) color, (h), (k) location, and (i), (l) radius. The color gradients are relatively artifact-free, but the AD gradients for position and radius contains image artifacts in the form of groups spurious outlier pixels, which the FD gradients do not contain. The FD gradients were produced using central differencing from renders with $4096$ SPP each; results were averaged over 16 such rounds of renders. In this way, the FD gradients are the average of 16 gradient images, each from a different round of renders.}
    \label{fig:cbox_ad_vs_fd}
\end{figure}

Figure~\ref{fig:cbox_ad_vs_fd} shows the AD and FD\footnote{We use the central difference method to obtain the gradients of the loss: $\nabla_{\btheta} \ell(\btheta_0) \approx \frac{\ell(\btheta_0 + \xi) - \ell(\btheta_0 - \xi)}{2\xi}$, with $\xi=0.01$m, where $\ell(\btheta_0)$ is the loss function. As noted in Figure~\ref{fig:cbox_ad_vs_fd}, the FD gradients are the average of several such gradient images.} gradients of this image with respect to the red ball's color (\ref{fig:ad_color_4096}, \ref{fig:fd_color_4096}), location (\ref{fig:ad_location_4096}, \ref{fig:fd_location_4096}), and radius (\ref{fig:ad_size_4096}, \ref{fig:fd_size_4096}), for $2^{12}=4096$ samples per pixel (SPP). The AD and FD gradients with respect to the red ball's color are both relatively noiseless. For the red ball's position and size, however, the AD gradients contain image artifacts in the form of groups spurious outlier pixels, as seen on the left side of \ref{fig:ad_location_4096} and below the red ball in \ref{fig:ad_size_4096}. The FD gradients for position and size have similar amounts of background noise to the AD gradients, but do not produce these image artifacts. 

This phenomenon is persistent with higher SPP values as well. Figure~\ref{fig:cbox_ad_vs_fd} shows the gradients for the ball's color (\ref{fig:ad_color_16384}, \ref{fig:fd_color_16384}), location (\ref{fig:ad_location_16384}, \ref{fig:fd_location_16384}) and size (\ref{fig:ad_size_16384}, \ref{fig:fd_size_16384}) for $2^{14} = 16384$ SPP. Again, we see similar outlier groups of pixels in the AD images which are not present in the FD images. Indeed, instead of becoming less apparent, the image artifacts seem to converge and stand out more in the higher SPP AD images, although the artifacts appear in different locations due to the use of different random seeds. Thus, for our use case, we deem FD gradients to be more appropriate.

A secondary reason for using FD gradients over differentiable rendering is that we observed differentiable rendering of multi-bounce photon paths has a much larger memory footprint and a $25\times$ to $30\times$ computational overhead as compared to rendering the actual plenoptic observations. Thus for problems with a small number of parameters ($<15$ or so), we were able to reduce computational load and memory by using FD gradients. 

It is worth noting that one potential solution to make use of differentiable rendering would be to render in multiple passes and calculate a final gradient as the mean or median of the individual AD gradients. Alternatively, we could adaptively increase the SPP values where visibility is an issue (from which the artifacts seem to derive). We did not pursue these avenues here. Such strategies may be useful when estimating a large number of parameters, for which finite differences does not scale well (for example, if trying to estimate a non-line-of-sight image in its entirety). It should be made clear, however, that using FD gradients is not superior \textit{in general} to differentiable rendering; for many problems involving accessing the gradient (such as performing gradient descent), a noisy gradient is expected and does not interfere with the result. Additionally, the outlier phenomenon described above seems to manifest only for gradients with respect to parameters which affect scene visibility, such as objects' position and sizes. Gradients with respect to parameters that do not affect visibility (such as color) did not produce the noted image artifacts, and are likely quite amenable to AD techniques (notwithstanding their computational and memory overhead).

%% file: rec_lb.tex
\section{Exact Renderer-Enabled Computation of Lower Bounds}\label{sec:exact_lower_bounds}

Here we present the first novel contribution of this paper: an approach to compute parameter lower bounds using rendering software, assuming the software produces the \textit{exact} intensities for the case of interest. Our approach employs the Hammersley-Chapman-Robbins lower bound (HCR-LB) \cite{chapman_robbins, hammersley}, which provides lower bounds on the variance of unbiased estimators. The HCR-LB and its variants have been used in constrained parameter estimation problems for sensor array signal processing, e.g., bearing estimation \cite{delay_est,source_loc,mcaulayRadar} and frequency estimation \cite{freq_est,gorman_vector_HCR}, to name a few. In this section, we will state the HCR-LB and discuss some of its salient aspects that make it ideally suited to this problem setting. 

\begin{lemma}[HCR Lower Bound]\label{lemma:HCR}

	Let $\btheta^*\in\Theta\subseteq\RR^J $ be any deterministic but unknown parameter, and let $\Yb_\Omega$ denote a set of noisy observations of the unknown parameter $\btheta^*$. Then the variance of any unbiased estimator of $\btheta^*_j$ obeys 
	
	\begin{align}
		\nonumber
		\Var(\btheta^*_j) \geq \HCR (\btheta^*_j),
	\end{align}
	where the lower bound is given by
	\begin{align}\label{eq:HCR_lemma}
		\HCR (\btheta^*_j) \hspace{0em}= \sup \limits_{\substack{\bDelta\neq \mathbf{0} \\ \btheta^*+\bDelta\in\Theta}} \hspace{0em} \frac{\bDelta^2_j}{\EE_{\Yb_\Omega \sim p(\Yb_\Omega; \btheta^*)}\sbr{\frac{p(\Yb_\Omega;\btheta^*+\bDelta)}{p(\Yb_\Omega;\btheta^*)} - 1}^2}, \hspace{1em}
	\end{align}
     for all $j=1,\dots,J$.	Here, $p(\Yb_\Omega;\btheta^*)$ and $p(\Yb_\Omega;\btheta^*+\bDelta)$ denote the pdfs (or pmfs) of the observations parametrized by $\btheta^*$ and $\btheta^*+\bDelta$, respectively, and the subscript on $\EE$ denotes expectation with respect to the random variable $\Yb_\Omega$ when drawn according to the distribution parameterized by $\btheta^*$.

\end{lemma}

The denominator in the RHS of (\ref{eq:HCR_lemma}) is the so-called $\chi^2$-divergence of $p(\Yb;\btheta^*+\bDelta)$ from $p(\Yb;\btheta^*)$ which essentially measures changes in the probability distribution functions when the true parameter $\btheta^*$ is perturbed by $\bDelta$. When a small change in the unknown parameter results in distinctly different sets of observations, resulting in a large $\chi^2$-divergence between the likelihoods, the HCR-LB is small, and vice versa. 

It is worth commenting on the relationship of the HCR-LB framework with the well known Cramer-Rao Lower Bound (CRLB) \cite{cramer_original, rao_original, rao_original2}. If we let $\bDelta_j\rightarrow \mathbf{0}$, the expression inside the supremum of (\ref{eq:HCR_lemma}) converges to the CRLB (if the corresponding limit exists). Thus when both bounds exist, the HCR-LB is at least as tight as the CRLB.  Unlike the CRLB, however, the HCR-LB makes no ``regularity" assumptions on the noise likelihood function and hence is applicable for a wider range of problems. In particular, the CRLB requires computing derivatives of the log-likelihood function (with respect to $\btheta$) and is not well defined in scenarios where the log-likelihood function is not differentiable (e.g., when the parameter space is a countable set). Even for simple scenes, due to the presence of occluding barriers and edges, sharp ``transition regions" may occur in the true plenoptic intensities $\Lb_{\btheta^*}$ as the underlying scene parameter $\btheta^*$ varies smoothly resulting in the log-likelihood being non-differentiable.  The HCR-LB does not require explicitly computing derivatives of the log-likelihood, and can be evaluated in our scenarios of interest by rendering or synthesizing $\Lb_{\btheta^*+\bDelta}$ for a suitably large collection of possible values of $\btheta^*$ and $\bDelta$ using a ray-tracing engine and then evaluating the functional form of the $\chi^2$-divergence. 

In addition to lower bounds on the variance of individual parameters $\btheta^*_j$, we can naturally extend the result in Lemma \ref{lemma:HCR}, to lower bound the MSE of estimators for a given value of $\btheta^*$ as follows.

\begin{corollary}[HCR Lower bound on the MSE]
\label{cor:HCR_MSE}

	Let $\btheta^*\in\Theta\subseteq\RR^J $ be any deterministic but unknown parameter, and let $\Yb_\Omega$ denote a set of noisy observations of the unknown parameter $\btheta^*$. Then the MSE of any unbiased estimator of $\btheta^*$ obeys 
	\begin{align}\label{eq:HCR_MSE}
		\MSE ({\btheta}^*) \geq\hspace{-0.5em} \sup \limits_{\substack{\bDelta\neq \mathbf{0} \\ \btheta^*+\bDelta\in\Theta}} \hspace{0.1em} \frac{ \norm{\bDelta}^2}{\EE_{\Yb_\Omega \sim p(\Yb_\Omega; \btheta^*)}\sbr{\frac{p(\Yb_\Omega;\btheta^*+\bDelta)}{p(\Yb_\Omega;\btheta^*)} - 1}^2} , 
	\end{align}
	where $p(\Yb_\Omega;\btheta^*)$ and $p(\Yb_\Omega;\btheta^*+\bDelta)$ denote the pdfs (or pmfs) of the observations parametrized by $\btheta^*$ and $\btheta^*+\bDelta$, respectively.
\end{corollary}

The proof of Corollary \ref{cor:HCR_MSE} appears in Appendix \ref{proof:HCR_MSE}. In the following sub-sections, we instantiate Lemma \ref{lemma:HCR} and provide functional expressions for the HCR-LB under some common noise models, namely, Poisson and additive white Gaussian noises. 

\subsection{HCR Lower Bound for Poisson Noise}\label{subsec:HCR_Poisson}
The Poisson distribution is commonly used to characterize noise which is discrete or quantized in nature, e.g., when the imaging device counts the number of photons incident on the detector over a certain window of time. Here, noisy plenoptic observations $\Yb_{\bomega}$ drawn independently from a Poisson distribution with rates given by the true plenoptic intensities $\Lb_{\btheta^*}(\bomega)$ are modeled as $\Yb_{\bomega}\overset{\rm ind}{\sim}\text{Poisson}(\Lb_{\btheta^*}(\bomega)),\ \forall \bomega\in\Omega$. If we specialize the HCR-LB given in (\ref{eq:HCR_lemma}) for this setting, for all $j = 1,2,\dots,J,$ we get
\begin{eqnarray}\label{eq:HCR-Poisson}
	\HCR_{\rm Poisson} (\btheta^*_j) =\sup \limits_{\substack{\bDelta\neq \mathbf{0} \\ \btheta^*+\bDelta\in\Theta}} \hspace{0.2em} \frac{\bDelta^2_j}{\exp \cbr{\sum\limits_{\bomega\in\Omega} \frac{\rbr{\Lb_{\btheta^*+\bDelta}(\bomega) - \Lb_{\btheta^*}(\bomega)}^2}{\Lb_{\btheta^*}(\bomega)}}-1},
\end{eqnarray}
where all the true plenoptic intensities in (\ref{eq:HCR-Poisson}) can be obtained from the rendering engine.

\subsection{HCR Lower Bound for Additive White Gaussian Noise}\label{subsec:HCR_Gaussian}
Noisy plenoptic observations with additive white Gaussian noise (AWGN) can be described by $\Yb_{\bomega} = \Lb_{\btheta^*}(\bomega) + \epsilon_{\bomega}$, where the noise $\epsilon_{\bomega}\overset{\rm i.i.d}{\sim}\cN(0, \sigma^2),\ \forall\ \bomega\in\Omega$. We assume that the noise variance $\sigma^2$ is known \emph{a priori}. 
If we specialize the HCR-LB given in (\ref{eq:HCR_lemma}) for this setting, for all $j = 1,2,\dots,J,$ we get 
\begin{align}\label{eq:HCR_AWGN}
	\HCR_{\rm AWGN} (\btheta^*_j) :=\hspace{-0.2em} \sup \limits_{\substack{\bDelta\neq \mathbf{0} \\ \btheta^*+\bDelta\in\Theta}} \hspace{0em}  \frac{\bDelta^2_j}{\exp \rbr{ \frac{\| \Lb_{\btheta^*}  - \Lb_{\btheta^*+\bDelta} \| ^2}{\sigma^2} }-1}, 
\end{align}
where all the true plenoptic intensities in (\ref{eq:HCR_AWGN}) can again be obtained from the rendering engine.

\subsection{Localizing Information Content in Plenoptic Observations}\label{subsec:FD-FI}
In addition to computing lower bounds, we can compute the pixel-wise Fisher Information (FI) as a proxy for localized information in the observations.  Indeed, since we assume that our observations are statistically independent, the total FI associated with the observations is the aggregate of the FI of the individual pixel-wise observations.  Here, the FI of a particular pixel $\bomega$, for any given parameter value $\btheta^*_j$, can be expressed as
\begin{align}\label{eq:FIM}
	\cI_{\bomega} (\btheta^*_j) = \EE_{Y_{\bomega} \sim p(Y_{\bomega}; \btheta)} \sbr{\left. \rbr{ \frac{\partial}{\partial \btheta_j} \log p (Y_{\bomega}; \btheta) }^2 \right\vert_{\btheta = \btheta^*} }, 
\end{align}
for $j=1,\dots,J$, where we implicitly assume that the partial derivative of the log-likelihood function is well-defined at $\btheta = \btheta^*$, and the expectation is with respect to the pdf (or pmf) of the observations $p (Y_{\bOmega}; \btheta^*)$. If we instantiate (\ref{eq:FIM}) for the AWGN and Poisson noise models considered above, for all $j=1,2,\dots,J$, we get
\begin{align}\label{eq:FIM2}
	\cI_{\bomega} (\btheta^*_j) =  \left\{ \begin{array}{rl} 
		\frac{1}{\Lb_{\btheta^*} (\bomega)} \rbr{\frac{\partial \Lb_{\btheta^*}(\bomega) }{\partial \btheta_j} }^2 & \mbox{Poisson noise} \\
		\frac{1}{\sigma^2} \rbr{\frac{\partial \Lb_{\btheta^*}(\bomega) }{\partial \btheta_j} }^2 & \mbox{AWGN w/variance }\sigma^2
	\end{array}\right..
\end{align}

In the absence of access to the gradients of the plenoptic observations, as here, we can still \emph{approximate} the pixel-wise Fisher information for different noise models by approximately computing (\ref{eq:FIM2}) using finite differences (FD), which we refer to as (pixel-wise) FD-FI. It should be noted here that the FD-FI is not available in general for every scene and parameter of interest (e.g. if we wish to estimate the number of objects in a scene). When the log-likelihood is not differentiable for parameters of interest, it is impossible to compute the FD-FI. It is a significant advantage, then, that the HCR-LB has no such requirement of differentiability, and can thus be used in these situations. We have decided to include the FD-FI in the examples below since the log-likelihood is differentiable with respect to the specific parameters of interest considered below (position, radius), and the FD-FI is a useful demonstrative tool.

\subsection{An Illustrative Example: Real Hallway Model}\label{subsec:scene_def}

\begin{figure}[] 
	\centering
	\subfloat[][]
	{
        \includegraphics[width=0.425\columnwidth]{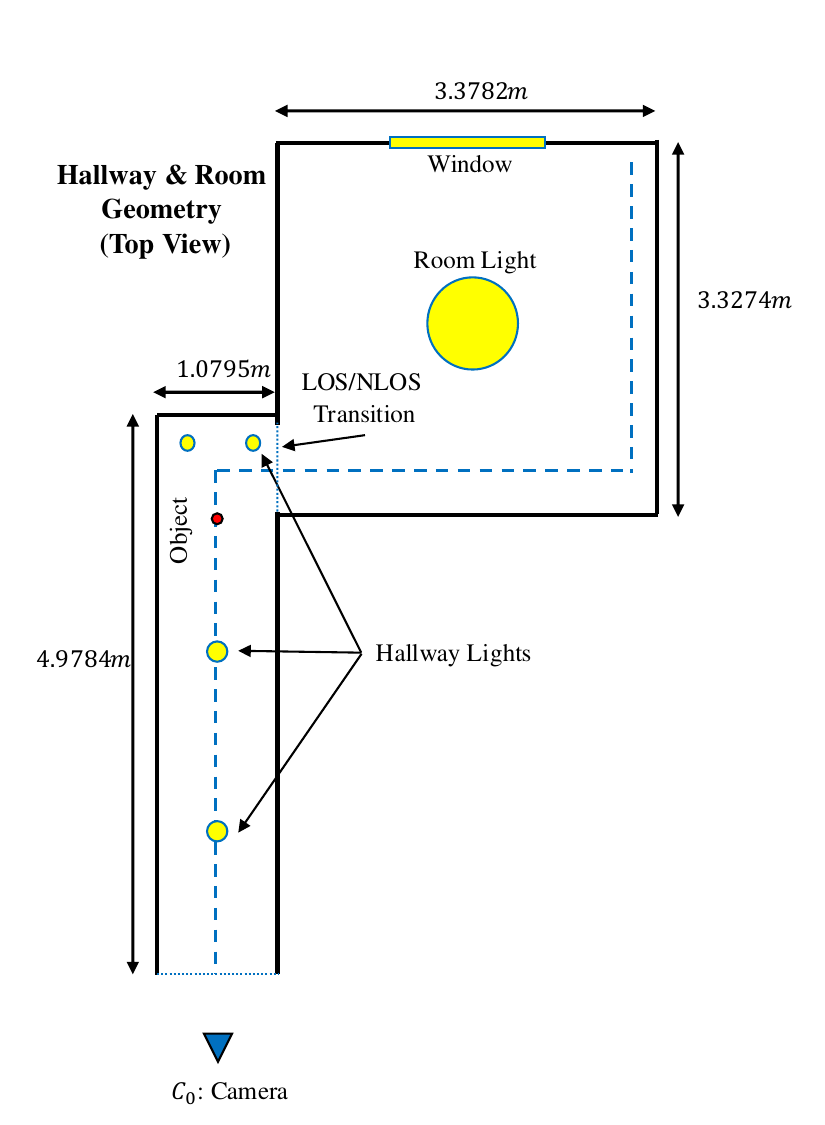}
		\label{fig:top_view_parkview}		
	} \hspace{2em}
	\subfloat[][]
	{
		\includegraphics[width=0.2185\columnwidth]{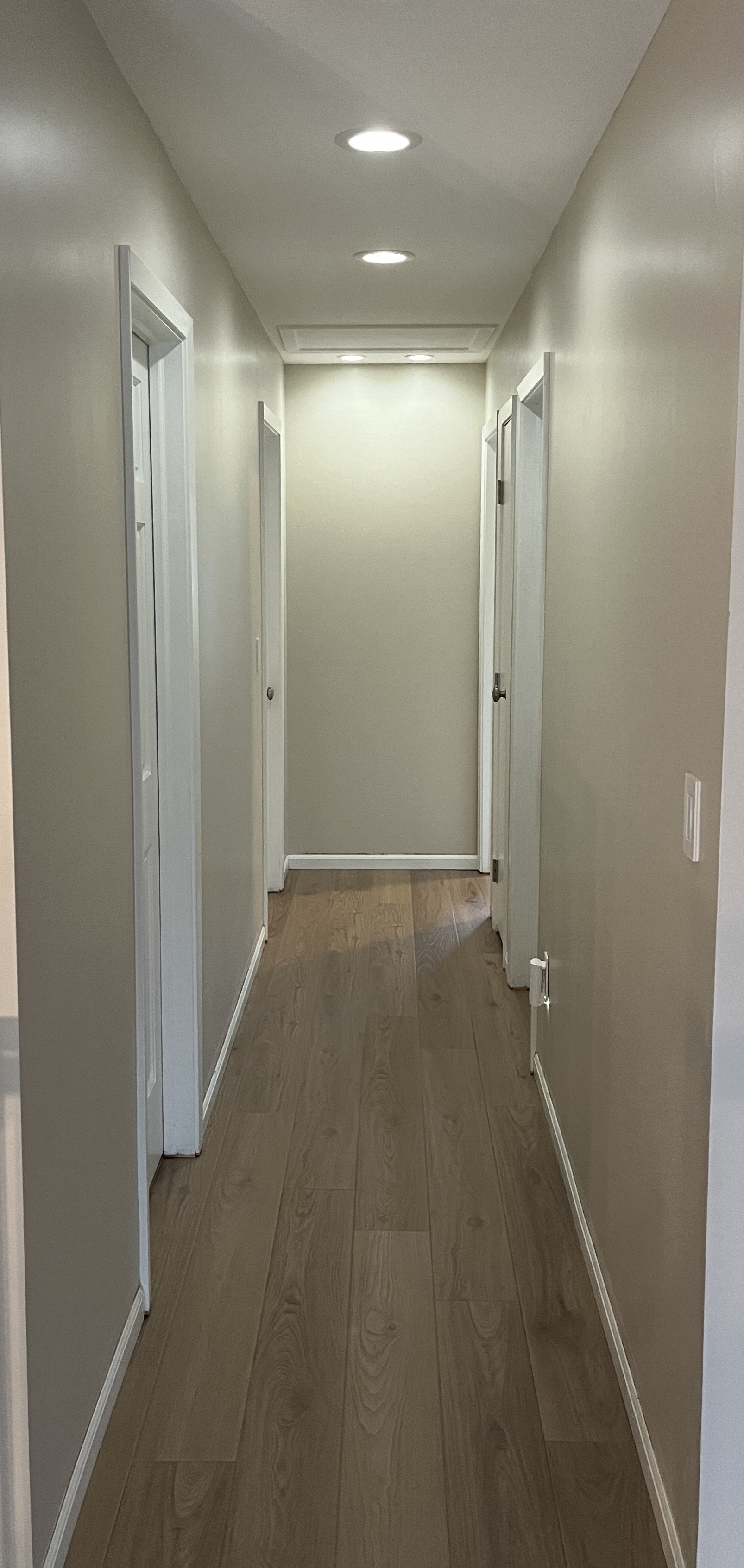}
		\label{fig:ground_truth_parkview}
	} 
	\subfloat[][]
	{
		\includegraphics[width=0.23\columnwidth]{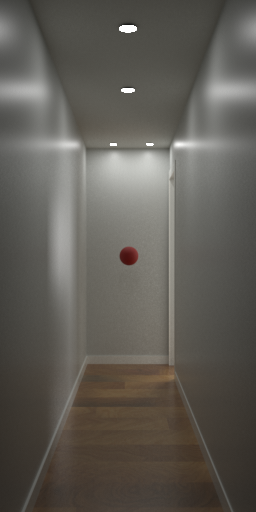}
		\label{fig:rgb_scene_parkview}
	}	
	\caption{Real hallway example scene: (a) A simplified hallway and room layout with dimensions marked, inspired by the real hallway image (b). The hallway in (a) is $4.9784\rm{m}$ long, the room has dimension $3.3274\rm{m}\times3.3782\rm{m}$, and both are $2.4384\rm{m}$ tall. The hallway is illuminated with 4 circular ceiling lights, the two closer of which have radius $4.445\rm{cm}$, and the two farther have radius $3.175\rm{cm}$.  The room is illuminated with one circular ceiling light with radius $20.32\rm{cm}$, and a window to the outdoors which is $1.3716 \rm{m} \times 1.1557\rm{m}$. The camera $C_0$ is located $1.5108\rm{m}$ outside the hallway. The location and radius of a red spherical ball constitute the unknown scene parameter $\btheta^*$. Note that (a) and (c) include only the room at the end of the hallway on the right (no other rooms are modeled). Panel (c) depicts an image of the hallway scene of (a) rendered using {\tt Mitsuba 3} \cite{jakob2022mitsuba3}.}
	\label{fig:scene_defn_parkview}
\end{figure}

We demonstrate the utility of our lower bounding framework using a scene that was modeled after the real hallway depicted in Figure~\ref{fig:ground_truth_parkview}. Since we assume so far that the renderer produces the exact intensities of interest for the scene, it is informative to choose a realistic scene for numerical experiments, since this will demonstrate the real-world utility of our approach.

The modeled scene used here has dimensions as identified in Figure~\ref{fig:top_view_parkview}. Consider a red spherical object in this hallway, and assume we are interested in estimating various parameters related to the object, e.g., its location, radius, etc., which when bundled together constitutes the scene parameter vector $\btheta^*$. \Cref{fig:scene_defn_parkview} shows how the plenoptic observations $\Lb_{\btheta^*}$ can be rendered using a ray-tracing engine for a given parameter value $\btheta^*$, which in this case comprises the ball location and radius. Given the ball location and radius, and \emph{a priori} knowledge of the scene layout, we can use the ray-tracing software to (approximately) solve the rendering equation (\ref{eq:rendering_eq}). A rendered RGB image of the scene with the red ball in the line-of-sight (LOS) regime is shown in Figure~\subref*{fig:rgb_scene_parkview}. If we restrict the ball to a constant height and let it translate along the blue dotted line in \Cref{fig:top_view_parkview}, then the ball is in the LOS regime in the hallway, switches between the LOS and NLOS regime at the transition point marked in that figure, and is fully in the NLOS regime once inside the room at the end of the hall. We shall use this hallway setup to compute lower bounds in this section. Note that in the modeled hallway, only the room at the end of the hallway on the right is preserved; the real hallway is connected to several other rooms not modeled in the render. In addition, light switches, the room door, outlets, and door hardware were not modeled in the rendered scene.

We produced rendered images with \texttt{Mitsuba 3}, using the \texttt{perspective} sensor model which simulates an idealized perspective camera with an infinitely small aperture (and thus an infinite depth of field with no optical blurring) \cite{jakob2022mitsuba3}. In the rendered hallway scene in Figure~\ref{fig:rgb_scene_parkview}, the reflectance properties of the walls and floor were chosen to be as close as possible to those of the paint and floor material of the actual hallway shown in Figure~\ref{fig:ground_truth_parkview}. Here, the walls, floor, and ceiling are modeled using the Mitsuba BSDF plugin \texttt{roughplastic}, which realistically models a rough dielectric material with internal scattering through the use of a microfacet model. The \texttt{roughplastic} BSDF combines specular and diffuse reflectance, and which type of reflectance dominates depends on a user-specified surface roughness parameter $\alpha$;  low values ($\alpha = 0.001 - 0.01$) correspond to very smooth materials, and larger values $\alpha = 0.1 - 0.3$) correspond to rougher materials. The ceiling and trim colors match the manufacturer-specified RGB triplets for the actual paint colors used, and the floor texture is an overlay of an image of wood grain. Further, the wall color was chosen to be different than the true paint color in order to keep the RGB channels more equal, and avoid biasing reflectance information due to the red color of the ball. 

The walls have $\alpha=0.12$; the ceiling has $\alpha=0.28$; the trim has $\alpha=0.08$; and the floor has $\alpha = 0.07$, with interior index of refraction (IOR) 1.5 and exterior IOR of 1, to model the reflectance of the top layer of the actual flooring material which is clear textured vinyl. The two overhead lights in the center of the hallway are circular with radius $4.445$cm and radiance $102.6$ $\rm{W\cdot sr^{-1}m^{-2}}$ (chosen to approximately equal the emitted light from 80 W light bulbs), and the two overhead lights at the back of the hallway are circular with radius $3.175$cm and radiance $151$ $\rm{W\cdot sr^{-1}m^{-2}}$ (chosen to approximately equal the emitted light from 40 W light bulbs). The overhead room light is circular with radius $20.32$cm and radiance $18.5$ $\rm{W\cdot sr^{-1}m^{-2}}$ (chosen to be approximately equivalent to the light emitted by three 100 W light bulbs), and the window is $1.3716\rm{m} \times 1.1557 \rm{m}$ with radiance $20$ $\rm{W\cdot sr^{-1}m^{-2}}$ (chosen to simulate daylight). All lights and the window emit white light (uniformly over all wavelengths). We used the default \texttt{hdrfilm} plugin which avoids post-processing (such as gamma correction) and records linear radiance values produced by the rendering engine. 

The rendering engine \texttt{Mitsuba 3} does not model exposure in the traditional sense of shutter speed or ISO/ASA for film plugins. Instead, the rendered pixel values represent the integral of all incoming radiance, integrated over the pixel area as shown in the rendering equation in (\ref{eq:rendering_eq}). This, theoretically, should result in a perfectly noise-free image which is perfectly ``exposed''. For comparison, the image of the real hallway in Figure~\ref{fig:ground_truth_parkview} was taken using a digital phone camera with a focal length of $5.1 \text{mm}$, an aperture of f$/1.6$ and exposure time of $1/60$\textsuperscript{th} of a second.

\subsection{Numerical Results: Lower Bounds for the Model of Real Hallway Scene}\label{subsec:HCR_results_exact}
\begin{figure}[h]
    \centering
    \includegraphics[scale=0.5]{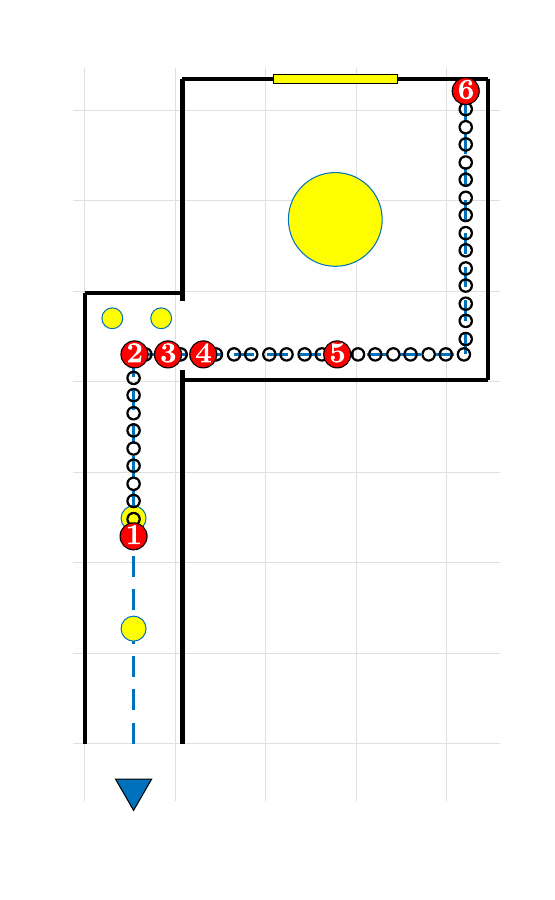}
    \caption{Ball location manifold in the scene modeled after a real hallway. Ball locations along the manifold shown in black (approximately every $20$cm). The red marked ball locations show, respectively: (1) The start of the manifold (position of the ball closest to the sensor). (2) The last point of front-to-back displacement of the ball in the hallway, where the ball starts moving left-to-right. (3) The last point where the ball is fully in the sensors LOS. (4) The first point where the ball is fully outside of the sensor's LOS. (5) The left-to-right midpoint of the room (where the ball is located for radius HCR-LBs). (6) The last point in the manifold of ball locations. }
    \label{fig:ballManifold}
\end{figure}

We compute lower bounds here for the hallway and room scene above. We assume that noisy plenoptic measurements are made by an imaging device located at $C_0$, which collects RGB images of size $256\times 512$. With this setup, we study the fundamental limits of two different \emph{scalar} estimation problems: 
\begin{itemize}
	\item Estimating 1D location of the ball (with a fixed radius of $10$cm) along the hallway, into the room, and inside the room (following the blue dotted line in Figure~\ref{fig:top_view_parkview}).
	\item Estimating radius (size) of the ball located in the room at the midpoint of the room wall, on the blue dotted line (linear location 427, as explained in the remainder of this section).
\end{itemize}
Here, 1D location refers to translation of the ball through the hallway at constant height along the blue dotted line in Figure~\ref{fig:top_view_parkview}, from a top-down view of the hallway. With a given starting point in the hallway, we can measure the linear displacement along the blue dotted line as the 1D location. The manifold of possible ball locations is shown in black in Figure~\ref{fig:ballManifold}, with some important positions denoted in red. 

\subsubsection{Results and Discussion}

For these estimation problems, we assume that the scene geometry and the BRDFs of the surfaces are known, i.e., the only unknown in the scene is either the location of the ball or its radius. For the problem of estimating the ball location, we discretize the $1$-D manifold shown along the blue dotted line in Figure~\ref{fig:ballManifold}; the ball can be located at any of $860$ equally-spaced points along this manifold (each $1$cm apart), starting at position $(1)$ and ending at position $(6)$ as shown in the figure. The black circles in Figure~\ref{fig:ballManifold} show the possible ball positions, down-sampled to approximately every $20\rm{cm}$ for clear visibility. Ball locations are numbered front to back and left to right from $1$ (denoted by $(1)$ in Figure~\ref{fig:ballManifold}: this is the closest point to the sensor, in the LOS region in the middle of the hallway) to $860$ (denoted by $(6)$ in Figure~\ref{fig:ballManifold}: this is the top-right of the manifold, the farthest point away from the sensor, in the NLOS region). For size estimation problem, we consider $141$ possible values for ball radii in the range $1$cm to $15$cm (with $0.1$cm increments), at position $427$ in the room (denoted by $(5)$ in Figure~\ref{fig:ballManifold}: this is the left-to-right midpoint of the room, in the NLOS-only region). We use the {\tt Mitsuba 3} renderer \cite{jakob2022mitsuba3} to synthesize physically accurate plenoptic samples $\Lb_{\btheta^*}$, for different values of the unknown parameter $\btheta^*$ (and $\bDelta$), and numerically evaluate the HCR bound given by \Cref{eq:HCR-Poisson,eq:HCR_AWGN}. 
In addition to numerically evaluating the HCR-LB, we use finite differences on the rendered data, as discussed in Section \ref{subsec:FD-FI}, to (approximately) compute pixel-wise Fisher information (\ref{eq:FIM2}), which is referred to as finite difference-Fisher information (FD-FI). 
All the scenes were rendered with $16384$ SPP using \texttt{Mitsuba 3}'s path tracer (\texttt{path} integrator) on a Linux Ubuntu computer with 2 NVIDIA RTX Quadro 8000 GPUs with $48$ GB RAM, and 2 NVIDIA TITAN V GPUs with $12$ GB RAM. It took approximately $5.5$ minutes to render each scene. 

\begin{figure*}[ht]
	\subfloat[][]
	{		
		\includegraphics[width=0.3\columnwidth]{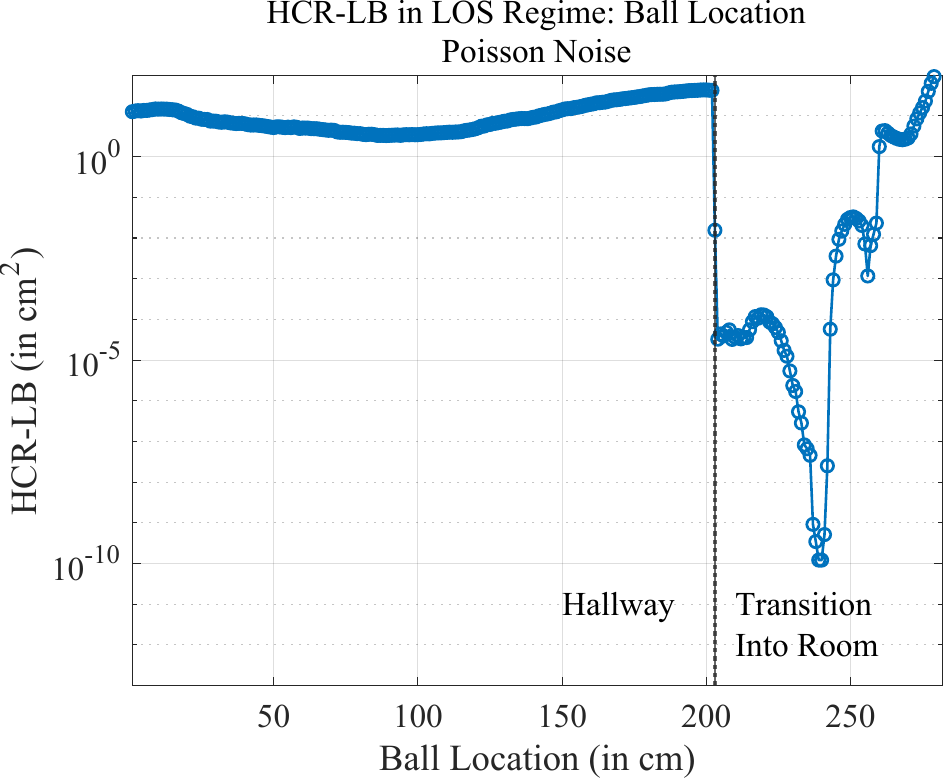}
		\label{fig:Poisson_HCR_los}
	}	
	\hfill
	\subfloat[][]
	{		
		\includegraphics[width=0.3\columnwidth]{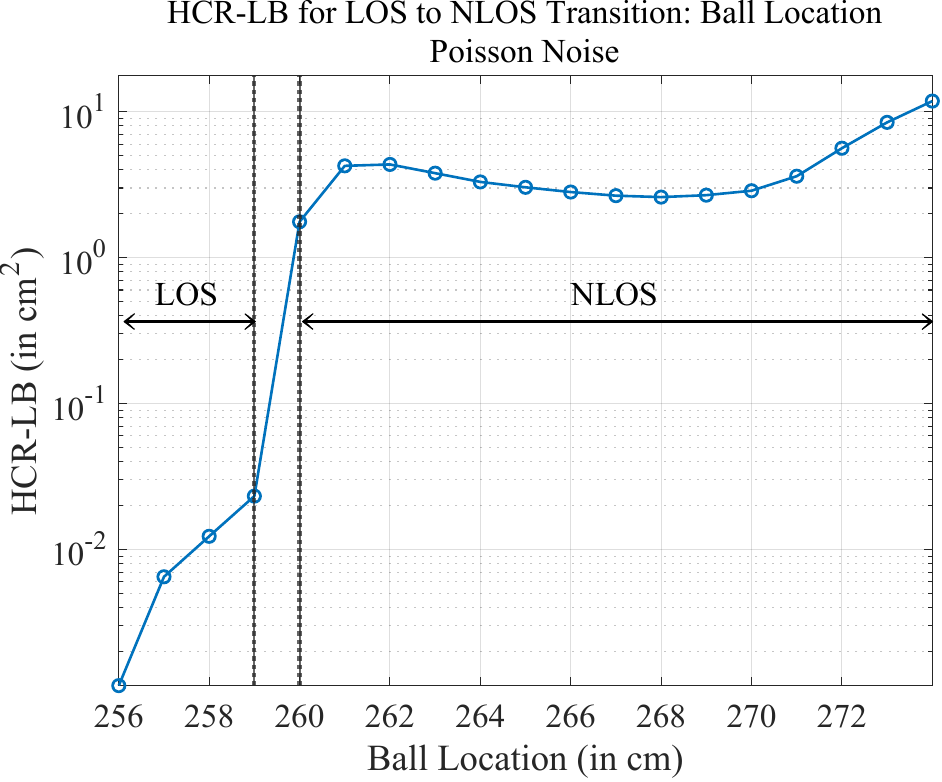}
		\label{fig:Poisson_HCR_transition}
	}
	\hfill
	\subfloat[][]
	{
		\includegraphics[width=0.3\columnwidth]{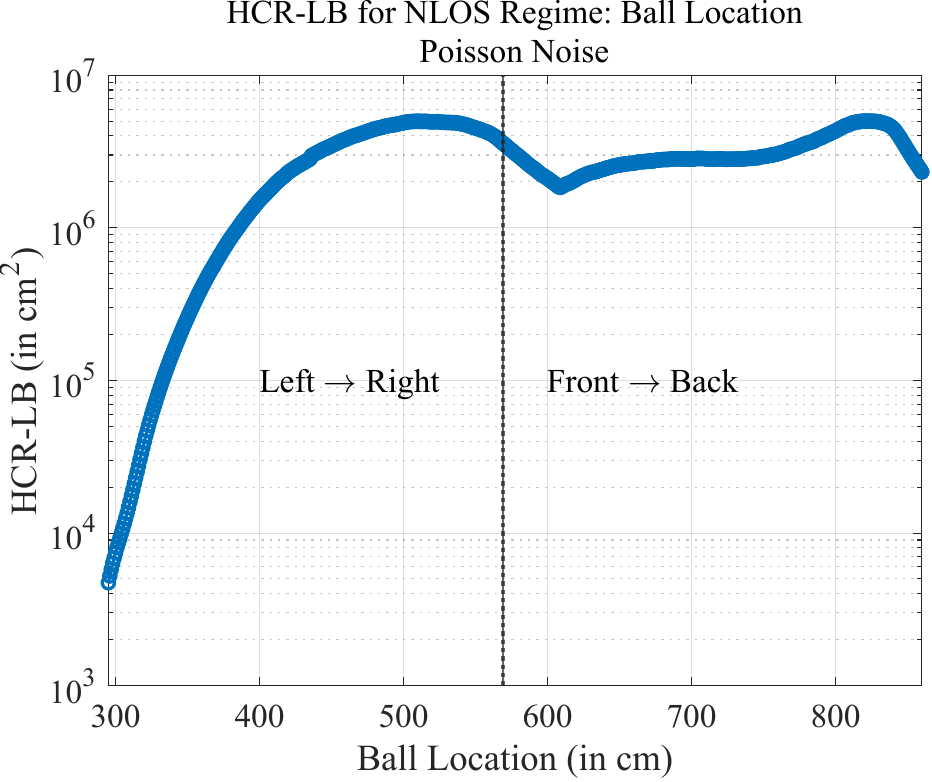}
		\label{fig:Poisson_HCR_nlos}
	}
	
	\subfloat[][]
	{		
		\includegraphics[width=0.3\columnwidth]{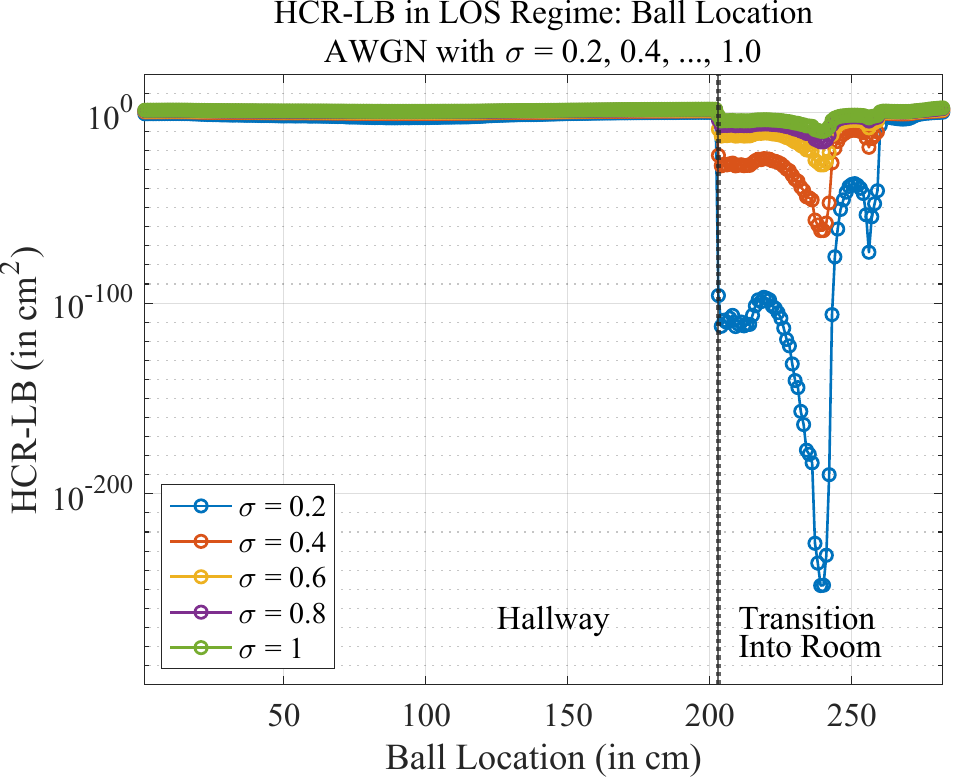}
		\label{fig:AWGN_HCR_los}
	}	
	\hfill
	\subfloat[][]
	{		
		\includegraphics[width=0.3\columnwidth]{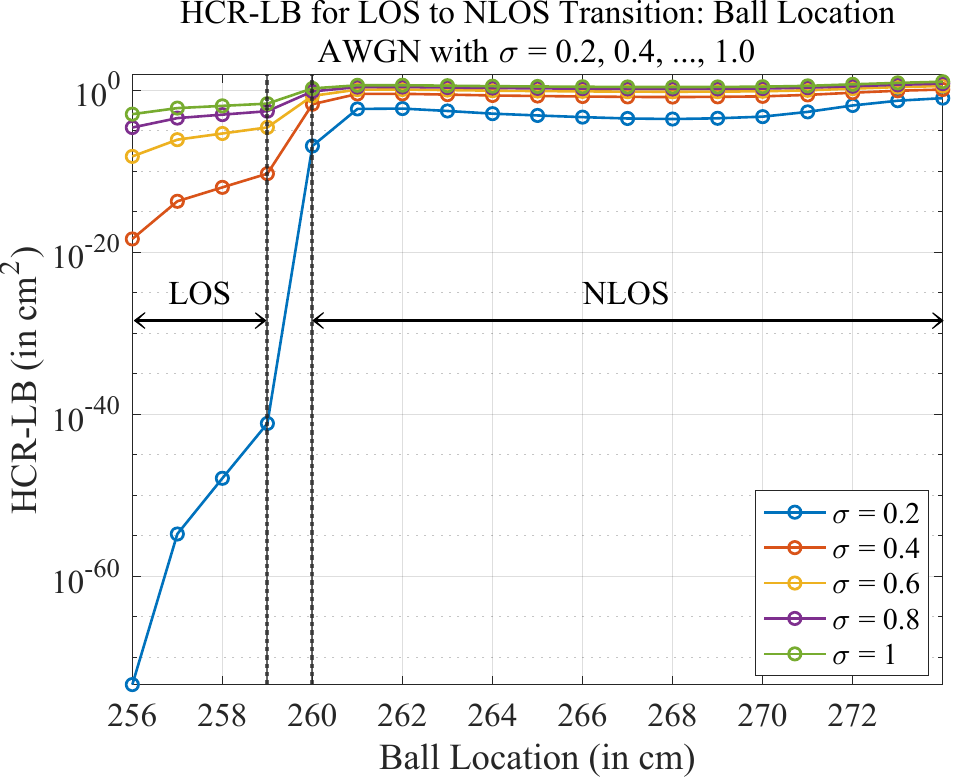}
		\label{fig:AWGN_HCR_transition}
	}
	\hfill
	\subfloat[][]
	{
		\includegraphics[width=0.3\columnwidth]{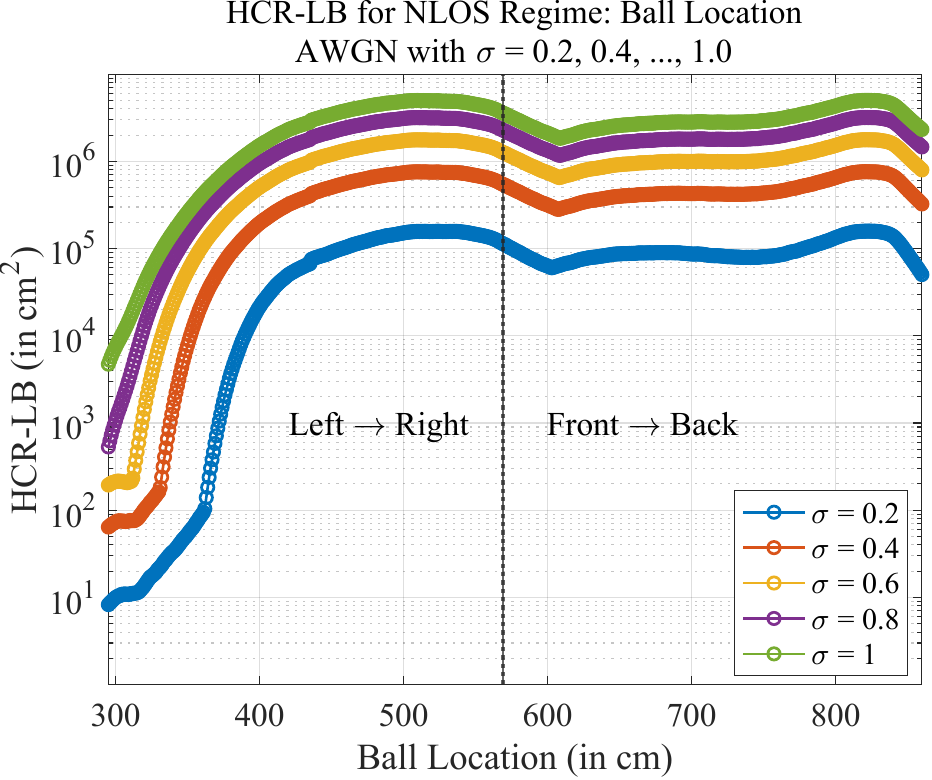}
		\label{fig:AWGN_HCR_nlos}
	}
	\caption{HCR-LB for ball location estimation: (a)-(c) for Poisson Noise, and (d)-(f) AWGN with different values of $\sigma$. Panels(a),(d): LOS region - the HCR-LB is small throughout, and the HCR-LB drops significantly when the ball starts moving to the right at the back of the hallway. Panels (b),(e): Transition from LOS to NLOS - sharp increase in the HCR-LB when the ball moves away from LOS. Panels (c),(f): NLOS region - the HCR-LB in is much higher, indicating the potential hardness of the estimation problem.}
	\label{fig:HCR_results}
\end{figure*}

The HCR-LB under Poisson noise for estimating the ball location is shown in \Cref{fig:Poisson_HCR_los,fig:Poisson_HCR_transition,fig:Poisson_HCR_nlos}. Similarly, the HCR-LB under AWGN with different levels of noise variances $\sigma^2$, is shown in \Cref{fig:AWGN_HCR_los,fig:AWGN_HCR_transition,fig:AWGN_HCR_nlos}. Notwithstanding the fact that higher noise variances in the AWGN case lead to harder estimation tasks, and the general expectation that LOS estimation is far easier than NLOS estimation (an insight to which the lower bounds perhaps lend some additional credence), perhaps the key takeaway in each of the settings here is that the ``hardness'' of position estimation in NLOS regimes is not necessarily uniform, and may in fact exhibit some interesting nuances.  Indeed, in NLOS regimes there may be lighting and geometric configurations and considerations that make estimation comparatively easier or more difficult depending on the ball location.  We explore this in more detail later in this section.

\begin{figure}
    \centering
    \subfloat[][]{
    \includegraphics[width=0.45\columnwidth]{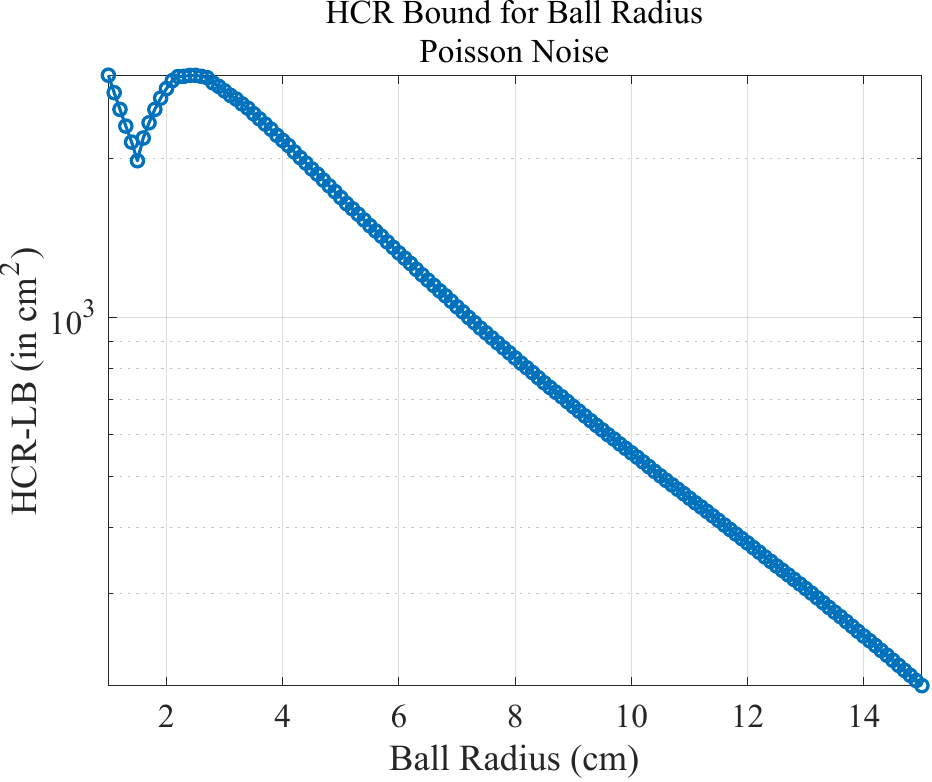}
		\label{fig:Poisson_HCR_radius}
    }
    \subfloat[][]{
    \includegraphics[width=0.45\columnwidth]{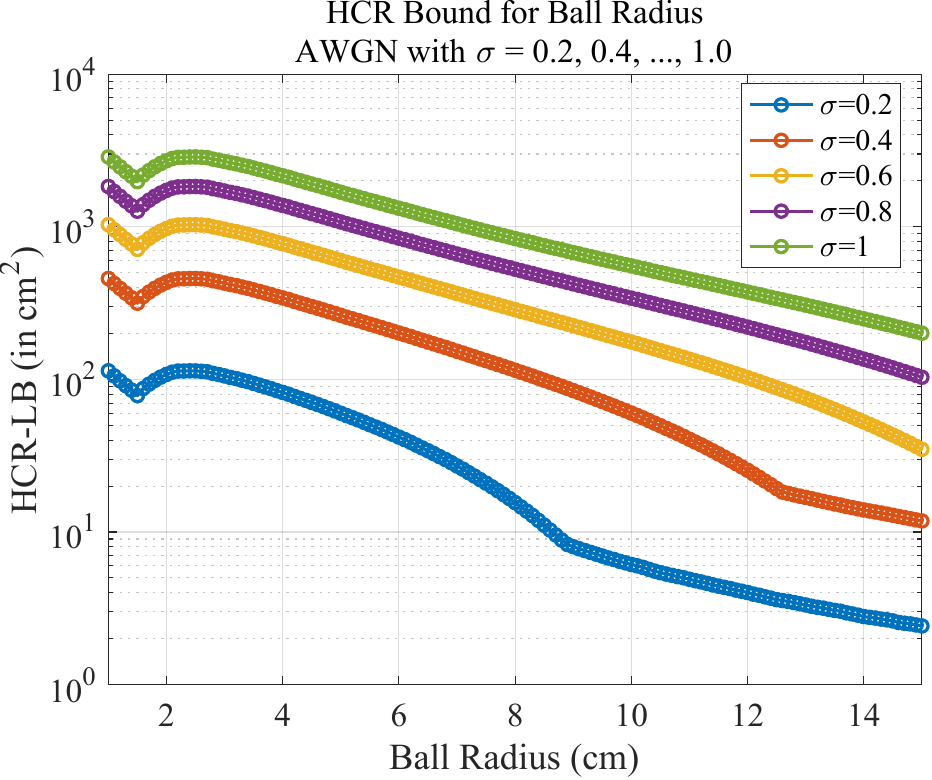}
		\label{fig:AWGN_HCR_radius}
    }
    \caption{HCR-LB for ball radius estimation, with ball located at position 427, denoted by (5) in Figure~\ref{fig:ballManifold}. Panel (a): Poisson noise model; Panel (b) AWGN with different variances. The overall trend observed is that the HCR-LBs generally decrease with increasing size (radius) of the ball.}	
	\label{fig:HCR_radii}
\end{figure}

The HCR-LBs for estimating the ball radius at a fixed position (location 427, denoted by $(5)$ in Figure~\ref{fig:ballManifold}, in the left-to-right center of the room, in the NLOS regime), under each of the noise models, are given in Fig.~\ref{fig:HCR_radii}.  Here, the general trend is that (NLOS) radius-estimation generally becomes easier the bigger the ball is.  This seems fairly intuitive, as bigger objects would cast bigger shadows, and exhibit more interactions with the light field(s) in the scene.  That said, there seems to be a subtle ``dip'' in the HCR-LBs when the radius is approximately $1.3$cm.  We conjecture this is also due to some complex interactions associated with the scene and lighting geometry.

For the location estimation problem, the HCR-LB is (unsurprisingly) small ($\ll 1$) when the ball is in the sensor's LOS moving from the front to the back of the hallway. The lower bound drops sharply when the ball begins its left-to-right translation (with respect to the camera) into the room (see \Cref{fig:Poisson_HCR_los,fig:AWGN_HCR_los}). This is expected, as detecting an object is easier when it moves horizontally as compared to when it moves away from the camera. 
The HCR-LB increases by a few orders of magnitude when the ball goes away from LOS -- location $260 \mapsto$ location $261$ (see  \Cref{fig:Poisson_HCR_transition,fig:AWGN_HCR_transition}), and continues to increase sharply for ball locations further into the room (see \Cref{fig:Poisson_HCR_nlos,fig:AWGN_HCR_nlos}). Inside the room (translating left to right and then front to back), the actual value of the bound gets very large $\sim10^6$cm$^2$, which indicates that accurately locating the ball far away from the LOS region is (again, unsurprisingly) a much more difficult estimation problem.

For the radius estimation problem, the scale of the results in each of the cases indicate that the radius estimation problem is challenging. Indeed, variance lower bounds on the order of $10^{0.5}-10^{3.5}$ suggest minimum standard errors $\sigma$ on the order of $\approx 1.7-56$cm.  These are, at best, on the order of the ball radii, and can often be much larger.

Some recent works \cite{corner_cam, light_field_from_shadows} have proposed that sharp edges and occlusions in a scene can help recover NLOS imagery. These sharp occlusions act like a pin-hole camera, projecting information about NLOS regions of a scene on to visible regions like walls and floors. The authors of \cite{corner_cam} refer to this imaging phenomenon as the ``corner-camera.'' Our pixel-wise FD-FI images are able to provide some evidence for this effect, showing regions around the corners containing higher amounts of information about the hidden object location than other regions. This effect can be clearly seen from the video of the pixel-wise FD-FI and HCR-LB as the ball moves from location 1 to 860, provided in the supplementary material (See Appendix \ref{vid:HCR-LB} for details).
Looking at the FD-FI images, we also believe that some of the dips in the HCR-LB while the ball is in the room (in \Cref{fig:Poisson_HCR_nlos,fig:AWGN_HCR_nlos}) result from ``corner-camera''-type effects.

\begin{figure}[]
    \subfloat[][]
    {
		\includegraphics[width=0.5\columnwidth]{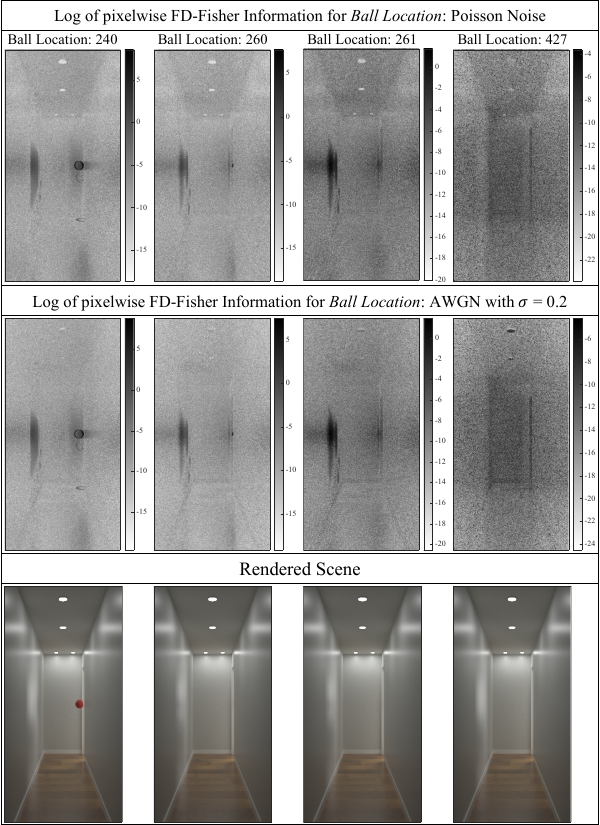}
        \label{fig:FI_location}
        }
    \subfloat[][]
    {
        \includegraphics[width=0.5\columnwidth]{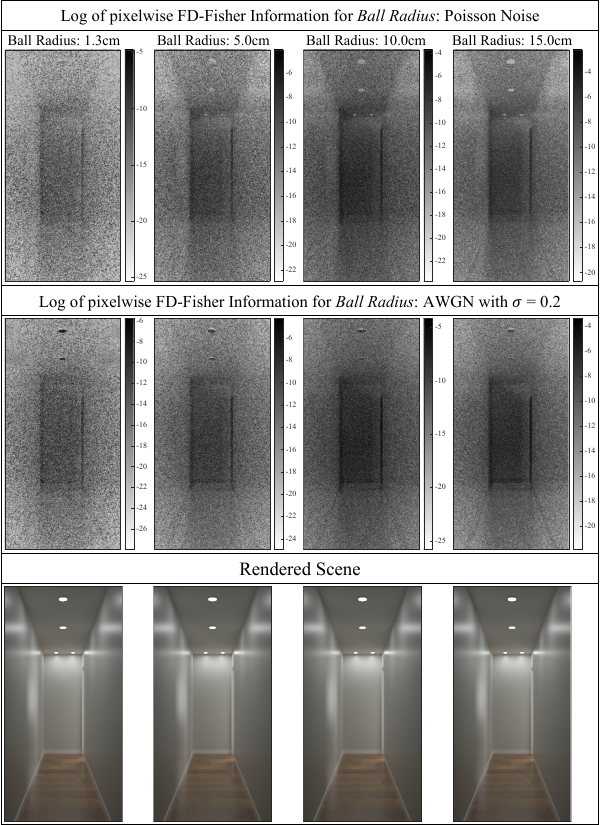}
        \label{fig:FI_radius}
        }
	\caption{Pixelwise FD-Fisher information (FD-FI) for: \ref{fig:FI_location} \textit{ball location} and \ref{fig:FI_radius} \textit{ball radius}, obtained by taking the mean of contributions from each RGB channel. Note that FD-FI images are on a log scale, and absolute scaling differs from image to image. Darker regions $\Rightarrow$ more informative. Pixelwise FD-FI shows where and how information about the parameter of interest is localized in our observations. These images highlight subtle details about the scene parameters which are not visible from the nominal RGB images (bottow-row) obtained from the rendering software. Pixelwise FD-FI shown for Poisson noise (top row) and AWGN with $\sigma=0.2$ (second row). For \textit{ball location} FD-FI \ref{fig:FI_location}, $4$ different locations are considered (from left to right): completely in LOS (position (3) in \Cref{fig:ballManifold}), just \emph{inside} LOS (between (3) and (4) in \Cref{fig:ballManifold}), just \emph{outside} LOS (position (4) in \Cref{fig:ballManifold}), left-to-right midpoint of room (position (5) in \Cref{fig:ballManifold}). For \textit{ball radius} FD-FI \ref{fig:FI_radius}, the ball is stationary at location 427 (left-to-right midpoint of the room, (5) in \Cref{fig:ballManifold}), and $4$ different radii are considered: $1.3$, $5.0$, $10.0$, and $15.0\text{cm}$. Note that different regions in the scene are more informative than others for different ball locations.}	
	\label{fig:FI_both}
\end{figure}

The pixelwise FD-FI images in Figure \ref{fig:FI_location} 
provide valuable insights on where and how information about the location parameter of interest is distributed among the observed samples. As one might expect, when in LOS, the edges of the ball provide most information about its location. When the ball moves away from LOS, information about its location is conveyed by shadows cast by the ceiling room light and window onto the floor and wall, and other \emph{indirect} photons from the back wall and the floor. These subtle details are however not visible in the RGB images in bottom row of Figures~\ref{fig:FI_location}.
The ability to localize information content in indirect photons is extremely useful for those who are interested in developing NLOS imaging algorithms. The FD-FI images can be used as a guide to develop clever imaging modalities that collect observations with ``high information content'' about the NLOS scene parameters.

Likewise, pixelwise FD-FI images for the radius estimation problem in Figure~\ref{fig:FI_radius} show that regions of high information are mainly on the back wall and regions of the door frame which arise from multi-bounce photons, and the number of informative pixels increases for larger radii. We can see that the regions of high information differ significantly between Figure~\ref{fig:FI_location} and Figure~\ref{fig:FI_radius}. We conclude that these informative regions/observations depend both on what the parameter of interest is (see difference between Figures~\ref{fig:FI_location} and \ref{fig:FI_radius}) and the \emph{particular value} of the parameter (see differences among images in each of \Cref{fig:FI_location,fig:FI_radius}). Another interesting point to note is that while the informative regions change significantly for the type of parameter under consideration and also for different values of a given parameter, they are nearly identical for the two noise models considered here.

\subsection{Numerical Results: Plenoptic Information Localization}\label{subsec:plenoptic_info}

\subsubsection{Setup} A natural extension of the above work is to examine the effect that the sensor position has on the information content of the plenoptic data. Since the plenoptic function describes the light field at each point in space and wavelength, we can sample it over a variety of sensor positions to get spatially diverse samples. With this spatially sampled plenoptic data, we can then compute the pixelwise FD-FI as defined in the previous section, for a specific position of the red ball. Here we place the red ball at the boundary of the LOS and NLOS regions (position: $260$, between (3) and (4) in Figure~\ref{fig:ballManifold}) and vary the sensor position in the $x$ and $z$ directions while keeping the sensor pointed at the same point near the back of the hallway. In this way we render different viewpoints of the ball: although the ball is not in the sensor's LOS at location $260$ from the \textit{standard} viewpoint, when we adjust the sensor position to the left, part of the ball comes into the sensor's LOS. Conversely, when we move the sensor to the right, we can see more of the reflection of the ball on the left hallway wall due to the light coming from inside the room. 

We construct a $31 \times 31$ grid of sensor positions in the $xz$-plane, with the standard viewpoint at the center and offset sensor positions from there. We use a step size of $2.54 \rm{cm}$ in each direction, so that the maximum and minimum offsets in the $x$ and $z$ directions is $\pm 40.64 \rm{cm}$. In this way we render $31^2 = 961$ different viewpoints with gradients at $2^{14} = 16384$ samples per pixel; here rendering each image with its FD gradient took approximately 17 minutes on the RTX Quadro 8000 GPUs, and approximately 60 minutes on the TITAN V GPUs. Rendering was split between the two sets of GPUs, and rendering the entire set of $961$ images and gradients for the ball's location and radius took approximately $1230$ GPU-hours in total. Since the jobs were split between the sets of GPUs, this took approximately 2 weeks.

\begin{figure}[t]
	\centering
	\subfloat[][]
	{		
		\includegraphics[width=0.35\columnwidth, align=c]{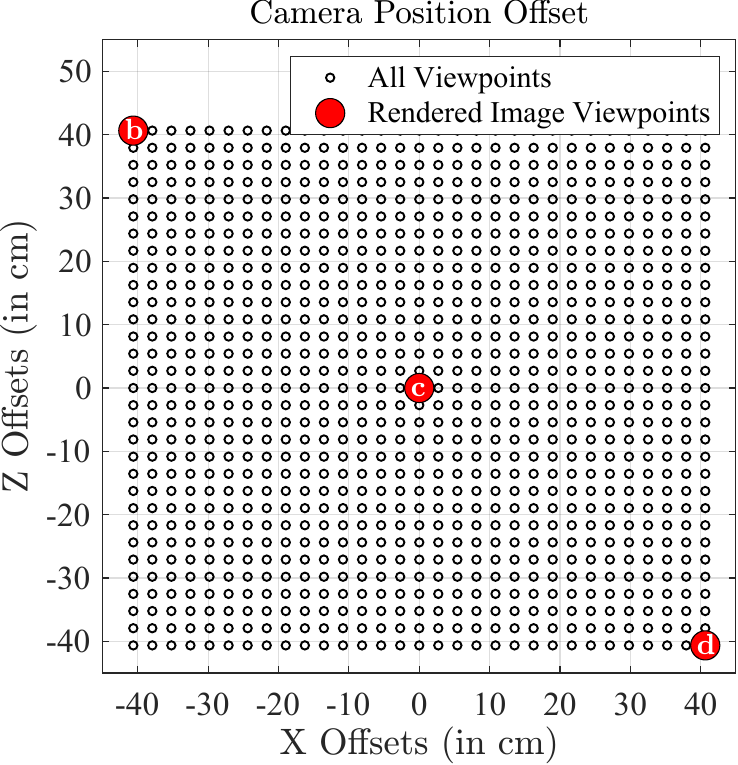}
		\label{fig:sensor_offset_grid}
	}	
	\subfloat[][]
	{		
		\includegraphics[width=0.15\columnwidth, align=c]{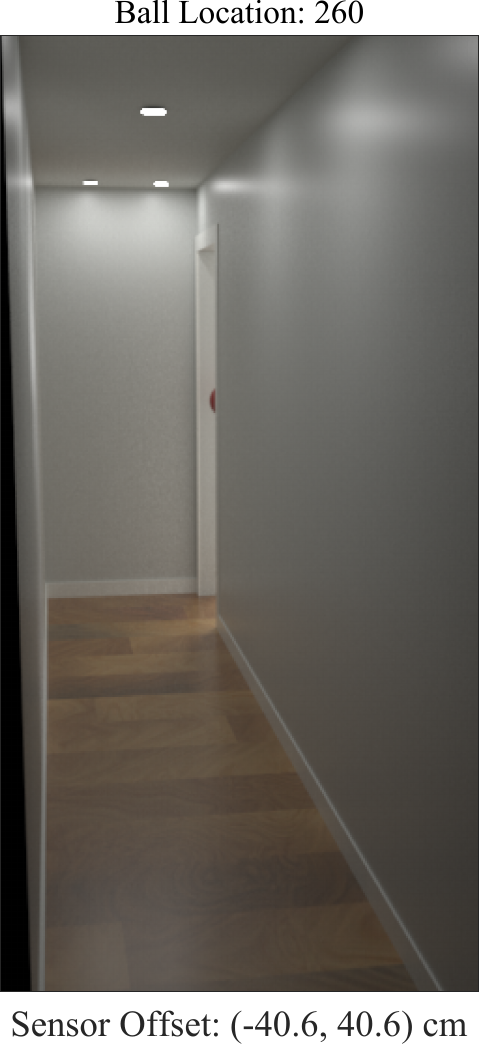}
		\label{fig:top_left}
	}
	\subfloat[][]
	{		
		\includegraphics[width=0.15\columnwidth, align=c]{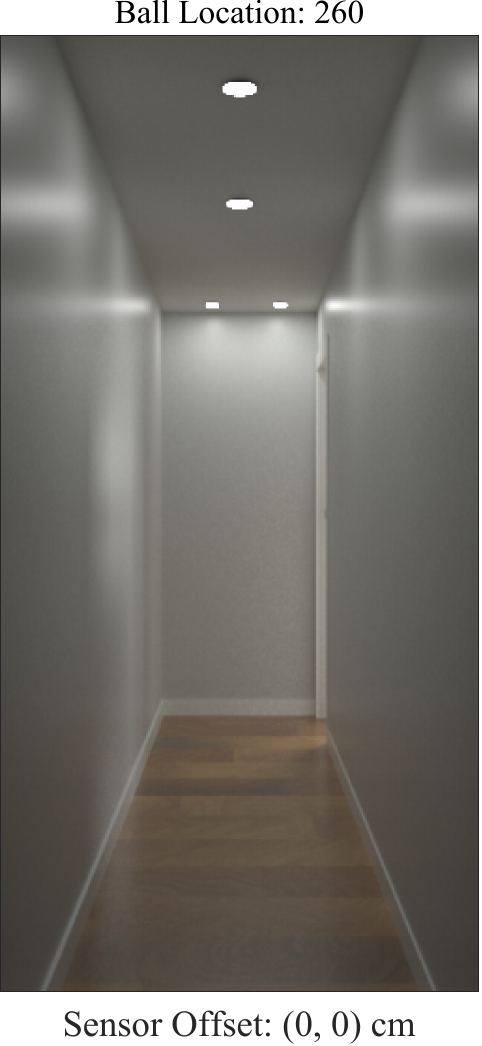}
		\label{fig:standard_view}
	}
	\subfloat[][]
	{		
		\includegraphics[width=0.15\columnwidth, align=c]{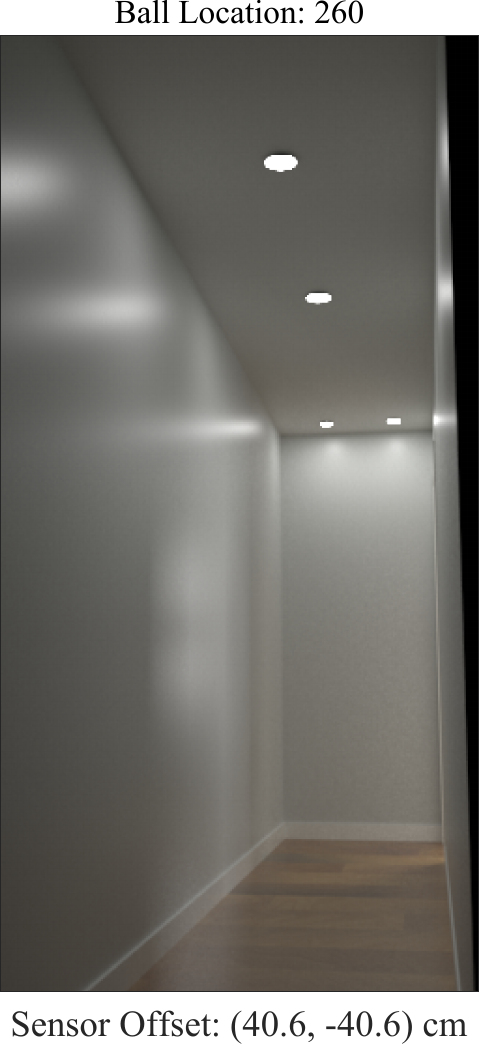}
		\label{fig:bottom_right}
	}
    \caption{(a) shows the $31 \times 31$ sensor offset grid with step size $2.54\rm{cm}$ in the $x$ and $z$ directions, for a maximum shift of $\pm40.64\rm{cm}$ in either axis. Rendered images with the ball at position $260$ (between (3) and (4) in Figure~\ref{fig:ballManifold}; this is the last point where any portion of the ball is in the LOS from the sensor's standard viewpoint) are as follows: (b) shows the rendered image from the sensor with offset $(-40.6, 40.6)\rm{cm}$ (top-leftmost in the grid); (c) shows the standard viewpoint; (d) shows the image from the sensor with offset $(40.6, -40.6)\rm{cm}$ (bottom-rightmost in the grid).}
    \label{fig:sensor_grid_example}
\end{figure}

We repeat the experiment with gradients for the ball's location and radius, and compute gradients with respect to each parameter for each viewpoint. Figure~\ref{fig:sensor_offset_grid} shows the grid of sensor offsets (relative to the standard viewpoint at the center) in $x$ and $z$, and Figures~\ref{fig:top_left}-\ref{fig:bottom_right} show three sample renders from different viewpoints (which are noted in Figure~\ref{fig:sensor_offset_grid}).

\subsubsection{Results and Discussion}

\begin{figure}[t]
    \centering
    \subfloat[][]
    {
        \includegraphics[width=0.155\linewidth]{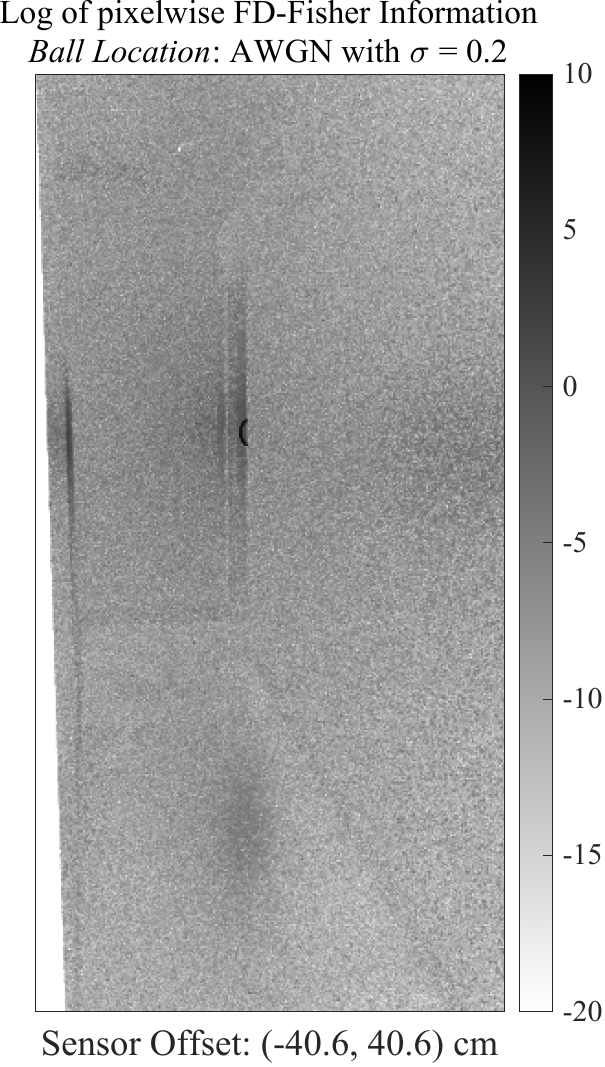}
        \label{fig:top_left_awgn_pos}
    }
    \subfloat[][]
    {
        \includegraphics[width=0.155\linewidth]{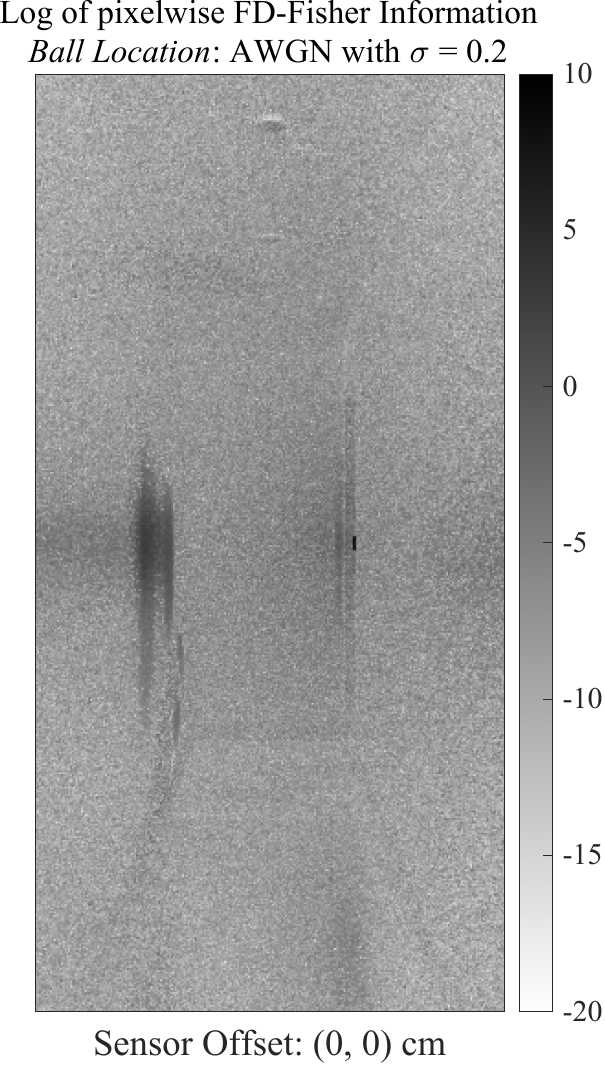}
        \label{fig:standard_awgn_pos}
    }
    \subfloat[][]
    {
        \includegraphics[width=0.155\linewidth]{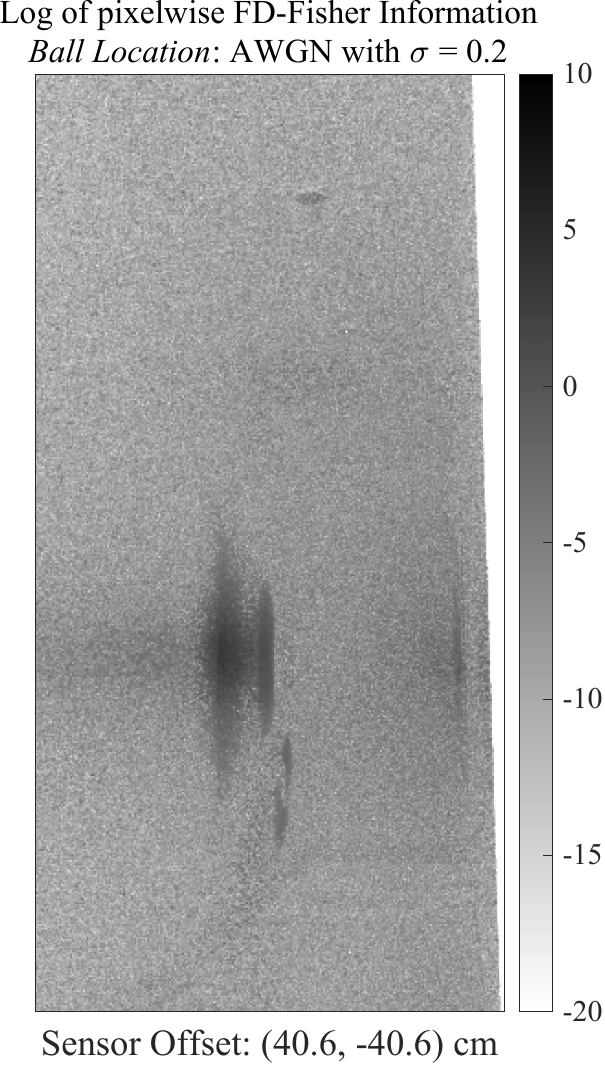}
        \label{fig:bottom_right_awgn_pos}
    } \\
    \subfloat[][]
    {
        \includegraphics[width=0.155\linewidth]{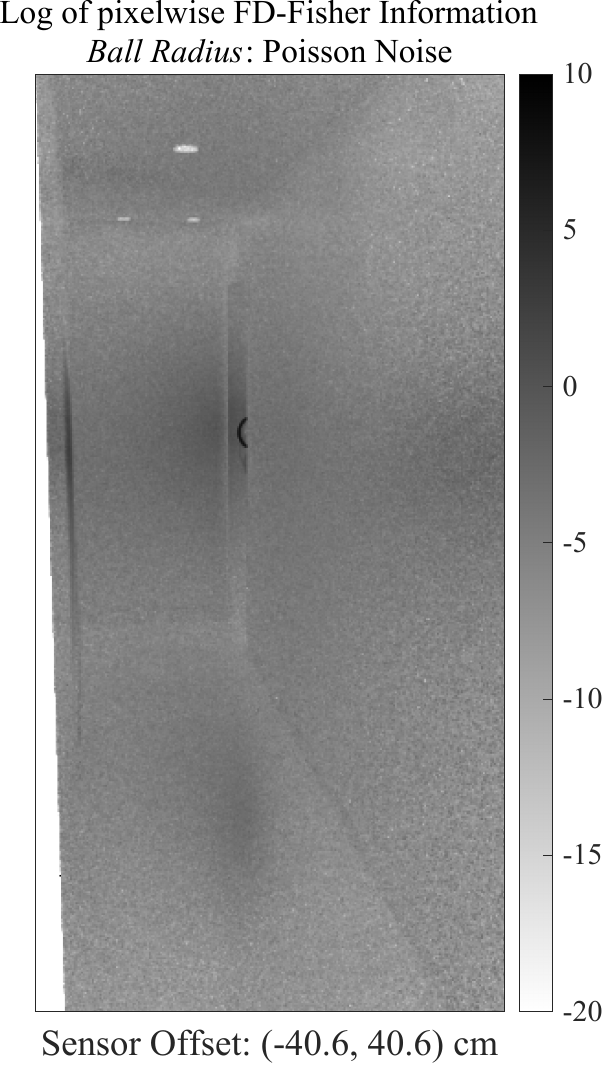}
        \label{fig:top_left_poisson_rad}
    }
    \subfloat[][]
    {
        \includegraphics[width=0.155\linewidth]{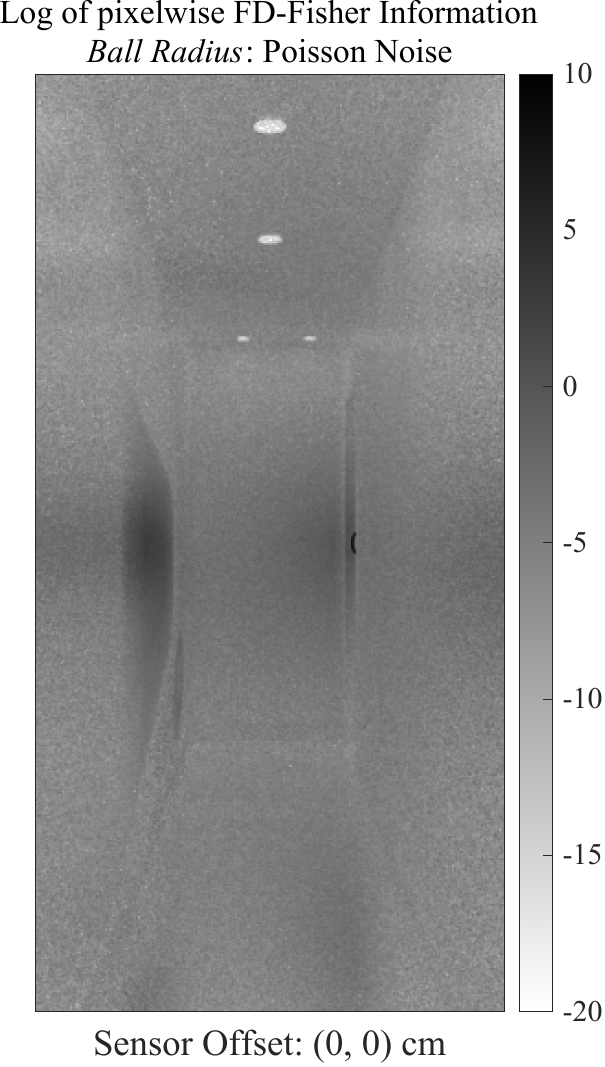}
        \label{fig:standard_poisson_rad}
    }
    \subfloat[][]
    {
        \includegraphics[width=0.155\linewidth]{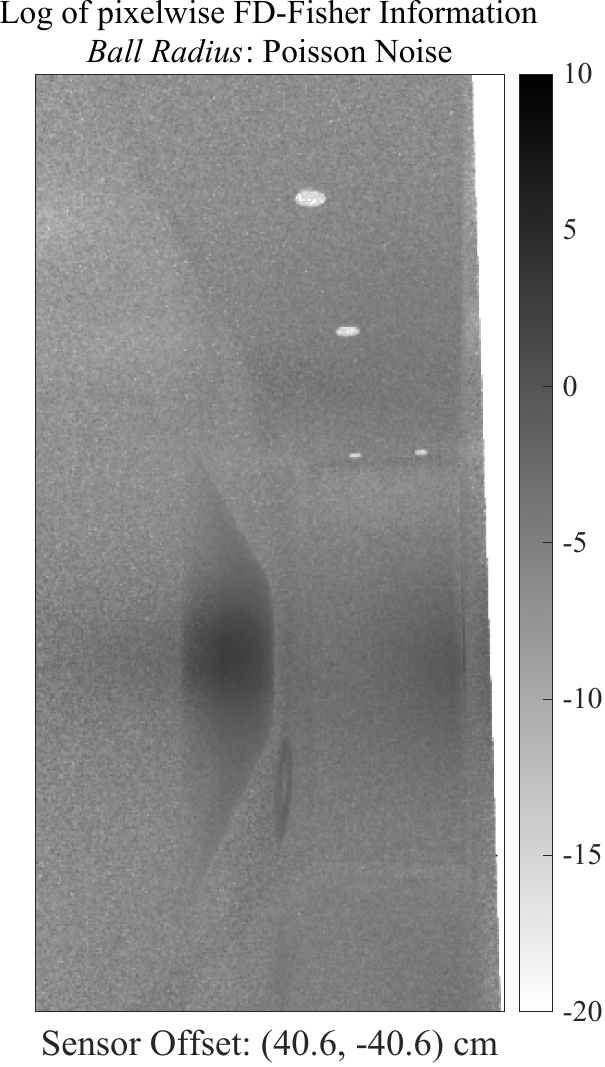}
        \label{fig:bottom_right_poisson_rad}
    }
    \caption{Pixelwise FD-Fisher Information for the multiple viewpoints denoted in Figure~\ref{fig:sensor_offset_grid}: (a), (d) sensor offset $(-40.64, 40.64)\rm{cm}$; (b), (e) sensor offset $(0, 0)\rm{cm}$; (c), (f) sensor offset $(40.64, -40.64)\rm{cm}$. Top row: FD-FI for ball location with AWGN with $\sigma = 0.2$. Note that the highest information content is found in (a) when the boundary of the ball is visible, but then switches to be the left wall of the hallway in (b) and (c), as this is where most reflections containing information about the ball's position reside. Bottom row: FD-FI for ball radius with Poisson noise. Here again, the highest information content is found along the ball's boundary in (d), and on the left wall of the hallway in (e) and (f), followed closely by the ``corner camera'' penumbra on the hallway floor.}
    \label{fig:pos_rad_3views}
\end{figure}

\begin{figure}[h]
	\centering
	\subfloat[][]
	{		
		\includegraphics[width=0.24\columnwidth]{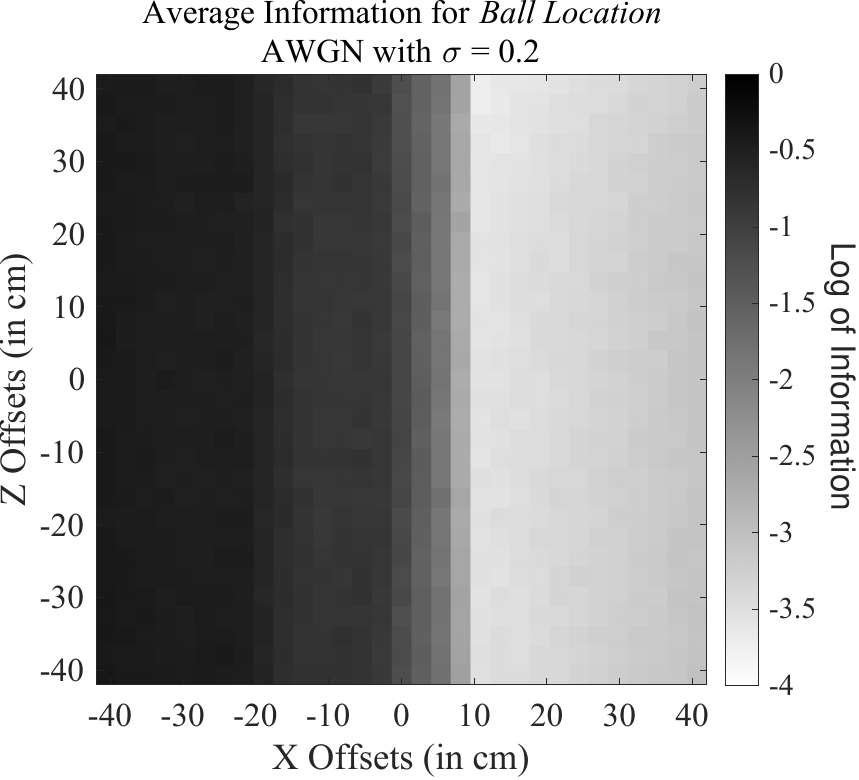}
		\label{fig:pos_awgn_multiview}
	}	
	\subfloat[][]
	{		
		\includegraphics[width=0.24\columnwidth]{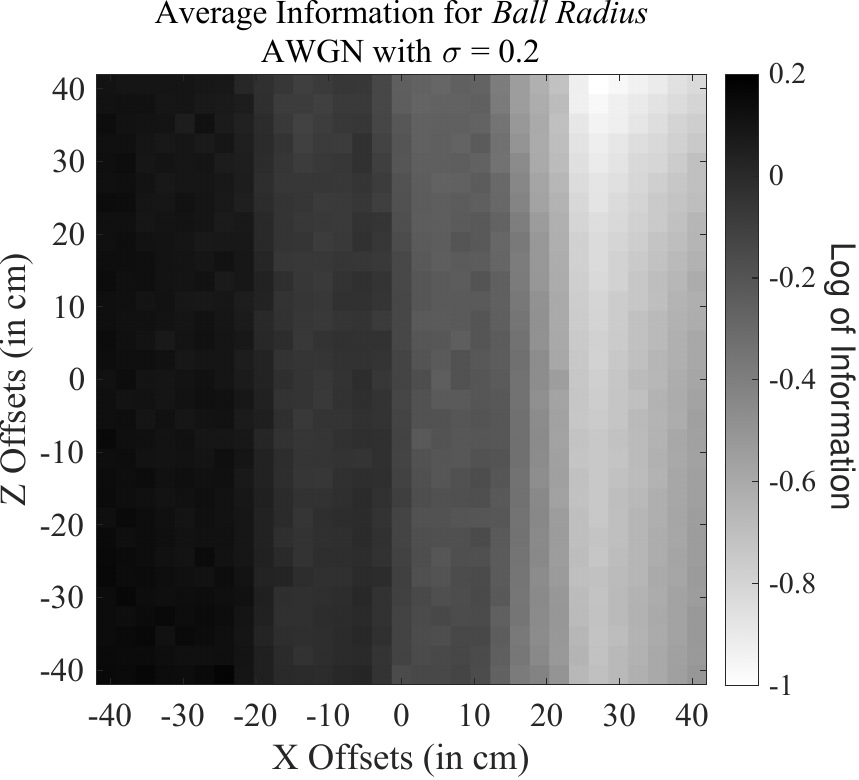}
		\label{fig:rad_awgn_multiview}
	} \\
	\subfloat[][]
	{		
		\includegraphics[width=0.24\columnwidth]{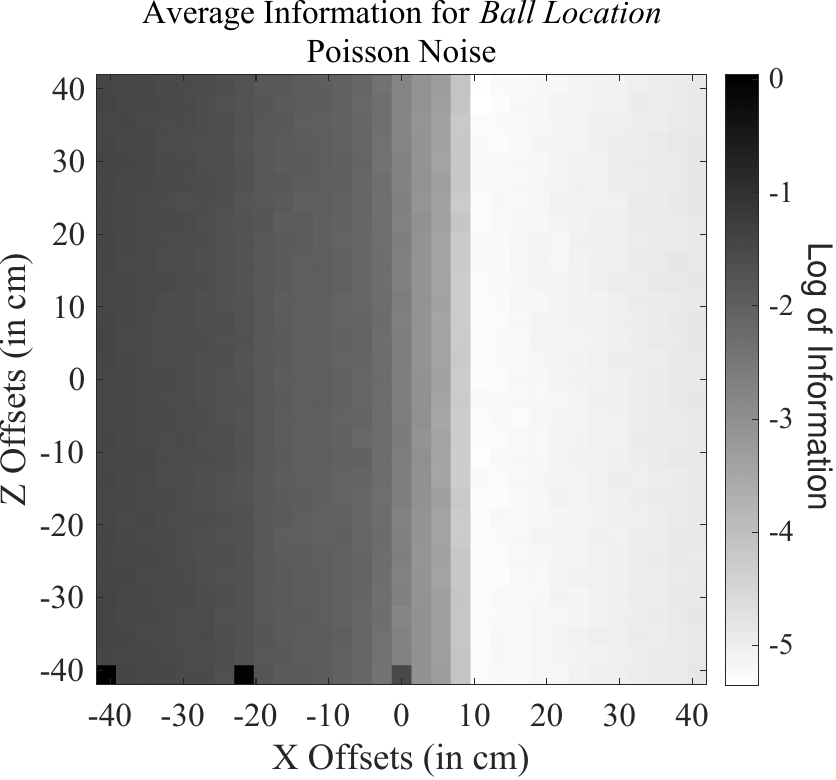}
		\label{fig:pos_poisson_multiview}
	}
	\subfloat[][]
	{		
		\includegraphics[width=0.24\columnwidth]{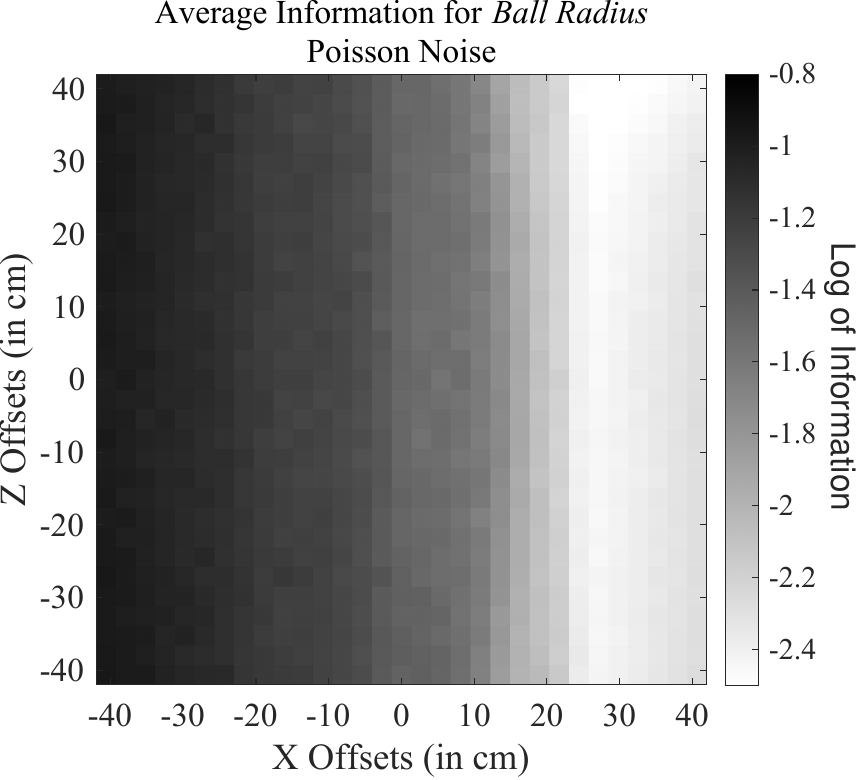}
		\label{fig:rad_poisson_multiview}
	}
    \caption{Average information for each view. Top row: AWGN with $\sigma = 0.2$, for (a) ball location, and (b) ball radius. Note the presence of several transition regions in (a) around $x\approx-20\rm{cm}$ and then $x\approx10\rm{cm}$, and in (b) at $x\approx-20\rm{cm}$, $x\approx0\rm{cm}$, and $x\approx20\rm{cm}$; when $x > 10\rm{cm}$, none of the ball is visible and most of the right wall of the hallway is occluded from the camera as well. Bottom row: Poisson noise for (c) ball location and (d) ball radius. Here, there is a single transition region in (c) around $x\approx10\rm{cm}$, but 2 transition regions in (d) at $x\approx0\rm{cm}$ and $x\approx10\rm{cm}$. The several very large pixels on the bottom row of (c) are areas where much of the red ball is visible.}
    \label{fig:multiview_grids}
\end{figure}

Figure~\ref{fig:pos_rad_3views} shows the pixelwise FD-Fisher Information for the three sensor offsets in \Cref{fig:sensor_grid_example} with the red ball just outside of the camera's LOS in the doorway (between (3) and (4) in Figure~\ref{fig:ballManifold}). We show FD-FI images for ball location with AWGN (top row), and ball radius with Poisson noise (bottom row). Unsurprisingly, when the red ball is visible, the pixels with the highest information content occur at the ball's boundary, as seen in \Cref{subsec:HCR_results_exact}. When the ball is outside of the LOS of the sensor (due to the sensor position offset, not ball motion), the highest-information regions are where the most reflections concerning the ball's location and radius are concentrated: the left wall of the hallway (due to the room lights casting shadows) and the floor of the hallway by the door to the room (due to the ``corner camera'' effect of the doorway). 

Figure~\ref{fig:multiview_grids} shows the spatial average over the FD-FI images for each sensor offset: for each of the four images, each pixel is the average of an entire FD-FI image. We show these average images for AWGN with $\sigma=0.2$ (top row) and Poisson noise (bottom row). Again unsurprisingly, the average information in the FD-FI images is higher when the ball is partially or completely in the camera's LOS: in all 4 average information images in \Cref{fig:multiview_grids}, there is a transition region around $x\approx10\rm{cm}$ where the sensor transitions from being parallel to the right wall of the hallway and the ball is totally obscured from view. However, it is interesting to note that there appear to be several other vertical ``transition regions'' in the images, between areas of higher and lower average information. 

For the AWGN noise model (top row): in \Cref{fig:pos_awgn_multiview} we see a region of high information from $x\approx-40\rm{cm}$ until $x\approx-20\rm{cm}$, when the average information decreases slightly. Then, there is another transition at $x\approx0 - 10\rm{cm}$, after which the average information is more or less constant. Likewise in \Cref{fig:rad_awgn_multiview}, we see similar transitions at $x\approx-20\rm{cm}$, $x\approx0\rm{cm}$, and $x\approx20\rm{cm}$. These transitions may be due to different types of reflections dominating the information in these regions: from $x\approx-40 \to x\approx-20\rm{cm}$, the sensor can see the red ball; from $x\approx-20 \to x\approx0\rm{cm}$, the sensor can see high information regions near the ball in the doorway, and on the left wall, floor, and back wall of the hallway; however, from $x\approx20 \to x\approx40\rm{cm}$, the sensor can only see high information regions on the floor and walls of the hallway. We see less pronounced transitions in the Poisson noise case: in \Cref{fig:pos_poisson_multiview} there appears to be a single transition around $x\approx 10 \rm{cm}$ with two smoother transition regions around $x\approx -5 \rm{cm}$ and $x\approx -20 \rm{cm}$, and in \Cref{fig:rad_poisson_multiview}, there is one transition around $x\approx 22 \rm{cm}$, with smoother transition regions at $x\approx 10 \rm{cm}$ and $x\approx -5 \rm{cm}$. Consistent with our intuition, there is virtually no variation in information in the $z$-offset direction for either noise model.

In the next section we examine the effects of rendering errors on the computed lower bounds.

%% file: inexact_renderings.tex
\section{Extensions: Computing Lower bounds with Inexact Rendering}\label{sec:rendering_errors}

The framework above assumes that the rendering engine yields the \emph{exact} plenoptic intensities for a given set of scene parameter values.  However, rendering engines are not perfect and the rendered images used to compute the bounds above could be erooneous with per-pixel error that depends inversely on the number of Monte-Carlo samples used by the renderer to solve the rendering equation (\ref{eq:rendering_eq}). This section contains the second major novel contribution of this paper: analyzing the inconsistencies that arise when using plenoptic data with rendering errors to compute lower bounds of the form of \cref{eq:HCR-Poisson,eq:HCR_AWGN}. We also provide a simple algorithm to estimate intervals for the true HCR-LB when using inexact renderings. 

The errors induced by light-transport (or rendering) algorithms have been previously studied by Arvo et al. \cite{arvo1994error}, whose authors identify different sources of error in the rendering algorithm and provide theoretical bounds. More recently, Celarek et al. \cite{celarek2019error} propose a methodology to numerically estimate rendering errors in terms of MSE and also per-pixel standard deviation using multiple short renderings, and Iseringhausen et. al. \cite{iseringhausen_non-line--sight_2020} provide numerical bounds on the error introduced by rendering approximations. While this work aims to use the error estimates to compare various rendering methodologies, our aim here is to assess and quantify how rendering errors percolate into our HCR-LB framework. 

It is further useful to highlight the distinction between the rendering errors considered in this section and the different noise models for the plenoptic observation in \Cref{subsec:HCR_Poisson,subsec:HCR_Gaussian}. While the Poisson and AWGN noise models describe the noise/uncertainty in the observations of an imaging system for scene parameter estimation, this section analyzes the effect of using inaccurately rendered data to compute the HCR-LB presented in \Cref{subsec:HCR_Poisson,subsec:HCR_Gaussian}.

For some value of the scene parameter $\btheta$, let us denote the true plenoptic intensities as $\Lb_{\btheta}$, and the output of the rendering algorithm computed using $N$ samples per-pixel as, $\tilde{\Lb}^{(N)}_{\btheta} = \Lb_{\btheta} + \cE_{\btheta}^{(N)}$, where $\cE^{(N)}_{\btheta}$ is the (additive) rendering error. In general, the rendering error is not actually additive Gaussian, but when rendering with a large number of samples this is approximately the case due to the central limit theorem. In our analysis, we assume that for all $\btheta\in\Theta$, and $\bomega\in\Omega$, the rendering algorithm satisfies the following assumptions: 
\begin{enumerate}[label=\bf A.\arabic*]
	\setlength\itemindent{1em}
	\item \label{as:unbiased} The rendering algorithm is unbiased, i.e.  $\EE [\tilde{\Lb}^{(N)}_{\btheta} (\bomega)] = \Lb_{\btheta} (\bomega)$, or equivalently $\EE [\cE^{(N)}_{\btheta} (\bomega)] = 0$.
	\item \label{as:decay_rate} Most weighted sums of pixel-wise error variance decays at the rate proportional to $ N^{-1} $.
	Formally, for weights $W_{\bomega} \sim \text{Uniform}\:[0,{\rm L_{max}}]$, where $\rm L_{max}$ is an absolute constant, we assume that
	$$\sum\limits_{\bomega\in\Omega} W_{\bomega} \cdot \Var( \cE^{(N)}_{\btheta} (\bomega) ) = C_{\btheta} (\bomega) \cdot N^{-1},$$ with high probability\footnote{The probability is with respect to the randomness in the weights $W_{\bomega}$ and not the rendering errors.}, for some scene-dependent constant $C_{\btheta} (\bomega)>0$. 
	\item \label{as:bdd_moments} The higher-order moments of the relative per-pixel error are $o(N^{-1})$; in particular, we assume $\exists\ N_0, \delta>0$, s.t. for all $N>N_0$ and $k>2$
    \begin{eqnarray}
        \ \EE \left[ \left| \frac{\cE_{\btheta}^{(N)}(\omega)}{\Lb_{\btheta}(\omega) }  \right|^k \right] = \cO(N^{-(1+\delta)} ) \nonumber
    \end{eqnarray}
	\item \label{as:indep_errors} The pixel-wise errors for different scene parameter values are statistically uncorrelated. 
\end{enumerate}

\begin{remark}\label{rem:unbiased}
	Assumption \ref{as:unbiased} is a mild assumption and is satisfied by a lot of modern rendering algorithms. In particular, we use {\tt Mitsuba 3}'s \cite{jakob2022mitsuba3} path tracer, {\tt Mitsuba}'s \cite{Mitsuba} bi-directional path tracer (BDPT) and {\tt Redner}'s \cite{redner} path tracer, which are unbiased, for rendering all our scenes.
\end{remark}

\begin{remark}\label{rem:decay_rates}
	A sufficient condition for Assumption \ref{as:decay_rate} is that the error variance of each pixel, $ \Var ( \cE^{(N)}_{\btheta} (\bomega) ) = \Theta(N^{-1})$ hold for all ${\bomega\in\Omega}$. While such behavior is expected from standard Monte-Carlo based algorithms, modern rendering algorithms use other techniques like importance sampling to improve the performance. Hence the $N^{-1}$ rate might not hold pixel-wise, but is typically valid when averaged over all pixels. 
\end{remark}
We present simulations in Appendix \ref{app:var_validation} to empirically validate Assumption \ref{as:decay_rate}. In particular, we show empirically that the weighted sum of pixel-wise error variance decays at $1/N$ for the example scene used in this paper.
For rendering algorithms with different rates of convergence, we could simply assume that $$\sum\limits_{\bomega\in\Omega} W_{\bomega} \cdot \Var( \cE^{(N)}_{\btheta} (\bomega) ) = C_{\bomega}\cdot N^{-p},$$ where the degree of decay $p>0$ depends on the nature of the rendering algorithm used. The decay rate $p$ may be known \emph{a priori} or could be estimated empirically from simulations similar to that presented in Appedix \ref{app:var_validation}. Our analysis naturally extends for any $p>0$.

\begin{remark}\label{rem:bdd_moments}
	Assumption \ref{as:bdd_moments} is automatically satisfied if the rendering errors are bounded in magnitude, which they typically are for most rendering algorithms used in practice, and when Assumption \ref{as:decay_rate} holds.  This criterion suggests that for ``sufficiently large" values of $N$, the higher-order error terms are ``relatively small."
\end{remark}

\begin{remark}\label{rem:indep_errors}
	Assumption \ref{as:indep_errors} holds when the renderer uses different (random number) seeds for generating its samples.
\end{remark}

We first look at the functional form of the HCR-LB presented in \Cref{subsec:HCR_Poisson,subsec:HCR_Gaussian}. We can see that under both Poisson noise and AWGN models, the HCR-LB for $\btheta^*_j$ takes the form
\begin{align}\label{eq:lambda_and_HCR}
	\HCR (\btheta^*_j) = \sup \limits_{\substack{\bDelta\neq \mathbf{0} \\ \btheta^*+\bDelta\in\Theta}} \hspace{0.2em}  \frac{\bDelta^2_j}{\exp \cbr{ \lambda(\Lb_{\btheta^*}, \Lb_{\btheta^*+\bDelta}) }-1},
\end{align}
where $\lambda(\Lb_{\btheta^*}, \Lb_{\btheta^*+\bDelta}) $ captures the dependence on the rendered plenoptic intensities. It is easy to see that, 
\begin{align}\label{eq:lambda}
	\lambda(\Lb_{\btheta^*}, \Lb_{\btheta^*+\bDelta}) & =\begin{cases} 
		\lambda_P \triangleq \sum\limits_{\bomega\in\Omega} \frac{( \Lb_{\btheta^*}(\bomega)  - \Lb_{\btheta^*+\bDelta} (\bomega) )^2}{\Lb_{\btheta^*}(\bomega) }     &  \mbox{for Poisson noise}   \\
		\lambda_G \triangleq \frac{\| \Lb_{\btheta^*}  - \Lb_{\btheta^*+\bDelta} \| ^2}{\sigma^2}    & \mbox{for } \mathcal{N}(0, \sigma^2) \mbox{ noise} 
	\end{cases} 
\end{align}
We suppress the dependence of $\lambda$ on $\btheta^*$, $\bDelta$, and the forward model for the sake of brevity. Such dependences will be made explicit whenever additional clarity is required. 

When we use inaccurately rendered plenoptic data $\tilde{\Lb}^{(N)}_{\btheta^*}$ and $\tilde{\Lb}^{(N)}_{\btheta^*+\bDelta}$, where we use $N$ samples per-pixel for rendering, we end up with $\tilde{\lambda}^{(N)}$ and obtain an erroneous estimate of the HCR functional (the function being maximized in the HCR-LB),
\begin{align}\label{eq:HCR_functional}
	f(\tilde{\lambda}^{(N)}; \btheta^*, \bDelta) = \frac{\bDelta^2_j}{  \exp \left( \tilde{\lambda}^{(N)} \right) -1}.
\end{align}

We first characterize the effect of rendering error in computing $\lambda$ using the following claims, and then use that to analyze the effect on the HCR functional (\ref{eq:HCR_functional}), and the overall HCR-LB. The proofs of Theorems \ref{claim:Poisson} and \ref{claim:Gaussian} appear in Appendices \ref{proof:Poisson_claim} and \ref{proof:Gaussian_claim} in the supplementary material, respectively.

\begin{theorem}[Effect of rendering error on $\lambda_P$]\label{claim:Poisson}
	Let us denote the value of $\lambda_P$ computed using the (inexactly) rendered data with $N$ samples per-pixel as $\tilde{\lambda}_P^{(N)}$. Then we have
	\begin{align}\label{eq:lambda_P}
		\tilde{\lambda}_P^{(N)} = \lambda_P + \frac{C_P}{N} + \eta_P, 
	\end{align}
	where $C_P\geq0$ is a constant that depends only on scene parameters, $|\EE [\eta_P]| = \cO( N^{-(1+\delta)} )$, and $\Var(\eta_P) = \cO(N^{-1})$, and $\delta>0$ is the constant appearing in Assumption \ref{as:bdd_moments}.
\end{theorem}

\begin{theorem}[Effect of rendering error on $\lambda_G$]\label{claim:Gaussian}
	Let us denote the value of $\lambda_G$ computed using the (inexactly) rendered data with $N$ samples per-pixel as $\tilde{\lambda}_G^{(N)}$. Then we have
	\begin{align}\label{eq:lambda_G}
		\tilde{\lambda}_G^{(N)} = \lambda_G + \frac{C_G}{N} + \eta_G,
	\end{align}
	where $C_G\geq0$ is a constant that implicitly depends only on scene parameters and AWGN variance $\sigma^2$, $\EE [\eta_G] = 0$, and $\Var(\eta_G) = \cO(N^{-1})$. 
\end{theorem}

\Cref{claim:Poisson,claim:Gaussian} imply that $\tilde\lambda$ overestimates the true $\lambda$ in expectation, where the estimation bias shrinks at the rate of $\Theta(N^{-1})$. While the $\frac{C_P}{N}$ and $\frac{C_G}{N}$ terms denote the (positive) \emph{bias}, the $\eta_P$ and $\eta_G$ terms codify the \emph{variance} of the computed $\tilde\lambda$'s. Note as well that $\eta_P$ introduces a small amount of bias, unlike $\eta_G$, due to the characteristics of Poisson noise. Although both bias and the variance terms exhibit the same rates of $\cO(N^{-1})$, there is a \emph{bias-variance} tradeoff in the computed $\tilde{\lambda}$ that depends on the scene parameters $\btheta^*$ and $\btheta^* + \bDelta$. In order to better understand this \emph{bias-variance} tradeoff, let us take a closer look at the expression of $\tilde\lambda$. We use the AWGN case here for the sake of exposition, but the tradeoff holds similarly for the Poisson noise model as well. 

Using (\ref{eq:lambda}) we can write,
\begin{align}
	\tilde{\lambda}_G^{(N)} = \lambda_G + & \underbrace{ \frac{1}{\sigma^2} \| \cE_{\btheta^*} - \cE_{\btheta^*+\bDelta} \|^2}_{\frac{C_G}{N}} + \underbrace{ \frac{2}{\sigma^2} (\Lb_{\btheta^*} - \Lb_{\btheta^*+\bDelta} )^T \cdot (\cE_{\btheta^*} - \cE_{\btheta^*+\bDelta} ) }_{\eta_G}.
\end{align}

Let us see how $\frac{C_G}{N}$ and $\eta_G$ behave under 2 different scenarios:
\begin{enumerate}[label=\underline{\bf Case (\arabic*):}]
	\setlength\itemindent{2.1em}
	\item \label{case:var}
		$\| \Lb_{\btheta^*} - \Lb_{\btheta^*+\bDelta} \| \gg \| \cE_{\btheta^*} - \cE_{\btheta^*+\bDelta} \|$:\\
		In this scenario, $|\frac{C_G}{N}| \ll |{\eta_G}|$, and we can write $\tilde{\lambda}_G^{(N)} \approx \lambda_G + \eta_G$. Since $ \EE [\eta_G] = 0$ (from Assumption \ref{as:unbiased}), we can see that the computed value $\tilde\lambda_G^{(N)}$ is (approximately) unbiased, i.e., $\EE [\tilde{\lambda}_G^{(N)}] \approx \lambda_G$. Furthermore, we have $\Var ( \tilde\lambda_G^{(N)} ) \approx \Var (\eta_G) = \cO(N^{-1})$ from Assumption \ref{as:decay_rate}. In other words, this is a regime where the (zero-mean) \emph{variance} term dominates the overall error. This typically happens when:
		\begin{itemize} 
			\setlength\itemindent{1em}
			\item $\| \bDelta \|$ is large, or
			\item the parameters of interest are in LOS, which means that even a small perturbation of the true parameter $\btheta^*$ would result in a large difference in the plenoptic observations, or
			\item the value of $N\gg 1$, which would result in the rendering errors being very small.
		\end{itemize}
	\item \label{case:bias}
		$\| \Lb_{\btheta^*} - \Lb_{\btheta^*+\bDelta} \| \ll \| \cE_{\btheta^*} - \cE_{\btheta^*+\bDelta} \|$:\\
		In this scenario, $|\frac{C_G}{N}| \gg |\eta_G|$, and we can write $\tilde{\lambda}_G^{(N)} \approx \lambda_G + \frac{C_G}{N} > \lambda_G$. This in turn means that the HCR functional is underestimated, i.e., $f(\tilde{\lambda}^{(N)}_G; \btheta^*, \bDelta,) < f(\lambda_G; \btheta^*, \bDelta)$. Furthermore, we have $\EE [\frac{C_G}{N}] = \Theta(N^{-1})$ and $\Var (\frac{C_G}{N}) = \cO( N^{ -(1 +\delta) })$ from Assumptions \ref{as:decay_rate} and \ref{as:bdd_moments}, respectively. In other words, this is a regime where the (non-negative) \emph{bias} term dominates the overall error. This happens when:
		\begin{itemize} 
			\setlength\itemindent{1em}
			\item $\| \bDelta \|$ is small, or
			\item the parameters of interest are in NLOS, which means that even large perturbations of the true parameter $\btheta^*$ would yield very similar plenoptic observations, or
			\item the value of $N$ is small, which would result in larger rendering errors. 
		\end{itemize}
\end{enumerate}
Furthermore, since the overall HCR-LB is achieved by maximizing the HCR functional $f(\lambda;\btheta^*, \bDelta)$, the supremum typically occurs for small values of $\lambda$ or when $\| \Lb_{\btheta^*} - \Lb_{\btheta^*+\bDelta} \|$ is small (corresponding to Case (2) above). Thus the discussion above implies that the overall HCR-LB is usually underestimated, especially for NLOS imaging problems, when using inaccurately rendered plenoptic data to compute the bounds.

\begin{remark}[Effect of rendering error on the overall HCR lower bound]\label{rem:HCR}
	A direct implication of Theorems \ref{claim:Poisson} and \ref{claim:Gaussian} is that the HCR-LB for Poisson and AWGN noise models computed using plenoptic data rendered with $N$ samples per-pixel obeys, 
	\begin{align}
		\label{eq:rem:HCR_convergence}
		\lim\limits_{N\rightarrow\infty} \HCR_N(\theta^*_j) &\overset{a.s}{=} \HCR(\theta^*_j), \hspace{1em}\forall\ j = 1,2,\dots,J.
	\end{align} 
	Furthermore, if the supremum of the true HCR functional $f(\lambda; \btheta^*, \bDelta)$ occurs for $\| \bDelta \| \leq \epsilon$, then from the above discussion we can infer that there exists a constant $N_0(\epsilon)$ that depends on the scene parameters such that for all $N<N_0(\epsilon)$, with high probability,
	\begin{align}
	\label{eq:rem:HCR_bias}
		{ \HCR_N(\theta^*_j) } &\leq \HCR(\theta^*_j) \hspace{1em}\forall\ j = 1,2,\dots,J.
	\end{align}
	The value $N_0(\epsilon)$ essentially determines how the number of samples per-pixel affects the bias-variance tradeoff described above. 
\end{remark}

Equation (\ref{eq:rem:HCR_bias}) simply follows from the observation that for small values of $N$, the bias-terms $\left(\frac{C_P}{N} \text{ and } \frac{C_G}{N}\right)$ dominate the variance terms ($\eta_P$ and $\eta_G$ respectively) so that $\tilde{\lambda}^N \geq \lambda$, which in turn means that $\HCR_N (\theta^*_j) \leq \HCR(\theta^*_j)$.

\subsection{Estimating HCR Lower Bounds with Inexact Rendering}\label{subsec:HCR_estimation}
Our error analysis above shows how rendering error manifests itself in the computation of the HCR-LB. One important takeaway from this analysis is that in order to get a good estimate of the HCR-LB, one should use as many samples per-pixel as are feasible to render the scenes. However, this might result in a computational bottleneck as ray-tracing is a highly time and memory intensive process. For example, rendering some of the high-resolution images in Section \ref{subsec:HCR_results_exact} until convergence using {\tt Mitsuba 3} \cite{jakob2022mitsuba3} took several minutes, and we had to render hundreds of such images for the location and radius estimation examples, leading to many hundreds of hours of GPU compute time. 

In this section, we describe a simple method to estimate upper and lower intervals for the true HCR-LB. For any given values of $\btheta^*$ and $\bDelta$, Theorems \ref{claim:Poisson} and \ref{claim:Gaussian} suggest that the relationship between the true and the observed $\lambda$s (for both Poisson and AWGN noise models) is given by,
\begin{align}
	\tilde{\lambda}^{(N)}(\btheta^*, \bDelta) = \lambda(\btheta^*, \bDelta) + \frac{C(\btheta^*, \bDelta)}{N} + \eta,
\end{align}
where $C(\btheta^*, \bDelta)\geq 0$ is the coefficient of the rate of decay of the bias, and $\eta$ models the higher-order (moments 3 and higher) terms of the rendering errors, with $\Var(\eta) = \cO(N^{-1})$. 
Thus, we can use a data-driven method to estimate the unknown $\lambda(\btheta^*, \bDelta)$ and $C(\btheta^*, \bDelta)$ by rendering the scenes with different values of $N$ and solving for the unknowns using a simple (weighted) least squares algorithm. 

In particular, if we render the scenes for parameter values $\btheta^*$ and $\btheta^*+\bDelta$ using $N_1, N_2, \dots, N_k$ samples per-pixel, for some $k\geq 2$, then we get the following system of linear equations,
\begin{align}\label{eq:lambda_eqn_system}
	\left[ \begin{array}{c} \tilde{\lambda}^{(N_1)}(\btheta^*, \bDelta) \\ \vdots \\ \tilde{\lambda}^{(N_k)}(\btheta^*, \bDelta) \end{array} \right] 
	=  \left[ \begin{array}{cc} 1 & \frac{1}{N_1} \\ \vdots & \vdots \\ 1 & \frac{1}{N_k} \end{array} \right] 
		\hspace{-0.2em}\left[ \begin{array}{c} \lambda(\btheta^*, \bDelta) \\ C(\btheta^*, \bDelta) \end{array} \right] 
		+ \left[ \begin{array}{c} \eta_1 \\ \vdots \\ \eta_k \end{array} \right] .
\end{align}
Equation (\ref{eq:lambda_eqn_system}) can be solved in closed form to obtain an unbiased estimate $\hat{\lambda}(\btheta^*, \bDelta)$ (approximately unbiased for the case of Poisson noise model) of $\lambda(\btheta^*, \bDelta)$. We can then use the estimated $\hat{\lambda}(\btheta^*, \bDelta)$ in (\ref{eq:lambda_and_HCR}) to get 
\begin{align}
	\hat{\HCR}(\btheta^*_j) := \sup \limits_{\substack{\bDelta\neq \mathbf{0} \\ \btheta^*+\bDelta\in\Theta}} \hspace{0.2em}  \frac{\bDelta^2_j}{\exp \cbr{ \hat{\lambda} (\btheta^*, \bDelta) }-1}, \hspace{1em}\forall\ j = 1,\dots,J,
\end{align}
where $\hat{\HCR}(\btheta^*_j)$, is a random variable whose randomness or stochasticity arises due to the rendering errors. 

Then we make the following claim whose proof appears in Appendix \ref{proof:HCR_estimate}.

\begin{claim}[Relationship between $\hat{\HCR}(\btheta^*_j)$ and $\HCR(\btheta^*_j)$]\label{claim:HCR_estimate}
	For any given parameter value $\btheta^*$, the HCR lower bound computed using unbiased estimates $\hat{\lambda} (\btheta^*, \bDelta)$ (for all $\bDelta$) is, in expectation, an upper bound on the true HCR lower bound, i.e.,
	\begin{align}\label{eq:claim:HCR_estimate}
	\EE \sbr{ \hat{\HCR}(\btheta^*_j) } \geq \HCR(\btheta^*_j),  \hspace{1em}\forall\ i = 1,2,\dots,J.
	\end{align}
\end{claim}

While Claim \ref{claim:HCR_estimate} gives an upper bound (in expectation) on the true HCR-LB, Remark \ref{rem:HCR} suggests that the HCR-LB computed directly using rendered plenoptic data is (typically) a lower bound on the true HCR-LB value. Thus, for any given scene and unknown parameter value, we can compute an upper and lower interval within which the true HCR-LB is expected to lie using a fixed computational/rendering budget. In the next section, we present experimental results for the problem of object localization, which demonstrates the utility of our error analysis framework.

It is worth commenting on the relationship between the $\lambda$s in our error analysis and the CR-LB. We can obtain the CR-LB for some scalar parameter $\btheta^*$ as
\begin{align}\label{eq:CR_lambda}
	\CR (\btheta^*) = \lim\limits_{\bDelta \rightarrow 0} \frac{\bDelta^2}{\lambda(\Lb_{\btheta^*}, \Lb_{\btheta^*+\bDelta})}.
\end{align}
From Remark \ref{rem:HCR} and the discussion above it, we can see that for small values of $\bDelta$, using rendered data (and $\tilde{\lambda}^N$) to compute the CR-LB (using finite differences), typically results in underestimating the CR-LB as well especially for NLOS parameter estimation problems.

\subsection{An Illustrative Example: Idealized Hallway Scene}\label{sec:ideal_hallway_def}

\begin{figure}[]
	\centering
	\subfloat[][]
	{
		\includegraphics[width=0.4\columnwidth]{figures/NLOS_lb/scene_defn/top_view_exp2.pdf}
		\label{fig:top_view_2}
	}
	\subfloat[][]
	{
		\includegraphics[height=0.3\columnwidth]{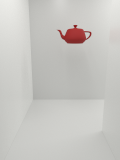}
		\label{fig:rgb_scene_0.2m}
	}
	\subfloat[][]
	{
		\includegraphics[height=0.3\columnwidth]{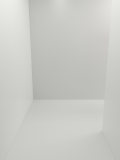}
		\label{fig:rgb_scene_0.9m}
	}
	\caption{Scene geometry for NLOS object localization. (a) Top-view of scene layout of a $\Pi$-shaped hallway with dimensions marked. Corridors A, B, and C are $2.5\rm{m}$, $3\rm{m}$, and $2.5\rm{m}$ long respectively, and $2\rm{m}$ tall. The hallway is illuminated with white ceiling lights with a radiance of $12\ \rm{W\cdot sr^{-1}m^{-2}}$. Instead of a red spherical object, we consider a red teapot, and the camera is placed in the middle of hallway A and captures RGB images. The \emph{scalar} parameter of interest $\btheta^*$ is the horizontal displacement of the teapot from intersections of corridors A and B. RGB images rendered using \texttt{Redner} for different values of $\btheta^*$ are shown in (b) and (c). $65536$ samples-per pixel were used, and it took around $3.3$ minutes to render each scene; (b) $\btheta^* = 0.2\rm{m}$: teapot fully in LOS; (c) $\btheta^* = 0.9\rm{m}$: teapot just moved completely away from LOS.}	
	\label{fig:scene_defn_2}
\end{figure}

In contrast to the scene modeled after the real hallway which was used in the experiments in the preceding section, we here define an idealized and simplified scene in order to clearly show the effect of rendering error on the computation of lower bounds. Here, the scene consists of a $\Pi$-shaped hallway with dimensions as marked in Figure \ref{fig:top_view_2}. The inner walls of this hallway are painted with a diffuse eggshell paint, and their BRDF is modeled as above using the \texttt{roughplastic} plugin in {\tt Mitsuba} \cite{Mitsuba,walter_microfacet_2007}. We consider the problem of estimating the location of a red teapot (downloaded from Morgan McGuire's website \cite{Teapot_McGuire2017}) that is constrained to lie in a straight line in Corridor B as shown in Figure \ref{fig:top_view_2}. The distance of the teapot from the intersection of corridors A and B is the \emph{scalar} parameter of interest $\btheta^*$. The camera is located at the center of corridor A as shown in Figure \ref{fig:top_view_2} and captures RGB images of size $160\times120$. Using the insights on Fisher information from Section \ref{subsec:HCR_results_exact}, we see that regions on the floor near the corner have more information about hidden objects than others. Hence we point the camera slightly towards the floor instead of looking straight at the back wall. The ceiling lights are $0.1\rm{m}\times 1.5\rm{m}$ with radiance set to $12\:\rm{W \cdot sr^{-1}m^{-2}}$, emitting white light (uniformly over all wavelengths)\footnotemark. We use {\tt Redner}'s \cite{redner} default path-tracer to render the scenes. \Cref{fig:rgb_scene_0.2m,fig:rgb_scene_0.9m} are the rendered images for $\btheta^* = 0.2\rm{m}$ and $0.9\rm{m}$ respectively using $65536$ SPP. 

\footnotetext{The rendered images using this illumination and camera setup used here are roughly similar to using a commercially-available $2000\:\rm{lm}$ ceiling light with an exposure time of $1/120$ seconds.}

\subsection{Numerical Results: NLOS Object Localization Lower Bounds with Inexact Rendering}\label{subsec:error_analysis_results}
Here, we show the effect of rendering errors on the lower bound computation for object localization, using our framework which now assumes the rendering is inexact. In this section and the section that follows, we use the idealized hallway scene.

\subsubsection{Setup}
We compute the HCR-LB for $100$ teapot locations from $0.7\rm{m}$ to $1.69\rm{m}$ at $1\rm{cm}$ increments. For obtaining $\hat{\HCR}$ from inexact renderings discussed in Section \ref{subsec:HCR_estimation}, we render all the $100$ scenes with $10$ different values of samples per-pixel, $N=2048, 3072,\dots, 11264\:(=11\cdot1024)$ SPP. For computing the HCR-LB directly from the rendered plenoptic data as discussed in Section \ref{sec:exact_lower_bounds}, we use $N_{\rm eff} = 65536$ SPP. We use the same rendering/computational budget for both methods, i.e., the overall rendering time using $N_{\rm eff} = 65536$ samples and rendering the scene $10$ times with $N=2048, 3072,\dots, 11264$, take approximately the same time (around $3.3$ minutes per teapot location).

\subsubsection{Results and Discussion}
\begin{figure}[]
	\centering
	\subfloat[][]
	{
		\includegraphics[height=0.24\columnwidth]{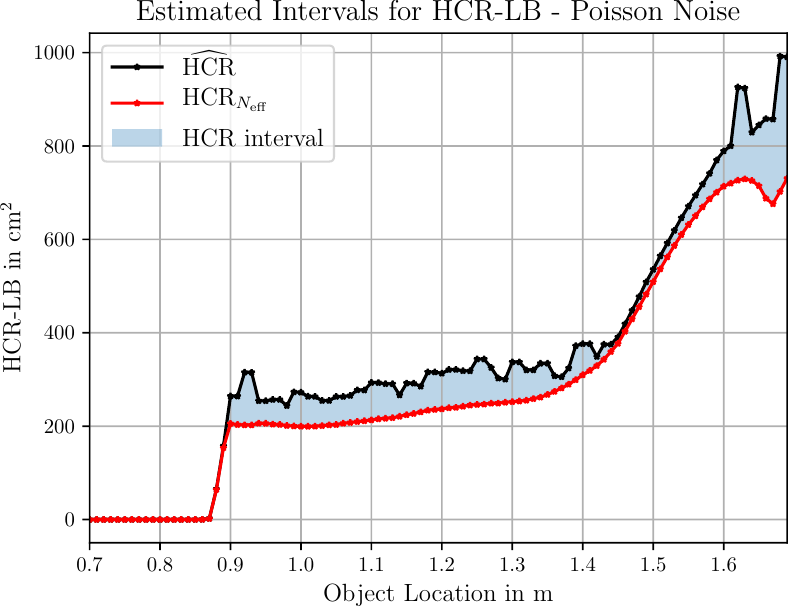}
		\label{fig:HCR_Poisson}
	}
	\subfloat[][]
	{
		\includegraphics[height=0.24\columnwidth]{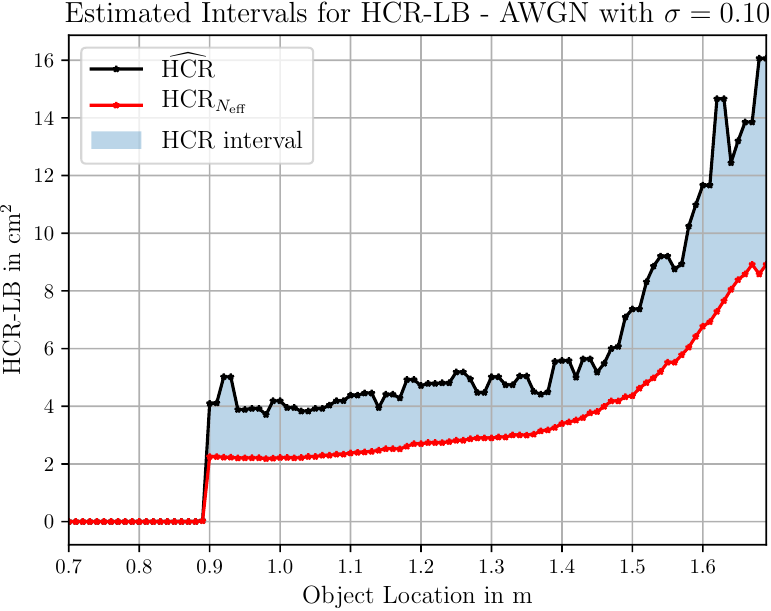}
		\label{fig:HCR_AWGN_0_1}
	}
	\subfloat[][]
	{
		\includegraphics[height=0.24\columnwidth]{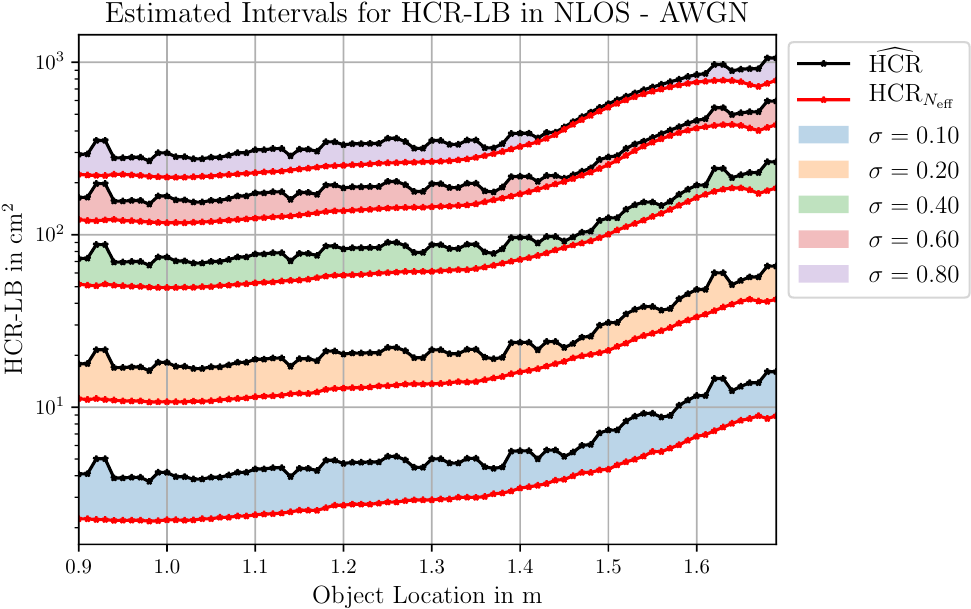}
		\label{fig:HCR_AWGN}
	}
	\caption{The HCR-LB for estimation of teapot location under AWGN and Poisson Noise. $\HCR_{N_{\rm eff}}$ (Red lines): The HCR-LB computed directly using rendered data with $N_{\rm eff} = 65536$ samples per-pixel; $\hat{\HCR}$ (Black lines): The HCR-LB estimated from rendering scenes with $N = 2048, 3072,\dots, 11264$. Due to rendering errors, typically we have $ \EE \sbr{ \hat{\HCR}(\btheta^*) } \geq \HCR(\btheta^*) \geq \HCR_{N_{\rm eff}}$. The region between $\hat{\HCR}$ and $\HCR_{N_{\rm eff}}$ denotes the interval within which the true HCR-LB is likely to lie. (a) The HCR-LB for Poisson noise model. (b) The HCR-LB for AWGN with $\sigma=0.1$. (c) The HCR-LB under AWGN when the teapot is not in LOS for $\sigma=0.1, 0.2, 0.4, 0.6,$ and  $0.8$.}	
	\vspace{-1.5em}
	\label{fig:HCR_est_plots}
\end{figure}

We can see from Figure \ref{fig:HCR_est_plots} that $\hat{\HCR}(\btheta^*) \geq \HCR_{N_{\rm eff}} (\btheta^*)$ for $\btheta^*\geq 0.9$m, which corresponds to the NLOS regime. We can also observe a sharp increase in lower bounds around $\btheta = 0.9\rm{m}$, which corresponds to the teapot moving away from the LOS of the camera. Such an increase is expected since localizing NLOS objects is a much harder problem as we saw in Section \ref{subsec:HCR_results_exact}. We also observe that the HCR-LB (both $\hat{\HCR}$ and $\HCR_{N_{\rm eff}}$) increases for both AWGN and Poisson Noise models as the teapot moves further away from LOS and into corridor B, which again is expected. 

It can also been seen that $\hat{\HCR}$ (black lines) are not as smooth as $\HCR_{N_{\rm eff}}$ (red lines). We believe that this is due to the fact that $\hat{\HCR}$ is a noisy estimate obtained from many sets of low SPP renderings and hence has higher variance than $\HCR_{N_{\rm eff}}$, which is computed using a single set of rendered images with higher SPP. 
Furthermore, it is worth emphasizing that $\hat{\HCR}$ shown as is does not necessarily upper bound the true HCR-LB but the relationship holds in expectation, i.e. $\EE [\hat{\HCR} (\btheta^*)] \geq \HCR(\btheta^*)$, where the expectation is taken with respect to rendering errors. Since computing $\EE[\hat{\HCR}]$ is computationally prohibitive, we use $\hat{\HCR}$ as a surrogate for the upper bound on the true HCR-LB in our discussions. Thus, it is important to keep in mind that the upper bounds (and hence the HCR intervals) shown here are \textit{approximate} and help us understand and interpret the fundamental limits of parameter estimation problems associated with plenoptic imaging systems. 

\begin{figure}[]
	\centering
	\subfloat[][]
	{
		\includegraphics[height=0.24\columnwidth]{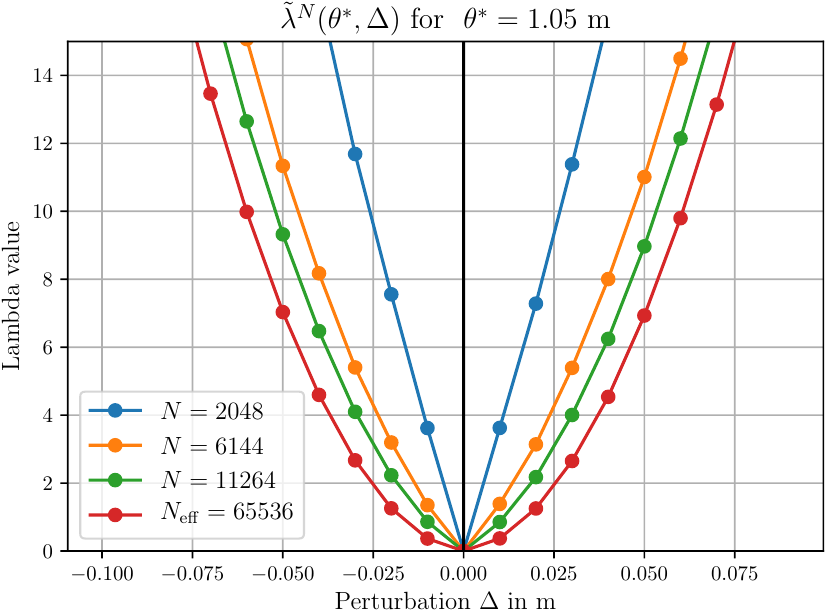}
		\label{fig:lambda_with_N}
	}
	\subfloat[][]
	{
		\includegraphics[height=0.24\columnwidth]{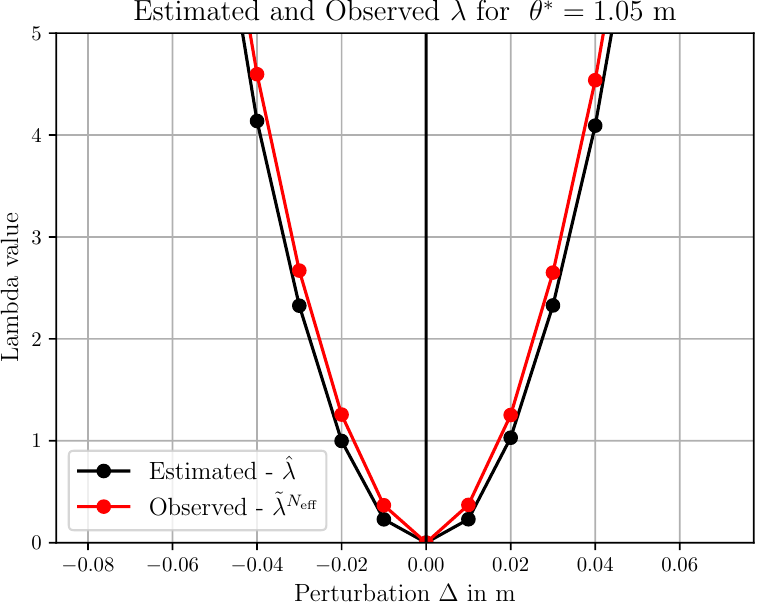}
		\label{fig:lambda_est}
	}
	\subfloat[][]
	{
		\includegraphics[height=0.24\columnwidth]{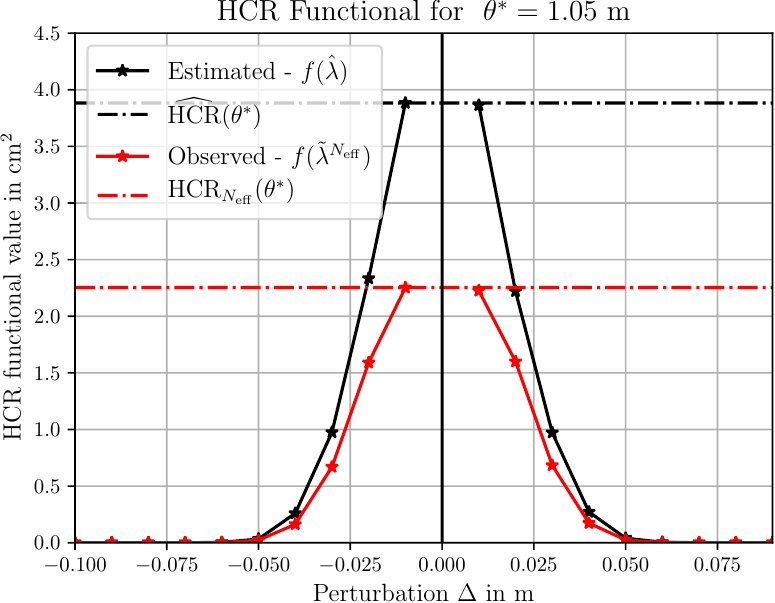}
		\label{fig:HCR_func_example}
	}
	\caption{Effect of samples per-pixel on $\lambda$ and the HCR functional $f(\lambda)$ for estimation of teapot location. Noise model: AWGN with $\sigma = 0.1$, true object location $\btheta^*=1.05{\rm m}$. (a) $\tilde{\lambda}^N$ in the neighborhood of $\btheta^*=1.05{\rm m}$ - shows how $\tilde{\lambda}^N$ decreases with $N$ uniformly for all values of $\bDelta$. (b) Plot of estimated and observed $\lambda$'s for $\btheta^*=1.05{\rm m}$ shows that $\tilde{\lambda}^{N_{\rm eff}} \geq \hat{\lambda}$, even with $N_{\rm eff} = 65536$ samples per-pixels. (c) HCR functional obtained from the estimated and observed $\lambda$s.}	
	\label{fig:lambda_est_plots}
	\vspace{-1em}
\end{figure}

\begin{figure}[]
	\centering
	\subfloat[][]
	{
		\includegraphics[height=0.24\columnwidth]{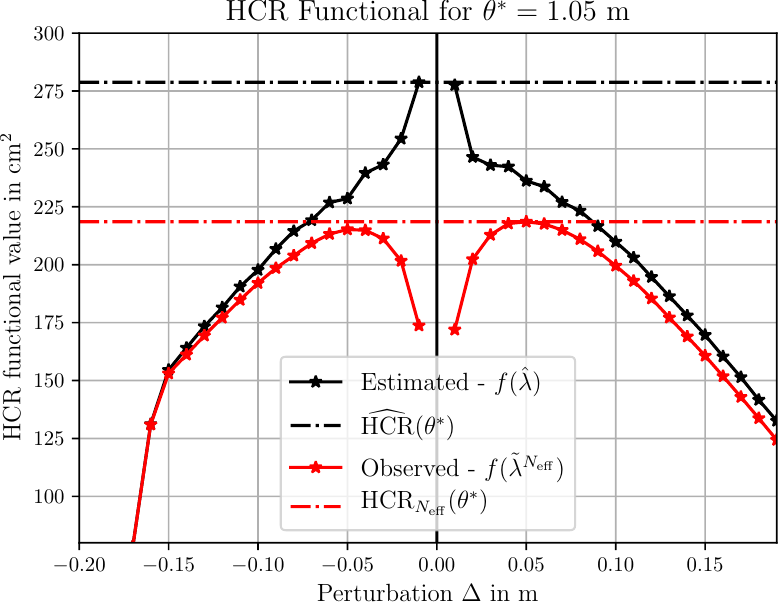}
		\label{fig:HCR_vs_delta_1}
	} 
    \subfloat[][]
	{
		\includegraphics[height=0.24\columnwidth]{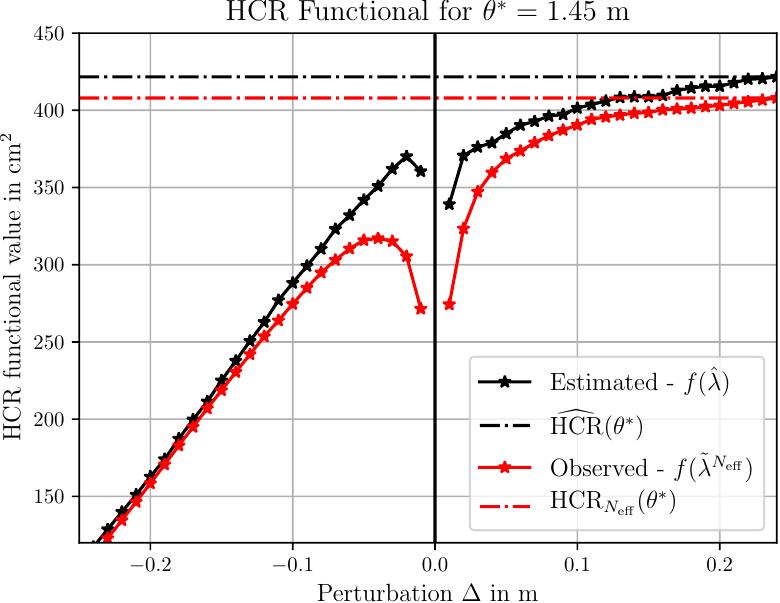}
		\label{fig:HCR_vs_delta_2}
	}
    \subfloat[][]
	{
		\includegraphics[height=0.24\columnwidth]{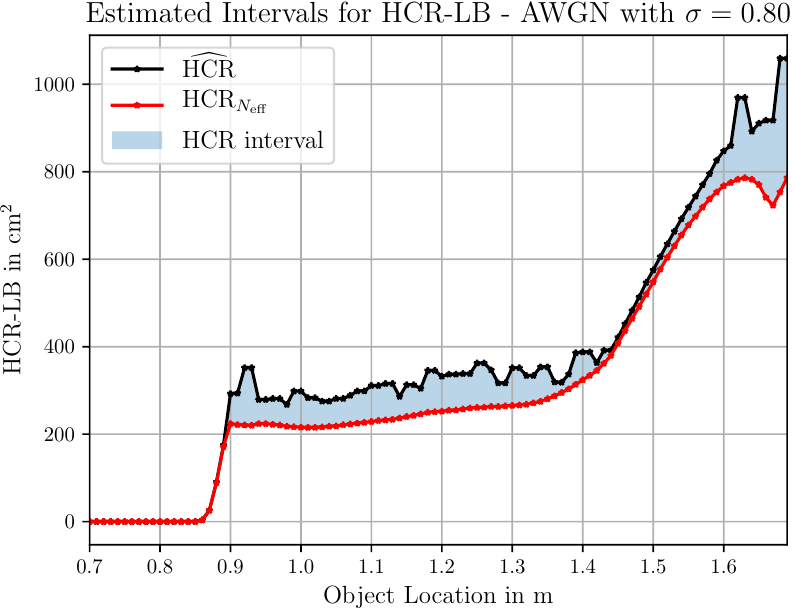}
		\label{fig:HCR_AWGN_0_8}
	} \\
	\subfloat[][]
	{
		\includegraphics[height=0.24\columnwidth]{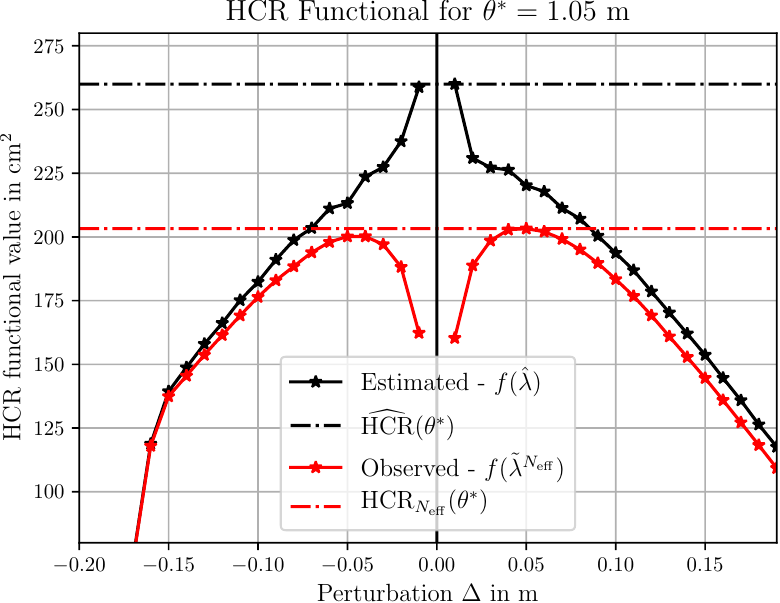}
		\label{fig:HCR_vs_delta_1P}
	} 
	\subfloat[][]
	{
		\includegraphics[height=0.24\columnwidth]{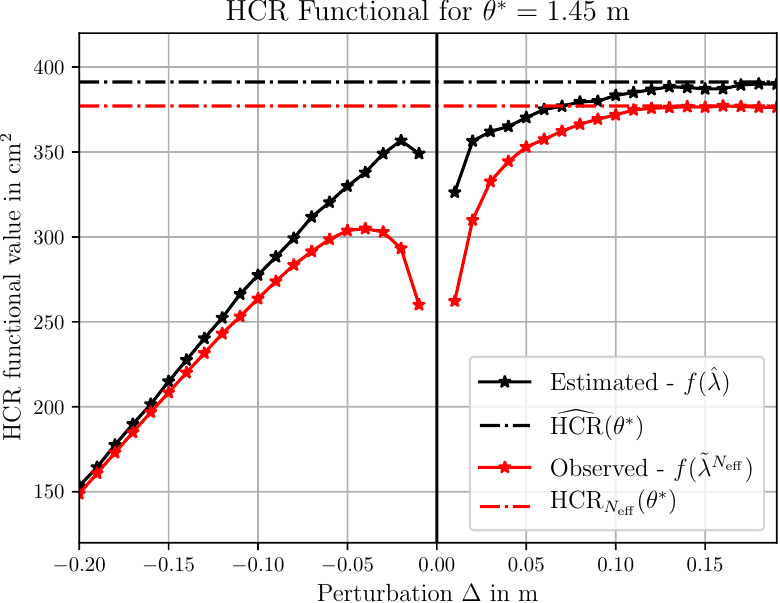}
		\label{fig:HCR_vs_delta_2P}
	} 
	\subfloat[][]
	{
		\includegraphics[height=0.24\columnwidth]{figures/NLOS_lb/Journal/bias_correction/HCR_Poisson.pdf}
		\label{fig:HCR_Poisson2}
	}
	\caption{Relationship between $\hat{\HCR}$ and $\HCR_{N_{\rm eff}}$ for different teapot locations: (a) - (c) for AWGN with $\sigma = 0.8$, (d) - (f) for Poisson Noise model. (a),(d) Maximum of the HCR functional occurs for $\Delta \rightarrow 0$ implying that $\hat{\HCR}$ is much larger than $\HCR_{N_{\rm eff}}$. (b),(e) Maximum of the HCR functional occurs for $ \| \Delta \| \gg 0$ implying that $\hat{\HCR}$ and $\HCR_{N_{\rm eff}}$ are approximately equal. (c),(f) The HCR-LB for $0.7{\rm m}\leq \btheta^* \leq 1.69{\rm m}$.}	
	\label{fig:HCR_vs_delta}
	\vspace{-1em}
\end{figure}

Figure \ref{fig:lambda_with_N} shows the effect of samples per-pixel $N$ on $\lambda$. As we increase $N$, we can see that $\tilde{\lambda}^N$ uniformly decreases across all values of $\bDelta$ for NLOS parameters, as predicted by our analysis. From Figure~\ref{fig:lambda_est}, we can see that $\tilde{\lambda}^{N_{\rm eff}} (\btheta^*, \bDelta) \leq \hat{\lambda} (\btheta^*, \bDelta) $ for all $\Delta$. Even though $\hat{\lambda}$ and $\tilde{\lambda}^{N_{\rm eff}}$ are close, Figure \ref{fig:HCR_func_example} shows that the difference in the HCR functional is significant around the neighborhood of $\bDelta = 0$, and the difference slowly tapers off for larger values of $\bDelta$. This implies that the overall HCR-LB will have larger uncertainty intervals when the maximum of the HCR functional occurs in the neighborhood of $\bDelta=0$ (see \Cref{fig:HCR_vs_delta_1,fig:HCR_vs_delta_1P}), and smaller intervals when the maximum occurs for $ \| \Delta \| \gg 0$ (see \Cref{fig:HCR_vs_delta_2,fig:HCR_vs_delta_2P}).

We can also see from Figure \ref{fig:HCR_AWGN} that the HCR intervals for AWGN model are relatively large for smaller values of $\sigma$ and decrease as we increase the noise level $\sigma$. This arises from the fact that as we increase $\sigma$, the location of the maximum of the HCR functional moves further away from $\bDelta = 0$, resulting in smaller HCR intervals as we see in Figure~\ref{fig:HCR_vs_delta}. Similar conclusions also hold for the HCR-LB under the Poisson noise model (see \Cref{fig:HCR_vs_delta_1P,fig:HCR_vs_delta_2P,fig:HCR_Poisson2}). 

From (\ref{eq:CR_lambda}), we can see that CR-LB is essentially the limit of the HCR functional for $\bDelta\rightarrow 0$. Then, Figure \ref{fig:HCR_vs_delta} would imply that the CR-LB computed with inexactly rendered data might have larger uncertainty intervals (in general) than the HCR-LB since, there could be instances where the supremum of the HCR functional is achieved away from the neighborhood of $\bDelta = 0$ in which case the uncertainty in the HCR-LB would be smaller (see \Cref{fig:HCR_vs_delta_2P,fig:HCR_vs_delta_2}).

The experimental results provided here validate our analysis of the effects of rendering errors on the lower bound computation and illustrate the utility of our HCR estimation framework outlined in Section \ref{subsec:HCR_estimation}, where we use multiple sets of erroneously rendered data with different numbers of samples per-pixel, to obtain intervals for the true HCR-LB.

\subsection{Numerical Results: Actual NLOS Object Localization using Maximum Likelihood Estimation, and Comparison to Lower Bound from Inexact Rendering}\label{subsec:MLE_exp}

While the HCR-LB provides lower bounds that are at least as tight as CR-LB, there are no guarantees on the existence of an unbiased estimator that achieves the lower bound. In order to show that the HCR-LB we derived here accurately depicts the true fundamental limits of scene parameter estimation, we compare our lower bounds assuming inexact rendering with the performance of the Maximum Likelihood (ML) Estimator. Maximum Likelihood Estimates (MLEs) are a good first choice because they are intuitive, simple to derive, \emph{asymptotically} unbiased (under modest assumptions), and are optimal for many simple estimation problems. While we make no claims about the optimality of MLEs, we show how they compare against the HCR-LB to give us a sense of how tight our lower bounds are. 

For concision, we consider only additive white Gaussian noise in this section. Under the additive white Gaussian noise model described in Section \ref{subsec:HCR_Gaussian}, we can obtain the MLE for $\btheta^*$, from noisy observations $\Yb_{\Omega}$ as, 
\begin{align}\label{eq:MLE_Gaussian}
	\hat{{\btheta}}_{\ML} (\Yb_\Omega) = \arg\min_{\btheta\in\Theta} \sum\limits_{\bomega\in \Omega}  \left(\Yb_{\bomega} - \Lb_{\btheta}(\bomega) \right )^2.
\end{align}
It is worth noting that ML estimation under Gaussian noise is equivalent to minimizing the $\ell_2$ loss. 

We use the optimization library in PyTorch \cite{pytorch} to solve (\ref{eq:MLE_Gaussian}), and obtain the ML estimates. In particular, we use the Adam optimizer \cite{adam}, where gradients of the loss function with respect to the scene parameters are obtained using finite differences (FD). 

\subsubsection{Setup}
\begin{figure}[]
	\centering
	\begin{tabular}{m{0.10\columnwidth} m{0.12\columnwidth} m{0.12\columnwidth} m{0.12\columnwidth} m{0.12\columnwidth}}
		 { \hspace{0em} \scriptsize $\btheta^*=0.8$m}
		& \hspace{0em} \scriptsize $\btheta^*=1.0$m
		& \hspace{0em} \scriptsize $\btheta^*=1.2$m
		& \hspace{0em} \scriptsize $\btheta^*=1.4$m
		& \hspace{0em} \scriptsize $\btheta^*=1.6$m\\
		 \includegraphics[width=0.18\columnwidth]{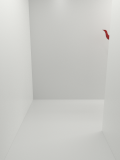} 
		& \includegraphics[width=0.18\columnwidth]{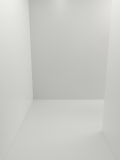} 
		& \includegraphics[width=0.18\columnwidth]{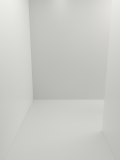} 
		& \includegraphics[width=0.18\columnwidth]{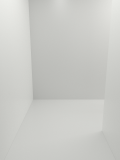} 
		& \includegraphics[width=0.18\columnwidth]{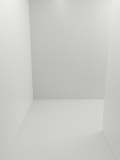} \\
		 \includegraphics[width=0.18\columnwidth]{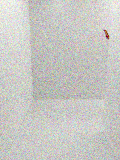} 
		& \includegraphics[width=0.18\columnwidth]{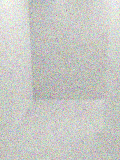} 
		& \includegraphics[width=0.18\columnwidth]{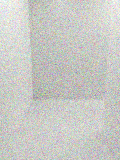} 
		& \includegraphics[width=0.18\columnwidth]{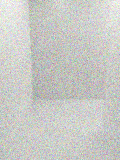} 
		& \includegraphics[width=0.18\columnwidth]{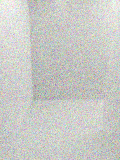} 		
	\end{tabular}
	\caption{\textbf{Top row:} Clean Images for different teapot locations rendered using 65536 samples per-pixel. 
		\textbf{Bottom row:} A single instance of noisy images corrupted by AWGN with $\sigma=0.1$. After the teapot goes completely out of LOS ($\btheta^*>0.9$m), it is very hard to discern any information about the teapot by simply looking at these images (both from clean and the noisy versions).}
	\label{fig:MLE_images}
	\vspace{-1em}
\end{figure}
We use the exact same setup used in Section \ref{sec:ideal_hallway_def}. We consider the problem of estimating the location of a red teapot that is constrained to lie in a straight line in Corridor B as shown in Figure \ref{fig:top_view_2}. The distance of the teapot from the intersection of corridors A and B is the \emph{scalar} parameter of interest $\btheta^*$. The camera is located at the center of corridor A as shown in Figure \ref{fig:top_view_2} and captures RGB images of size $160\times120$. We use {\tt Redner} \cite{redner} to render the true plenoptic observations using $N_{\rm eff} = 65536$ SPP and then synthetically add AWGN with $\sigma=0.1$ to generate noisy observations.  Figure \ref{fig:MLE_images} shows the rendered clean images and a single instance of noisy observations for 5 different teapot locations, $\btheta^* \text{ (in } {\rm m} \text{)} = 0.8, 1.0, 1.2, 1.4,$ and $1.5$. Except for $\btheta^*=0.8$m, all other locations correspond to the case where the teapot is completely out of LOS. 

\subsubsection{Results and Discussion}
We obtain the ML estimates by solving (\ref{eq:MLE_Gaussian}), where the derivative with respect to the $\ell_2$-loss is computed using FD. As opposed to the $65536$ samples per-pixel used render the clean (noiseless) images, we use low-SPP renderings to obtain fast (and noisy) gradients for the stochastic optimization procedure. We use $N=512$ SPP for the initial few iterations and then use $N=1024$ SPP towards the end. It is worth noting that it took approximately $3$ seconds and $6$ seconds to compute gradients with $512$ and $1024$ SPP, respectively, on an NVIDIA Quadro RTX 8000 GPU with 48GB of RAM. The overall runtime for (a single run of) the ML estimation algorithm varied from $28-40$ minutes depending the true location of the teapot. The algorithm converged faster when the teapot was closer to the beginning of the corridor.

\begin{figure}
	\centering
	\includegraphics[width=0.4\columnwidth]{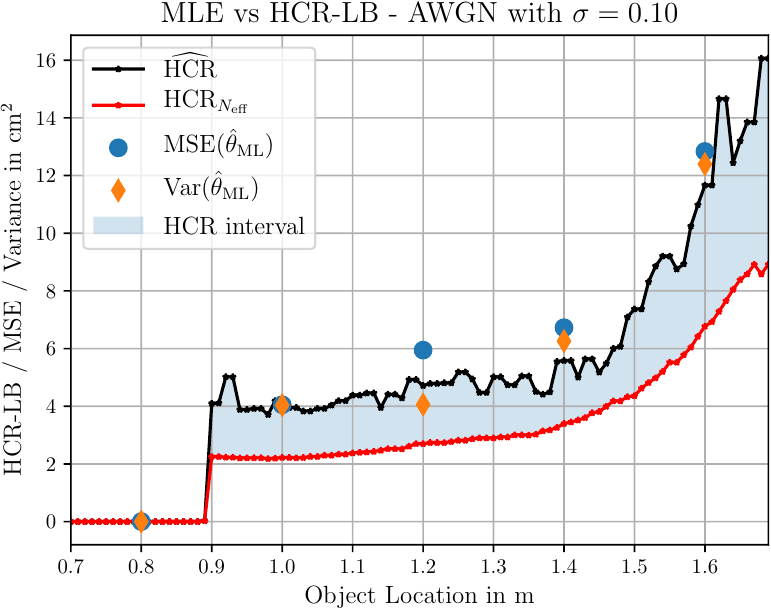}
	\captionof{figure}{Comparison of the HCR-LB and MLE for ball location estimation under AWGN with $\sigma=0.1$ shows that MLE achieves near-optimal performance for this problem and hence we conclude that our lower bounds are indicative of the true fundamental limits.}
	\label{fig:MLE_vs_HCR}
\end{figure}

We perform 30 independent runs of ML estimation, with different noise realizations, for each $\btheta^*$ to assess the performance of the MLE. The initialization point for every run was chosen uniformly at random in the interval $0.7{\rm m}$ to $1.7{\rm m}$. We report the average MSE and the variance of the ML estimates over the 30 runs along with the computed HCR-LB (both $\HCR_{N_{\rm eff}}$ and $\hat{\HCR}$) from the previous section, in Table \ref{tab:MLE_results}. Similar information is presented graphically in Fig.~\ref{fig:MLE_vs_HCR}; there, the estimated HCR interval (assuming inexact rendering) is also depicted.

\begin{table}[ht]
\centering
\renewcommand*{\arraystretch}{1.5}
\begin{tabular}{||c||c|c||c|c||}
			\hline
			$\btheta^*$ & $\HCR_{N_{\rm eff}} (\btheta^*)$ & $\hat{\HCR} (\btheta^*)$ 
			&  $\MSE( \hat{{\btheta}}_{\ML} )$ & $\Var( \hat{{\btheta}}_{\ML} )$ \\
			(in ${\rm cm}$) & (in ${\rm cm}^2$) & (in ${\rm cm}^2$) & (in ${\rm cm}^2$) & (in ${\rm cm}^2$) \\
			\hline
			$80$ & $4.17\times 10^{-214}$ & $1.30\times 10^{-216}$ &  $0.0024$ &  $0.0019$ \\
			\hline
			$100$ & $2.2201$ & $4.3282$ &  $4.0518$ &  $4.0372$  \\
			\hline
			$120$ & $2.6999$ & $4.9048$ &  $5.9466$ &  $4.0569$  \\
			\hline
			$140$ & $3.3964$ & $5.6872$ &  $6.7245$ &  $6.2610$  \\
			\hline
			$160$ & $6.7721$ & $11.6408$ &  $12.8361$ &  $12.3953$  \\
			\hline
		\end{tabular}
        \renewcommand*{\arraystretch}{1}
		\caption{Comparison of HCR lower bound and performance of MLE for AWGN with $\sigma=0.1$.}
		\label{tab:MLE_results}
\end{table}

First, it is worth noting that the average MSE and the variance of the MLEs are very close to each other for most values of $\btheta^*$. This implies that the ML estimator for this problem is (nearly) unbiased and hence the HCR-LB values are applicable to the ML estimates. While it can be seen that the average MSE is a bit higher than the variance for $\btheta^*=1.2$m, we believe that this difference would vanish as we perform more runs ($\gg30$) of the ML estimation. 

When the teapot is in LOS ($\btheta^*=0.8$m), we can see that the ML estimator (unsurprisingly) is able to estimate the object's location quite accurately. As the object translates further away from LOS the performance of the ML estimator decreases in tandem with the HCR-LB. Furthermore, the MSE (and the variance) of the MLEs are quite close to the HCR-LB intervals computed in the previous sections. Starting at random initializations from a $1$m wide interval, the ML estimates converge to within a few cm of the true teapot location. This is impressive and promising considering that it is nearly impossible to find any information about the hidden object directly by looking at the images in Figure \ref{fig:MLE_images}. On the other hand, the renderer-enabled ML estimator is able to make use of the information content in the very weak signals from the indirect photons to localize the object. 

While there are no guarantees about the existence of estimators that can achieve the HCR-LB for the general types of problems that can be addressed using our framework, we can see from this experiment that the performance of the ML estimator is indeed close to that of the HCR-LB. Also, the error rates of the MLEs exhibit similar behavior as a function of the true object location as the HCR-LB. This shows that the renderer-enabled HCR-LB framework proposed here yields lower bounds that seem to be indicative of the true underlying fundamental limits for scene parameter estimation problems in plenoptic imaging systems.

%% file: conclusion.tex
\section{Conclusion}\label{sec:conclusion}

We have presented here a framework to compute information-theoretic lower bounds for estimating scene parameters from noisy plenoptic data. Our approach employs the HCR-LB over the commonly used CR-LB, as the former is: (a) more amenable to our settings where computing partial derivatives (with respect to parameters of interest) might be infeasible, (b) applies to a wider range of problems, and (c) yields a bound that is at least as good as CR-LB when both bounds exist. 
Using computer graphics rendering packages, we overcome the difficulty of having to solve the forward model in closed form, and numerically evaluate the HCR-LB. 

Furthermore, we analyze the effects of rendering error on the computed HCR-LB and show that the rendering error typically introduces a bias in the computed HCR-LB values. In particular, we show that the HCR functional computed using erroneously rendered images underestimates the tightness of the bound, resulting in an overestimate of the true value especially for NLOS parameter estimation problems. We show that this bias vanishes at the rate of $\cO(N^{-1})$ for \emph{unbiased and progressive} renderers used here, where $N$ is the number of samples per-pixel used. Based on our error analysis, we also provide a simple method to estimate intervals for the true HCR-LB. Our error analysis automatically accounts for the error accrued in computing the CR-LB using finite differences (FD) and indicates that the uncertainty (or size) of the estimated intervals for the CR-LB would be at least as large as those for the HCR-LB. Thus, in addition to being at least as tight as the CR-LB in value, the HCR-LB is also at least as robust to rendering errors as the CR-LB. 

Our renderer-enabled lower-bounding framework has been used to compute lower bounds for a few illustrative NLOS imaging problems under two common noise models: Poisson noise and AWGN with different levels of noise variances. Additionally, we have computed and shown the pixel-wise finite difference Fisher information (FD-FI). This FD-FI data, while not always accessible when the log-likelihood is not differentiable, provides useful insights, especially for NLOS imaging problems, as they tell us which of the indirect photons/observations convey more information about the parameter(s) of interest. Additionally, the FD-FI data allows us to determine which spatial samples of the plenoptic function are most informative. We believe that these insights and tools can be used to develop novel adaptive sensing strategies for scene parameter estimation. Although we explore only spatial sampling of the plenoptic function in a classical RGB imaging system in this work, our estimation-theoretic framework readily generalizes to accommodate additional dimensions of the plenoptic function, e.g., polarization, motion in the scene, etc.

The potential benefits of our HCR-LB framework come at a cost of increased computational requirements. Computing the HCR-LB involves finding the supremum of the expression in (\ref{eq:HCR_lemma}) for general noise models. In this work, we compute the supremum by exhaustive search since the parameter space is small. However, if one is interested in problems with multiple parameters ($d\gg1$), the computational time for exhaustive search would increase exponentially making it prohibitive to apply in high-dimensional settings. For such settings, we could potentially explore either derivative free (zeroth-order) optimization methods, or recently developed differentiable renderers like Redner and \texttt{Mitsuba 3} \cite{redner, jakob2022mitsuba3} to compute derivatives and evaluate the HCR-LB, \Cref{eq:HCR-Poisson,eq:HCR_AWGN} using gradient-based optimization algorithms. However, as we have shown, the gradients generated by differentiable rendering engines are not always usable for this purpose, so care must be taken when selecting a gradient estimation method, and in some cases adaptive techniques must be used to avoid gradient artifacts. While it might be hard to obtain the global maxima of the HCR functional using gradient-based methods, convergence to any local maxima would yield a valid (but possibly loose) lower bound. It would be interesting to study the landscape of the HCR functional for such multi-parameter estimation problems. We defer these investigations to future work. 

Finally, we supplement our lower bounding framework by comparing the HCR-LB computed here with the performance of the Maximum Likelihood Estimator (MLE) for a simple but illustrative object localization problem. We see that the HCR-LB values closely match the behavior of the MLE, indicating that our framework is able to compute meaningful lower bounds that reflect the true fundamental limits of scene parameter estimation in plenoptic imaging systems. While (asymptotically) unbiased estimators like the MLE are useful in understanding the fundamental limits of parameter estimation, it is common to introduce bias into the estimates in the form of regularization to reduce the overall estimation error \cite{eldar2008rethinking}, especially in high-dimensional statistical inference. Generalizing our lower bounding framework for such \emph{biased} estimators would also be an interesting avenue for future work.

%% file: acknowledgment.tex
\section{Acknowledgment}
Mr. Liam Coulter acknowledges the support of the Raytheon Employee Scholars Program in pursuing his PhD. The authors also thank Prof. Gary Meyer and his students Prof. Michael Tetzlaff and Dr. Michael Ludwig, for useful discussions about ray-tracing algorithms and their help with the design of the $\Pi$-shaped hallway scene used in Section \ref{sec:rendering_errors}.

%% file: appendix.tex
\section{Appendix}

\subsection{Proof of Corollary \ref{cor:HCR_MSE}}
\label{proof:HCR_MSE}
	For unbiased estimators we have, $\MSE({\btheta}^*)  \triangleq \sum_{j=1}^{J} \Var(\btheta^*_j) $. We can further lower bound each of the variance terms in this summation using Lemma \ref{lemma:HCR} to obtain,
\begin{align*}
\MSE ({\btheta}^*) &\geq \sum\limits_{j=1}^{J} \sup \limits_{\substack{\bDelta\neq \mathbf{0} \\ \btheta^*+\bDelta\in\Theta}} \hspace{0.1em} \frac{\bDelta^2_j}{\EE_{\btheta^*}\sbr{\frac{p(\Yb_\Omega;\btheta^*+\bDelta)}{p(\Yb_\Omega;\btheta^*)} - 1}^2} \\
&\geq \sup \limits_{\substack{\bDelta\neq \mathbf{0} \\ \btheta^*+\bDelta\in\Theta}} \hspace{0.1em}  \frac{ \sum\limits_{j=1}^{J} \bDelta^2_j } {\EE_{\btheta^*}\sbr{\frac{p(\Yb_\Omega;\btheta^*+\bDelta)}{p(\Yb_\Omega;\btheta^*)} - 1}^2},
\end{align*}
where the last inequality follows from the fact that, $\sum\limits_j \sup\limits_x f_j(x) \geq \sup\limits_x \sum\limits_j f_j(x) $. \hfill \qedsymbol

\subsection{Empirical validation of Assumption \ref{as:decay_rate}}\label{app:var_validation}
We provide empirical evidence using simulations to validate Assumption \ref{as:decay_rate} here. For an example scene described below, we empirically study the relationship between the weighted sum of per-pixel variance of rendered images and the number of samples per-pixel $N$, used for rendering. It is worth noting that for \emph{unbiased} rendering algorithms, per-pixel variance of rendered images is equivalent to the per-pixel error variance.  i.e. $\Var( \Lb^{(N)} (\bomega) ) = \Var( \cE^{(N)} (\bomega) )$. Hence we use the variance of rendered pixel values in this section to empirically validate Assumption \ref{as:decay_rate}.

Our example scene for this simulation consists of a red teapot placed at the intersection of corridors A and B in the hallway scene described in Section \ref{sec:ideal_hallway_def}. 
We consider $11$ equally-spaced values of samples per-pixel $N$ ranging from $1024, 2048, \dots, 11264\ (= 11\cdot1024)$.
For each value of $N$, we render $100$ independent low resolution ($40\times 30$ pixel) RGB images of the same scene using \texttt{Redner} \cite{redner} and compute the per-pixel variance. 

\begin{figure}[h]
	\centering
	\includegraphics[width=0.8\columnwidth]{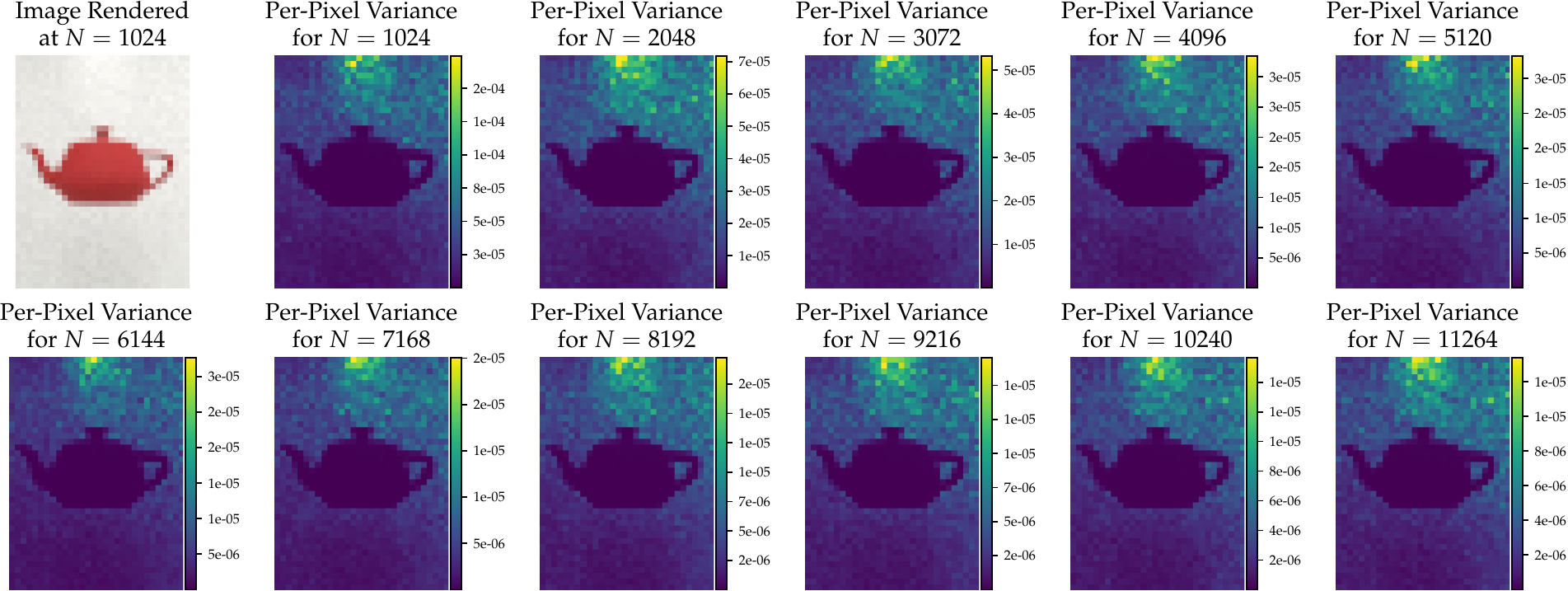}
	\caption{Top left: A single instance of the teapot image rendered with $1024$ samples. Other plots: Per-pixel variance (summed over the $3$ color channels) for the teapot image for different values of samples per-pixel $N$. It can be seen that per-pixel variance is not same across all pixels in the image. While the general pattern of pixel-wise variance is similar across different values of $N$, the magnitude of the variance decreases (as expected) with increasing samples.}	
	\label{fig:var_images}
\end{figure}

The first image (top-left) in Figure \ref{fig:var_images} shows a single instance of the rendered RGB scene with $N = 1024$ samples per-pixel. The other $11$ images in Figure \ref{fig:var_images} show the per-pixel variance (summed over the $3$ color channels) of the rendered images; note that the color bar magnitude limits vary across all the images.  The per-pixel variance is not uniform across the entire image: some regions of the image have smaller variance (pixels corresponding to the teapot) than others (pixels on the back wall above the teapot). However, we can also observe the general trend that pixel variances consistently decrease with increasing values of $N$. 

We consider weights $W_{\bomega} \sim \text{Uniform}\:[0,{\rm L_{max}}]$, where we set $\rm L_{max} = 12$, which is the radiance value of the lights in the scene. It is worth mentioning that the value of $\rm L_{max}$ was observed to have little to no effect on the results of this simulation. For different values of $N$, we then compute $\gamma^{(N)} := \sum\limits_{\bomega\in\Omega} W_{\bomega} \cdot \Var( \Lb^{(N)} (\bomega) )$. In order to find the exact rate of decay of $\gamma^{(N)}$ with respect to $N$, we fit parametric models given by, $\gamma^{(N)} = C_p\cdot N^{-p}$ for values of $p$ between $0.85$ and $1.15$ (in $0.001$ increments) and choose the value of $p$ that has the smallest $\ell_2$ error of fit, 
$$ p_{\rm opt} = \arg \min\limits_{0.85\leq p \leq 1.15} \| \gamma^{(N)} - C_p\cdot N^{-p} \|^2_2. $$

\begin{figure}[]
	\centering
	\subfloat[][]
	{
		\includegraphics[height=0.24\columnwidth]{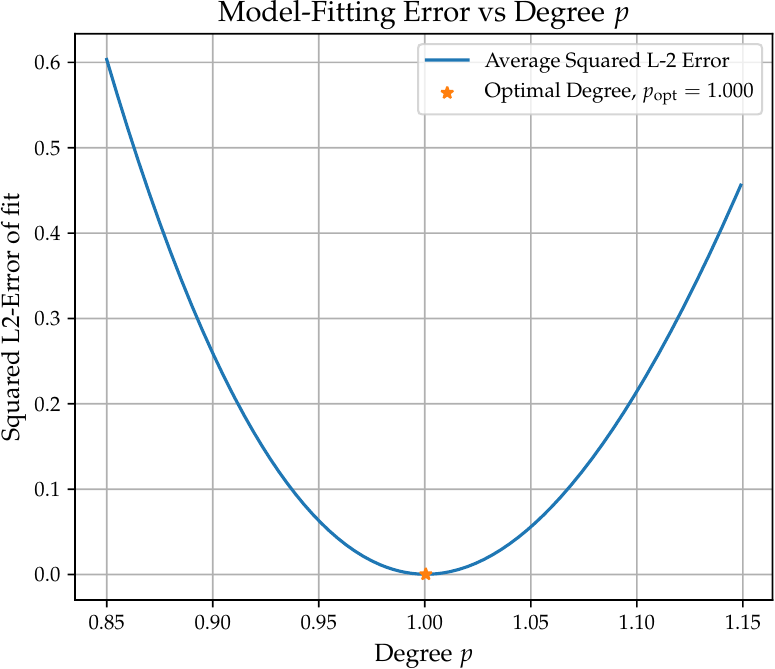}
		\label{fig:fit_vs_p}
	}
	\subfloat[][]
	{
		\includegraphics[height=0.24\columnwidth]{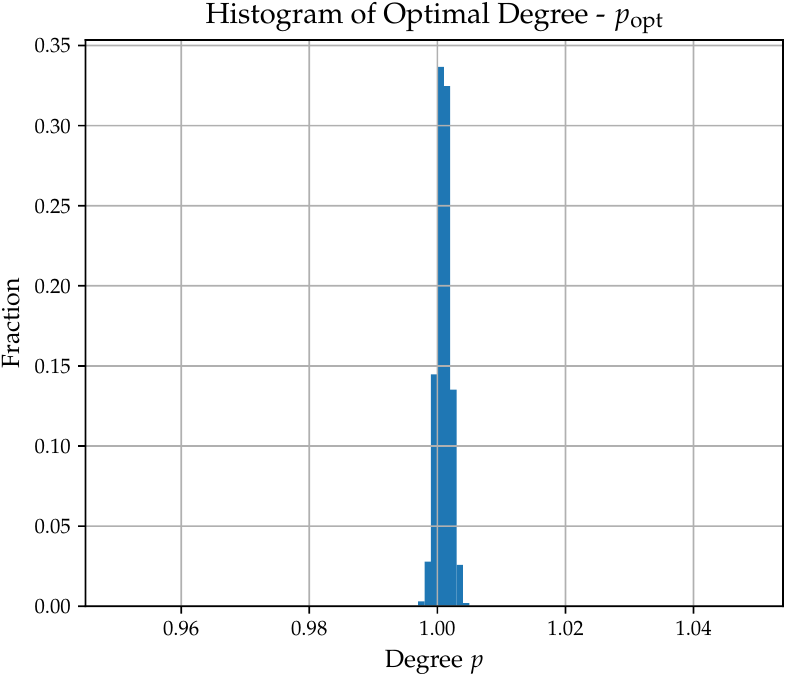}
		\label{fig:histogram_var_decay}
	} 
	\subfloat[][]
	{
		\includegraphics[height=0.24\columnwidth]{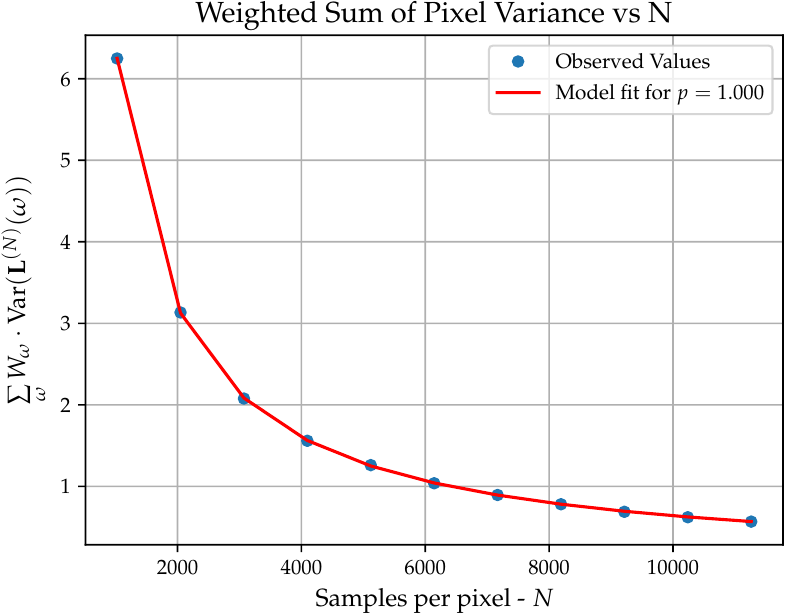}
		\label{fig:var_LS_fit}
	}
	\caption{Results from the simulation with $10^4$ independent draws of weights $W_{\bomega}$: (a) Average Squared L2-Error of fit vs Degree $p$; (b) Distribution of the optimal degree $p_{\rm opt}$ shows that most of it is concentrated around $p=1$; (c) A single instance of weighted sum of pixel variance ($\gamma^{(N)}$) along with the model fit using $p=1$.}	
	\label{fig:var_decay_results}
\end{figure}

We repeat the above process with $10^4$ independent draws of $W_{\bomega}$ to find the distribution of $p_{\rm opt}$. Figure \ref{fig:fit_vs_p} shows that the average fitting error is smallest for $p=1$, with an average squared $\ell_2$ error $=3.154\times 10^{-4}$. Furthermore, we can see from the distribution of $p_{\rm opt}$ in Figure \ref{fig:histogram_var_decay} that $p_{\rm opt}$ is highly concentrated around $p=1$. The mean, median and the mode of the distribution all occur at $p=1$. Finally, Figure \ref{fig:var_LS_fit} shows a single instance of $\gamma^{(N)}$ and the corresponding model fit using $p=1$, which shows how well the parametric model of $C_p\cdot N^{-1}$ fits the observed weighted sum of pixel variances. 

These simulations show that the weighted sum of pixel variances follow a $\Theta(N^{-1})$ parametric decay rate with high probability, thus validating Assumption \ref{as:decay_rate}. 

\subsection{Proof of Theorem \ref{claim:Poisson}}\label{proof:Poisson_claim}
Let us denote the true plenoptic intensities for parameter values $\Lb_{\btheta^*},\ \Lb_{\btheta^*+\bDelta}$ by $\Lb_1, \ \Lb_2$ respectively, for clarity and brevity of notation. 
We can then describe the corresponding (inexactly) rendered plenoptic values using $N$ samples per-pixel as, $\tilde{\Lb}^{(N)}_1 = \Lb_1 + \cE_1$ and $\tilde{\Lb}^{(N)}_2 = \Lb_2 + \cE_2$, where the dependence of the rendering noise on $N$ is implicit.

For the case of the HCR lower bound under Poisson noise (Section \ref{subsec:HCR_Poisson}), if we use the rendered plenoptic values, we end up with erroneous estimation of $\lambda_P$, given by
\begin{align}
\nonumber
\tilde{\lambda}_P &= \sum\limits_{\bomega\in\Omega} \frac{\left( \tilde{\Lb}_1^{(N)}(\bomega) - \tilde{\Lb}_2^{(N)}(\bomega) \right)^2}{\tilde{\Lb}_1(\bomega) } \\
\label{eq:tilde_lambda_P}
&=  \sum\limits_{\bomega\in\Omega} \frac{\left( \Lb_1(\bomega) - \Lb_2(\bomega) + \cE(\bomega) \right)^2}{ \Lb_1(\bomega) +\cE_1(\bomega) }, 
\end{align}
where we let $\cE := \cE_1 - \cE_2$. If the errors in the rendered image are \emph{relatively small}, i.e. $ |\cE_1(\bomega)| \ll \Lb_1(\bomega) $ for all $\bomega\in\Omega$, then we can use the series expansion to write 
\begin{align} \label{eq:approx}
\frac{1}{\Lb_1(\bomega) +\cE_1(\bomega)} =  \frac{1}{\Lb_1(\bomega)} \left(1 -  \frac{ \cE_1(\bomega) }{ \Lb_1(\bomega) } + \frac{ \cE_1(\bomega)^2 }{ \Lb_1(\bomega)^2 }  - \dots\right).
\end{align}
Substituting (\ref{eq:approx}) in (\ref{eq:tilde_lambda_P}), we get
\begin{align}
\label{eq:lambda_P_1}
\nonumber
\tilde{\lambda}_P &=  \sum\limits_{\bomega\in\Omega}  \left( \frac{\left\{ \big(\Lb_1(\bomega) - \Lb_2(\bomega)\big) + \cE(\bomega)  \right\} ^2}{ \Lb_1(\bomega) } \right) \cdot \left(1 -  \frac{ \cE_1(\bomega) }{ \Lb_1(\bomega) } + \frac{ \cE_1(\bomega)^2 }{ \Lb_1(\bomega)^2 }  - \dots\right)
\\\nonumber
& = \lambda_P 
+ \underbrace{\sum\limits_{\bomega\in\Omega}  \left( \frac{ \big(\Lb_1(\bomega) - \Lb_2(\bomega)\big) ^2}{ \Lb_1(\bomega) } \right) \sum\limits_{k=1}^{\infty} \left( \frac{ -\cE_1(\bomega) }{\Lb_1(\bomega) } \right)^k}_{ (\RNum{1}) } 
\\
& \hspace{5em} + \underbrace{\sum\limits_{\bomega\in\Omega}  \left( \frac{\left( \cE(\bomega)^2 + 2\:\big(\Lb_1(\bomega) - \Lb_2(\bomega)\big)\cE(\bomega) \right) }{ \Lb_1(\bomega) } \right) \sum\limits_{k=0}^{\infty} \left(\frac{-\cE_1(\bomega) }{\Lb_1(\bomega) } \right)^k}_{ (\RNum{2}) }.
\end{align}
If we further take a closer look at the term $(\RNum{1})$, we can see that
\begin{align}
\label{eq:mean_lambda_P_11}
\EE \left[ (\RNum{1}) \right] 
& = \sum\limits_{\bomega\in\Omega} \cbr{  \frac{ \left(\Lb_1(\bomega) - \Lb_2(\bomega) \right)^2}{ \Lb_1(\bomega)^3 }  \Var ( \cE_1(\bomega) )} +\  \cO(N^{-(1+\delta)}),
\end{align} 
where we have used the unbiasedness (Assumption \ref{as:unbiased}) and the bound on higher-order moments (Assumption \ref{as:bdd_moments}) of the rendering errors to get (\ref{eq:mean_lambda_P_11}). Using similar arguments, we can show that
\begin{align}
\EE \left[ (\RNum{2}) \right] 
&= \sum\limits_{\bomega\in\Omega} \left\{ \frac{1}{\Lb_1(\bomega)} \Var ( \cE(\bomega) )  - \frac{2 ( \Lb_1(\bomega) - \Lb_2(\bomega) ) }{\Lb_1(\bomega)^2} \EE [ \cE(\bomega) \cE_1(\bomega) ]  \right\} +\  \cO(N^{-(1+\delta)}) \nonumber 
\\
& = \sum\limits_{\bomega\in\Omega} \Bigg\{ \frac{\Var(\cE_1(\bomega) ) + \Var(\cE_2(\bomega) ) }{ \Lb_1(\bomega) }  - \frac{2 ( \Lb_1(\bomega) - \Lb_2(\bomega) ) }{\Lb_1(\bomega)^2} ( \Var (\cE_1(\bomega) ) - \EE [ \cE_1(\bomega) \cE_2(\bomega) ] ) \Bigg\} +\ \cO(N^{-(1+\delta)}) \label{eq:mean_lambda_P_22}
\\
& = \sum\limits_{\bomega\in\Omega} \left\{ \rbr{ \frac{ 2\Lb_2(\bomega) - \Lb_1(\bomega) }{\Lb_1(\bomega)^2} }  \Var ( \cE_1(\bomega) ) +\ \frac{\Var ( \cE_2(\bomega) )}{ \Lb_1(\bomega) } \right\} +\  \cO(N^{-(1+\delta)}), \label{eq:mean_lambda_P_23}
\end{align}
where we have used the fact that $\cE_1$ and $\cE_2$ are independent (Assumption \ref{as:indep_errors}) and zero-mean (Assumption \ref{as:unbiased}) to get $ \Var(\cE(\bomega) ) = \Var(\cE_1(\bomega) ) + \Var(\cE_2(\bomega) )$ and $\EE [ \cE_1(\bomega) \cE_2(\bomega) ] = \EE [ \cE_1(\bomega)]\cdot \EE [ \cE_2(\bomega)] = 0 $. Combining \Cref{eq:mean_lambda_P_11,eq:mean_lambda_P_23}, we get
\begin{align}
\EE \left[ (\RNum{1}) + (\RNum{2})\right] 
&= \sum\limits_{\bomega\in\Omega} \cbr{  \frac{ \Lb_2(\bomega)^2}{ \Lb_1(\bomega)^3 }  \Var (\cE_1(\bomega)) + \frac{\Var ( \cE_2(\bomega) )}{ \Lb_1(\bomega) } }  +\  \cO(N^{-(1+\delta)})
\nonumber \\ 
&= \frac{C_P}{N} \ +\  \cO(N^{-(1+\delta)}), \label{eq:mean_lambda_P_3}
\end{align}
for some scene-dependent constant $C_P\geq0$ and $\delta>0$, which is the constant appearing in Assumption \ref{as:bdd_moments}. Equation (\ref{eq:mean_lambda_P_3}) follows directly from Assumption \ref{as:decay_rate} about the rate of decay of weighted sums of pixel-wise error variance. 

Now we have the variance of $ (\RNum{1}) + (\RNum{2}) $, 
\begin{align*}
\Var \rbr{ (\RNum{1}) + (\RNum{2}) } = \EE \sbr{ \rbr{ (\RNum{1}) + (\RNum{2}) }^2} -  \cbr{ \EE \sbr{ (\RNum{1}) + (\RNum{2}) } }^2. 
\end{align*}
From (\ref{eq:lambda_P_1}), we can rewrite
\begin{align}
\label{eq:term_(1)}
(\RNum{1}) = \sum\limits_{\bomega\in\Omega} \frac{c_0(\bomega) \cE_1(\bomega) }{ \Lb_1(\bomega) + \cE_1(\bomega) },
\end{align} 
where we let $c_0(\bomega) := \frac{ \big(\Lb_1(\bomega) - \Lb_2(\bomega)\big) ^2}{ \Lb_1(\bomega) } $, and
\begin{align}
\label{eq:term_(2)}
(\RNum{2}) = \sum\limits_{\bomega\in\Omega} \frac{\left( \cE(\bomega)^2 + 2\:\big(\Lb_1(\bomega) - \Lb_2(\bomega)\big)\cE(\bomega) \right) }{ \Lb_1(\bomega) + \cE_1(\bomega) }.
\end{align} 
\sloppy If we use \Cref{eq:term_(1),eq:term_(2)} and the series expansion of $\cfrac{1}{ (\Lb_1(\bomega) + \cE_1(\bomega))^2 }$ to evaluate $\EE \sbr{ \rbr{ (\RNum{1}) + (\RNum{2}) }^2} = \EE \sbr{ (\RNum{1})^2 +(\RNum{2})^2 + 2\cdot(\RNum{1})\cdot(\RNum{2}) }$, we get
\begin{align}
\nonumber 
\Var \rbr{ (\RNum{1}) + (\RNum{2}) } &=  \sum\limits_{\bomega\in\Omega} \cbr{ c_1(\bomega) \Var( \cE_1 )  +  c_2(\bomega) \Var( \cE_2 )} +\ \cO(N^{-(1+\delta)}) - \cbr{ \EE \sbr{ (\RNum{1}) + (\RNum{2}) } }^2
\\ 
&= \cO(N^{-1}). \label{eq:var_lambda_P_2}
\end{align}
where $c_1(\bomega)$ and $c_2(\bomega)$ are some scene-dependent constants, and the final equality follows from Assumption \ref{as:decay_rate} and (\ref{eq:mean_lambda_P_3}). Combining \Cref{eq:mean_lambda_P_3,eq:var_lambda_P_2} with (\ref{eq:lambda_P_1}) we get the desired result in Theorem \ref{claim:Poisson}, thus completing the proof. \hfill \qedsymbol

\subsection{Proof of Theorem \ref{claim:Gaussian} }\label{proof:Gaussian_claim}
We use similar arguments as above to prove Theorem \ref{claim:Gaussian}. For the case of the HCR lower bound under AWGN noise (Section \ref{subsec:HCR_Gaussian}), if we use the rendered plenoptic values, we end up with erroneous estimation of $\lambda_G$, given by
\begin{align}
\label{eq:lambda_G_1}
\tilde{\lambda}_G &= \lambda_G + 
\underbrace{ \frac{1}{\sigma^2} \| \cE_1 - \cE_2 \|^2}_{\eta_1} + 
\underbrace{ \frac{2}{\sigma^2} (\Lb_1 - \Lb_2 )^T \cdot (\cE_1 - \cE_2 ) }_{\eta_2}.
\end{align}
If we further examine $\eta_1$, we can see that
\begin{align}
\nonumber
\EE [\eta_1]  &= \sum\limits_{\bomega\in\Omega} \frac{\Var ( \cE_1(\bomega) ) + \Var ( \cE_2(\bomega) )}{ \sigma^2} - \frac{2}{\sigma^2} \underbrace{\EE [\cE_1(\bomega) \cE_2(\bomega)]}_{=0} 
\\ \label{eq:mean_eta1}
&= \frac{C_G}{N},
\end{align}
for some scene-dependent constant $C_G\geq 0$, where the last equality follows from Assumption \ref{as:decay_rate}. Also, due to the unbiasedness of rendering errors, it is easy to see that $\EE [\eta_2] = 0$. Thus we have
\begin{align} \label{eq:mean_eta_G}
\EE [\eta_1 +\eta_2] = \frac{C_G}{N}.
\end{align}
As in the previous proof, we denote $\cE := \cE_1 -\cE_2$, and calculate the variance of $\eta_1 + \eta_2$, 
\begin{align}
\nonumber
\Var (\eta_1 + \eta_2) &= \EE [ (\eta_1 + \eta_2)^2 ] - \cbr{ \EE [\eta_1+\eta_2] }^2
\\\nonumber
&= \frac{1}{\sigma^4}\underbrace{\cbr{ \EE \sbr{ \| \cE \|^4} + 4 (\Lb_1 - \Lb_2 )^T \cdot \EE \sbr{ \| \cE \|^2\cdot \cE  }} }_{=\cO(N^{-(1+\delta)} )} + \underbrace{\frac{4}{\sigma^4} (\Lb_1 - \Lb_2 )^T \cdot \EE[\cE \cE^T] \cdot(\Lb_1 - \Lb_2 )}_{=\cO(N^{-1})} -\ \frac{C_G^2}{N^2}
\\
&= \cO(N^{-1}), \label{eq:var_lambda_G}
\end{align}
where we use Assumption \ref{as:bdd_moments} to bound the higher order moments of the rendering error by $\cO(N^{-(1+\delta)})$, and the independence of $\cE_1$ and $\cE_2$, and Assumption \ref{as:decay_rate} to get a handle on $\EE[\cE\cE^T]$. Combining \Cref{eq:mean_eta_G,eq:var_lambda_G} with (\ref{eq:lambda_G_1}), we get the desired result in Theorem \ref{claim:Gaussian}, thus completing the proof. \hfill \qedsymbol

\subsection{Proof of Claim \ref{claim:HCR_estimate}}
\label{proof:HCR_estimate}
We have
\begin{align}\label{eq:upper_bound_HCR}
\nonumber
\EE \sbr{ \hat{\HCR}(\btheta^*_j) } &= \EE \sbr{ \sup \limits_{\bDelta\neq \mathbf{0}} f\rbr{ \hat{\lambda}; \bDelta, \btheta^*} }\\
&\geq \sup \limits_{\bDelta\neq \mathbf{0}} \EE \sbr{ f\rbr{\hat{\lambda}; \bDelta, \btheta^*} },
\end{align}
where we use the fact that $\EE_X\sbr{\sup\limits_y G(X,y)} \geq \sup\limits_y \EE_X \sbr{ G(X,y) }$. Next we observe that the HCR functional $f( \lambda ) := \frac{\bDelta^2_j}{  \exp \left(  \lambda \right) -1}$ is a convex function of $\lambda$. Hence for every $\btheta^*$ and $\bDelta$, we can apply Jensen's inequality  to get,
\begin{align}\label{eq:proof:Jensens}
\hspace{-0.3em} \EE \sbr{ f\rbr{\hat{\lambda}; \bDelta, \btheta^*} } \geq  f \rbr{ \EE \sbr{ \hat{\lambda} } ; \bDelta, \btheta^* } 
= f\rbr{  \lambda ; \bDelta, \btheta^* }\hspace{-0.1em}, 
\end{align}
where the last equality in (\ref{eq:proof:Jensens}) follows from the fact that $\hat{\lambda}$'s are unbiased estimates. Combining \Cref{eq:upper_bound_HCR,eq:proof:Jensens} yields the desired result, thus concluding the proof. \hfill \qedsymbol

\subsection{Supplementary Video Results: Visualizing the Pixelwise Fisher Information and the HCR-LB}
\label{vid:HCR-LB}
For the ``ball location'' and ``ball radius'' estimation problems described in Section \ref{subsec:HCR_results_exact}, we provide a video showing how the HCR-LB and FD-FI under \emph{AWGN} evolve as the ball moves along the hallway, into the room, and through the room.

\vspace*{1.5em}
\noindent
{\bf
	Video file names:\\ {\tt HCR\_position\_video\_full.gif}
    \\ {\tt HCR\_radius\_video\_full.gif}
}

\vspace*{1.5em}
\noindent
{\bf Caption for videos:} The four panels, clockwise from the top left, correspond to; (Top-left) Layout of the hallway scene showing the true ball location, location of lights, and square root of the HCR-LB for the ball location and radius as a black circle around the ball, respectively. (Top-middle) Nominal (or noiseless) RGB image of the hallway rendered by {\tt Mitsuba 3}, showing what an RGB camera would \emph{ideally} see. (Top-right) Log of pixelwise finite difference-Fisher information (FD-FI) for the ball location or radius parameter, respectively. (Bottom) Log of HCR-LB (in $\log (\textrm{cm}^2)$) as a function of the ball location or radius, respectively. 

These videos demonstrates the utility of our renderer-enabled framework to compute scene parameter estimation lower bounds. The FD-FI images help us localize the ``interesting regions" in the scene and quantify the information content in indirect photons.